\documentclass[10pt,a4paper]{article}

\usepackage{jheppub}

\usepackage[utf8]{inputenc}
\usepackage[T1]{fontenc}
\usepackage{amsmath,amssymb,amsthm,amsfonts}
\usepackage{mathtools}
\usepackage{empheq}
\usepackage{bm}
\usepackage{hyperref}
\usepackage{xcolor}
\usepackage{enumitem}
\hypersetup{colorlinks=true, linkcolor=blue!60!black, citecolor=blue!60!black, urlcolor=blue!60!black}

\usepackage{tikz}
\usetikzlibrary{positioning,calc}

\newcommand{\R}{\mathbb{R}}

\newcommand{\E}{\mathbb{E}}
\newcommand{\Prob}{\mathbb{P}}
\newcommand{\Var}{\operatorname{Var}}
\newcommand{\Cov}{\operatorname{Cov}}
\newcommand{\sgn}{\operatorname{sgn}}
\newcommand{\erf}{\operatorname{erf}}

\newcommand{\Dz}{\mathcal{D}z}
\newcommand{\Drep}{\mathcal{D}}
\newcommand{\Rcal}{\mathcal{R}}
\newcommand{\Hcal}{\mathcal{H}}
\newcommand{\Zcal}{\mathcal{Z}}
\newcommand{\Acal}{\mathcal{A}}
\newcommand{\Bcal}{\mathcal{B}}
\newcommand{\Scal}{\mathcal{S}}
\newcommand{\Ecal}{\mathcal{E}}
\newcommand{\avg}[1]{\langle #1 \rangle}

\theoremstyle{plain}
\newtheorem{theorem}{Theorem}

\theoremstyle{definition}

\theoremstyle{remark}
\newtheorem{remark}[theorem]{Remark}

\allowdisplaybreaks

\title{Semi-supervised Hopfield model: Theoretical and Numerical results}
\author[1,a,b]{Linda Albanese} %
\author[2,a]{, Andrea Ladiana}%
\author[3,a]{, Andrea Lepre}
\affiliation[1]{Dipartimento di Matematica e Fisica "Ennio De Giorgi", Università del Salento, Italy}
\affiliation[2]{Dipartimento di Scienze di Base e Applicazioni all'Ingegneria, Sapienza Università di Roma, Italy}
\affiliation[3]{Dipartimento di Matematica, Sapienza Università di Roma, Italy}
\affiliation[a]{Istituto Nazionale di Alta Matematica Francesco Severi (INdAM), Roma, Italy}
\affiliation[b]{Istituto Nazionale di Fisica Nucleare, Sezione di Lecce, Italy.}

\abstract{In the daily practice of Machine Learning, fully labeled datasets are a luxury: labels demand expensive and time-consuming human annotation, whereas raw, unlabeled data can be harvested automatically and in bulk. Semi-supervised learning, where the network jointly exploits the few labeled and the many unlabeled examples at its disposal, is the standard answer to this asymmetry, yet a statistical mechanical theory of semi-supervised Hebbian learning is still lacking. In this paper we fill this gap for the Hopfield network: we prescribe a synaptic coupling given by the convex combination, weighted by a mixing parameter $\lambda \in [0,1]$, of the supervised and unsupervised Hebbian kernels built from the same archetypes, and we solve for the emergent computational capabilities of the resulting network.

A signal-to-noise analysis yields the one-step Mattis magnetization and the learning threshold, i.e. the minimum dataset size for stable retrieval. Using Guerra’s interpolation, we then derive the Replica Symmetric quenched pressure in the high-storage regime, treating the correlated disorder generated by the supervised and unsupervised channels through a particular eigen-channel decomposition. The resulting phase diagram shows that a mixed strategy outperforms both pure protocols. Finally, we prove that the quenched pressure is convex in $\lambda$, so thermodynamics cannot select an interior mixture: $\lambda$ is therefore a learning hyperparameter. 

All the analytical findings are successfully checked against extensive Monte Carlo simulations.}

\begin{document}
\maketitle

\section{Introduction}
In modern Machine Learning, labels are the expensive ingredient: annotating a dataset requires human expertise, time and money, whereas unlabeled data can be collected automatically and in massive quantities. As a consequence, the datasets a learning machine actually experiences are almost never fully labeled, nor completely unlabeled, and the standard answer of the Machine Learning community to this structural asymmetry is \emph{semi-supervised learning}, where the (few) labeled examples and the (many) unlabeled ones are jointly
exploited during training \cite{chapelle2006semi, vanengelen2020survey}, from
entropy-based regularization \cite{grandvalet2005semi} to deep generative
modeling \cite{kingma2014semi}. Despite its ubiquity in applications, this paradigm still lacks a transparent,
analytically solvable model within the Hebbian framework of associative
memories, where the interplay between the two channels of information is under
full mathematical control: providing such a model, and solving it, is the purpose
of the present paper. 

The natural framework for this program is the statistical mechanics of neural networks \cite{Amit, MPV}, which we now briefly outline.

Since Hopfield's seminal work on biologically inspired associative memories \cite{Hopfield}, the statistical mechanics of spin glasses \cite{ sherrington1975solvable, MPV} has proved to be a privileged route toward a quantitative understanding of information processing in neural networks \cite{Amit, Coolen, nishimori2001statistical, hertz1991introduction}. The theory developed in the eighties by Amit, Gutfreund and Sompolinsky (AGS) \cite{AGS} is a milestone in this field, which turned the Hopfield model, a fully-connected mean-field network whose pairwise couplings implement the celebrated Hebbian prescription, into a solvable statistical mechanical system, whose phase diagram sharply splits the space of the control parameters (noise level and storage load) into regions with different emergent computational capabilities.

Since then, the Hopfield model has become the fundamental \textit{associative memory} network, spawning countless variations on theme, including its dense generalizations with many-body couplings \cite{HopKro1, Krotov2018, Baldi, Gardner, Densesterne, Bovier, Fachechi1, Albanese2021}.

However, in its original formulation, the Hopfield network does not really \emph{learn}: the patterns are experienced just once and directly stored in the synaptic matrix, while in Machine Learning jargon \emph{learning} refers to the ability of the machine to build its own representation of unknown archetypes out of a corpus of their noisy examples. In recent years this conceptual gap has been bridged by turning the Hebbian storing prescription into genuine Hebbian \emph{learning} rules, split into a supervised setting, where a teacher groups the examples class by class, and an unsupervised one, where the network experiences the whole dataset with no label at all: see in particular \cite{prlmiriam, EmergencySN, AgliariDeMarzo} and \cite{albanese2024hebbian} for a derivation of these prescriptions from first principles via maximum entropy. Within these frameworks one can compute, beyond the storage capacity, the \emph{threshold for learning}, that is the minimal number of examples per archetype allowing the network to infer, and successively retrieve, the archetypes hidden in the dataset: conceptually, this is the closest statistical mechanical analogue of the complexity bounds of the sample of statistical learning theory \cite{shalev2014understanding}.

So far, however, the supervised and unsupervised protocols have essentially been investigated separately, one at a time, while, as stressed above, the operationally relevant scenario is the mixed one. Semi-supervision has of course not escaped the attention of the
statistical mechanics community, but it has been addressed along different routes.
In the framework of on-line learning, the generalization error of a semi-supervised
protocol has been obtained through a set of deterministic differential equations
for the order parameters, which connect continuously the supervised and the
semi-supervised prescriptions \cite{fujii2017statistical}, albeit for a
teacher-student perceptron trained by gradient-based rules.

A second, by now classical line of research addresses semi-supervised
\emph{clustering} and community detection: there the graph itself is the datum and
a fraction of the node labels is revealed to the observer, so that the spin glass
machinery delivers the distribution of the classifications rather than the sole
minimal cut \cite{getz2006semi}, the free energy of the constrained problem
together with the fate of its criticality \cite{versteeg2011statistical}, the
detectability thresholds of the semi-supervised stochastic block model
\cite{zhang2014phase} and the large deviations associated to the choice of the
revealed subset \cite{cui2022large}. Hopfield networks themselves have been
employed in this very spirit, the couplings encoding the similarity among the
items of a given dataset and the dynamics, restricted to the unlabeled nodes,
propagating the labels of the observed ones: a strategy which proved effective
even under strongly unbalanced labelings \cite{bertoni2011cosnet, frasca2013neural}
and which is still exploited in modern computer vision pipelines for automatic
image annotation \cite{tutevych2026efficient}. In all these settings the network
is a device for \emph{transductive inference} on a fixed dataset: the labels are
propagated among the very items whose mutual similarity defines the couplings, and
no archetype is reconstructed out of a corpus of its noisy examples.

In this work, instead, we consider a Hopfield network supplied with a dataset made of $M_L$ labeled and $M_U$ unlabeled noisy examples of $K$ random archetypes, and we prescribe a synaptic coupling given by the convex combination of the supervised and unsupervised Hebbian kernels, ruled by a mixing parameter $\lambda\in[0,1]$ whose extreme values recover the two pure protocols. As we shall prove, $\lambda$ must not be confused with the empirical fraction of labeled examples $M_L/(M_L+M_U)$, but it can be considered as a synaptic weight entering the learning rule, which the experimenter is free to tune. Its optimal value results from a non-trivial balance between the relative abundances of the two channels and the intrinsic quality of the examples, and it collapses onto the naive labeled fraction only in the limit of noiseless datasets.

Our results are both analytical and numerical. First, by a signal-to-noise analysis of the one-step dynamics at zero temperature, we compute in closed form the Mattis magnetization achieved after a single network update, whence we extract the threshold for learning $M^\star(\alpha,r,\lambda)$ as a function of the storage load $\alpha$, the dataset quality $r$ and the mixing $\lambda$; these predictions are in complete agreement with Monte Carlo simulations. Next, we solve the model at the replica symmetric level of description by means of Guerra's interpolation \cite{guerra_broken, GuerraNN, guerra2002thermodynamic}. This step requires some care: since the two channels are built out of the same archetypes, the quenched noises they generate are mutually correlated, the supervised channel can be gaussianized by the Central Limit Theorem, the unsupervised one by the universality of the quenched noise established by Carmona and Hu \cite{CarmonaWu, Genovese}, and the resulting structured disorder is treated exactly through a spectral decomposition into independent eigen-channels, each carrying its own eigenvalue and multiplicity. We thus obtain the quenched statistical pressure, the self-consistency equations for the order parameters and, in the zero-temperature limit, the phase diagram of the network: remarkably, at intermediate values of $\lambda$ the retrieval region is strictly larger than in both the pure protocols, namely semi-supervision genuinely pays. 

\par\medskip
Finally, we face the question of the optimal mixing. We prove that the quenched pressure is convex in $\lambda$, so that an interior stationary point, characterized by an exact equipartition of the energies extracted from the two channels, is a \emph{maximum} of the free energy: equilibrium thermodynamics, by itself, would push the synaptic rule toward the pure channels and never stop midway. This no-go result formally legitimates the operational reading of $\lambda$ as an externally tuned hyperparameter and shifts the optimality criterion from the free energy to the retrieval capacity, whose maximization yields a closed-form expression for $\lambda^\star$.

\par\medskip
The paper is structured as follows: in Sec.~\ref{sec:generalities} we introduce the model, together with its control and order parameters; in Sec.~\ref{sec:S2N} we develop the signal-to-noise analysis, deriving the one-step Mattis magnetization and the threshold for learning; in Sec.~\ref{sec:RS} we solve the model at the replica symmetric level via Guerra's interpolation, obtaining the quenched statistical pressure, the self-consistency equations and the phase diagrams; Sec.~\ref{ssec:opt-lambda} is devoted to the optimal mixing parameter, addressed both from the thermodynamic and from the retrieval perspective; finally, conclusions and outlooks are drawn in Sec.~\ref{sec:conclusions}. Technical details and lengthy computations are collected in the Appendices.

\section{Generalities} \label{sec:generalities}

Let us consider $N$ neurons $\boldsymbol{\sigma}=(\sigma_1,\dots,\sigma_N)\in\{-1,+1\}^N$ and a database $\Scal=\{\boldsymbol{\eta}^{\mu a}\}_{\mu=1, \hdots, K}^{a=1, \hdots, M}$, where $\eta_i^{\mu,a} \in \{-1, +1\}$. The elements of the dataset are assumed to be some noisy examples generated from $K$ archetypes $\boldsymbol{\xi}^\mu\in\{-1,+1\}^N$, $\mu=1,\dots,K$, whose entries are i.i.d.\ Rademacher random variables, namely
\begin{equation}\label{eq:xi-law}
\Prob(\xi_i^\mu=\pm 1)=\frac{1}{2},\qquad i=1,\dots,N,\;\;\mu=1,\dots,K.
\end{equation}

Therefore, the dataset is constructed so as to emulate a \textit{semi-supervised setting}, in which the former subset, made of $M_L$ elements, is \emph{labeled} and treated as the supervised channel, whereas the remaining $M_U$ elements, indexed by $c=1,\dots,M_U$, are \emph{unlabeled} and constitute the unsupervised channel, with $M=M_L+M_U$.

More precisely, we have that each element of our database can be written as 
\begin{equation}\label{eq:eta-def}
\eta_i^{\mu a}=\xi_i^\mu\,\chi_i^{\mu a},\qquad \Prob(\chi_i^{\mu a}=\pm 1)=\frac{1\pm r}{2},\quad r\in(0,1],
\end{equation}
where the binary random variables $\chi_i^{\mu a}$ fulfill
\begin{equation}\label{eq:chi-moments}
\E[\chi_i^{\mu a}]=r,\qquad \E[(\chi_i^{\mu a})^2]=1,\qquad \E[\chi_i^{\mu a}\chi_i^{\mu b}]=\delta_{ab}+(1-\delta_{ab})r^2,
\end{equation}
which yields, for the example matrix elements,
\begin{equation}\label{eq:eta-moments}
\E[\eta_i^{\mu a}]=0,\qquad \E[\eta_i^{\mu a}\eta_j^{\nu b}]=\delta_{ij}\,\delta_{\mu\nu}\bigl[\delta_{ab}+(1-\delta_{ab})r^2\bigr].
\end{equation}
We stress that for $r=1$, the dataset becomes a trivial set of copies of the archetypes\footnote{Another way to rewrite the elements of the dataset with respect to the archetypes is to consider an additive noise rather than a multiplicative one. This can be achieved by decomposing each variable $\chi_i^{\mu,a}$ into its expectation $r$ and a zero-mean fluctuation with variance $1-r^2$: $\chi_i^{\mu,a}=r + \tilde \chi_{i}^{\mu,a}$, $\mathrm{Var}(\tilde \chi_i^{\mu,a})=\mathbb{E}(\tilde \chi_{i}^{\mu,a})^2=1-r^2$, $i=1, \hdots, N$, $a=1, \hdots, M$, $\mu=1, \hdots, K$. Therefore, 
\begin{equation}\label{eq:eta-decomp}
\eta_i^{\mu a}= \xi_i^\mu(r + \tilde \chi_{i}^{\mu,a})=r\,\xi_i^\mu+\zeta_i^{\mu a},\qquad \zeta_i^{\mu a}:=\xi_i^\mu\tilde \chi_i^{\mu,a}.\
\end{equation}
This will be useful in the signal analysis of Appendix \ref{app:signal}. 
}.

\par\medskip
Following the seminal work by Hopfield \cite{Hopfield} and the Hebbian prescription, the coupling matrix $J_{ij}$ is given by a convex combination of the couplings induced by the labeled and unlabeled subsets of the dataset
\begin{equation}\label{eq:J-def}
J_{ij}^\lambda=\frac{1}{N}\sum_{\mu=1}^{K}\left[\;\frac{\lambda}{\Gamma_L}\sum_{a,b=1}^{M_L}\eta_i^{\mu a}\eta_j^{\mu b}\;+\;\frac{1-\lambda}{\Gamma_U}\sum_{c=1}^{M_U}\eta_i^{\mu c}\eta_j^{\mu c}\;\right] = \lambda J_{ij}^{ (L)} + (1-\lambda) J_{ij}^{(U)},
\end{equation}
with $\lambda\in[0,1]$ a mixing parameter tuning the relative weight of the two channels. The extreme cases $\lambda\in \{0,1\}$ correspond to the absence of one of the two groups inside the database.

The Hamiltonian is the standard quadratic form
\begin{align}\label{eq:H-def}
\Hcal_N(\boldsymbol\sigma|\Scal, \lambda)&=-\dfrac{1}{2}\sum_{i,j=1}^{N}J_{ij}^\lambda\,\sigma_i\sigma_j\notag \\
&= - \dfrac{1}{2N} \sum_{\mu=1}^{K} \sum_{i,j=1}^{N}\left[\;\frac{\lambda}{\Gamma_L}\sum_{a,b=1}^{M_L}\eta_i^{\mu a}\eta_j^{\mu b}\;+\;\frac{1-\lambda}{\Gamma_U}\sum_{c=1}^{M_U}\eta_i^{\mu c}\eta_j^{\mu c}\;\right] \sigma_i \sigma_j,
\end{align}
with $\Gamma_L$ and $\Gamma_U$ two normalization factors fixed as follows\footnote{The two factor differ because the two channels offer different
objects to be normalized. The labels allow the supervised channel to merge all the $M_L$
examples of a given archetype into a single collective variable $\sum_{a}\eta_i^{\mu a}$,
whose second moment follows from \eqref{eq:chi-moments},
$\E\bigl[\bigl(\sum_{a}\eta_i^{\mu a}\bigr)^{2}\bigr]=M_L^2r^2+M_L(1-r^2)=\Gamma_L$;
dividing by it gives the supervised block unit second moment. In the unsupervised channel no such collective variable can be
formed, each example entering the coupling through its own outer product, of second moment
$\E[(\eta_i^{\mu c})^2]=1$: the same prescription would give $\Gamma_U=M_U$ and would leave
the retrieval signal proportional to $r^2$, hence vanishing on poor datasets, so the
unsupervised block is normalized on its signal, see Appendix \ref{app:S2N} for more details.}:
\begin{equation}\label{eq:Gamma-def}
\;\Gamma_L:= M_L^2\Bigl[r^2+\tfrac{1-r^2}{M_L}\Bigr]
,\qquad \Gamma_U:= M_U\,r^2.\;
\end{equation}

\begin{figure}[t!]
\centering
\hspace*{2.5cm}
\includegraphics[width=0.7\linewidth]{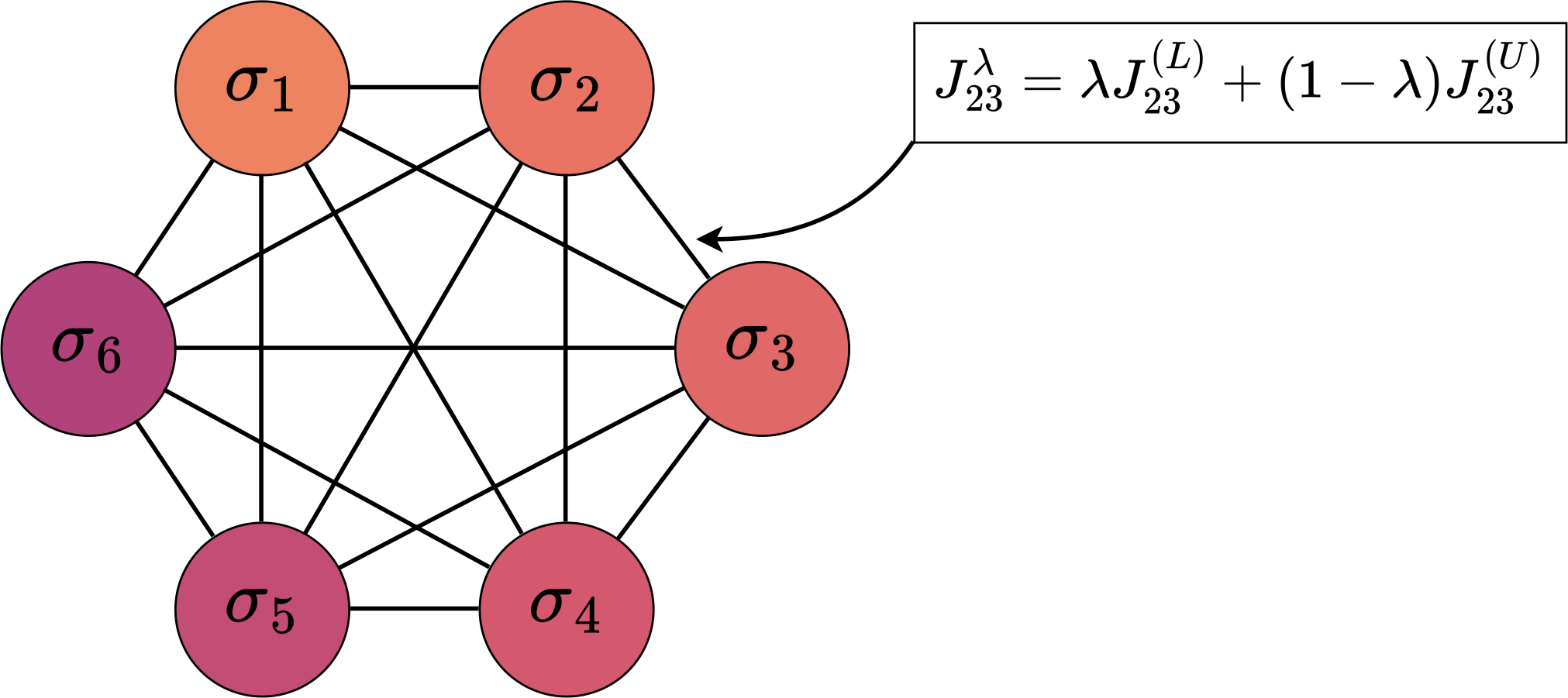}
\caption{Toy representation of the network whose Hamiltonian is defined in \eqref{eq:H-def} with $N=6$. Each neuron interacts with all the others through the effective synaptic coupling $J_{ij}^{\lambda}$ in \eqref{eq:J-def}, given by the convex combination of the supervised and unsupervised Hebbian contributions.}
\label{fig:mf_convex_network}
\end{figure}

A schematic representation of the network is reported in Fig. \ref{fig:mf_convex_network}. 

It is convenient to introduce now the \textit{load} of the network, namely the ratio between the number of patterns and the size of the network in the thermodynamic limit 
\begin{align}
    \alpha:=\lim_{N \to +\infty} \dfrac{K}{N}.
    \label{eq:load}
\end{align}

In this work, we are mainly interested in the \textit{high-storage regime}, namely when the load of the network $\alpha$ is a real finite number \cite{AGS}.  

\begin{remark}
    In the physical model under investigation self-interaction terms are excluded, that is, a neuron $\sigma_i$ interacts with any other neuron $\sigma_k$ with $k \neq i$. This should be accounted for in \eqref{eq:H-def} by inserting a corrective contribution that neutralizes diagonal terms. However, since this contribution is constant, here it is neglected in order to retain the notation simple.
    \\ In fact, for the unsupervised block:
    \begin{equation}
        J_{ii}^{(U)}=\frac{1-\lambda}{N\Gamma_U}\sum_{\mu,c}1=\frac{(1-\lambda)K M_U}{N\Gamma_U}=\frac{(1-\lambda)KM_U}{NM_Ur^2}\to\frac{(1-\lambda)\alpha}{r^2}, \qquad N \to +\infty.
    \end{equation}
  For the supervised it is convenient to take the expectation value:
    \begin{equation}
        \E[J_{ii}^{(L)}]=\frac{\lambda}{N\Gamma_L}\sum_\mu\E\Bigl[\left(\sum_a\eta_i^{\mu a}\right)^2\Bigr]=\frac{\lambda K}{N\Gamma_L}\Gamma_L=\frac{\lambda K}{N}\to\lambda\alpha, \qquad N\to +\infty,
    \end{equation}
    which are both constant and are therefore neglected.
\end{remark}

\begin{remark}
\label{rem:lambda}
    After the introduction of the model a remark regarding the role of $\lambda$ is in order. 

It is important to distinguish the upstream stage of data collection from the downstream stage of synaptic learning. If one were allowed to actively allocate a fixed budget of $M$ examples, the trivial strategy for maximising the retrieval capacity would be to collect only supervised examples, namely $M_L=M$ and $M_U=0$. Indeed, the supervised channel benefits from the noise reduction induced by the Central Limit Theorem (CLT), while avoiding the divergent interference term of order $O(r^{-2})$ associated with the unsupervised channel \cite{unsup, super}.

In the present setting, however, $M_L$ and $M_U$ are external constraints fixed by the available dataset \cite{chapelle2006semi, vanengelen2020survey}. This standard paradigm in semi-supervised learning reflects the practical reality that acquiring labels is typically an expensive and time-consuming process requiring human expertise, whereas unlabeled data can be collected automatically and in massive quantities. Therefore, the parameter $\lambda\in[0,1]$ should \emph{not} be interpreted as the empirical fraction of labeled data, for instance by imposing $\lambda=M_L/M$ (see for instance \cite{albanese2024hebbian}). Rather, $\lambda$ is a genuine synaptic weight entering the Hebbian prescription, which dynamically tunes the relative relevance assigned by the network to the supervised and unsupervised channels during learning\footnote{We stress that the presence of a free coefficient weighting the unsupervised
contribution against the supervised one is a recurrent feature of
semi-supervised algorithms, rather than an artifact of the present construction:
in entropy-regularized semi-supervised learning, for instance, the unlabeled
examples enter the cost function through an entropy term whose relative weight is
a hyperparameter tuned by the experimenter and not dictated by the labeled
fraction \cite{grandvalet2005semi}, exactly as $\lambda$ here.}.

Accordingly, in this manuscript we show that there exists an optimal value $\lambda^\star$ (see Section \ref{ssec:opt-lambda}) which does not merely reflect the relative proportion of labeled and unlabeled examples, but rather results from a balance between such proportions and the intrinsic quality of the dataset. This confirms the role of $\lambda$ as a crucial and tunable hyperparameter of the synaptic learning rule.
\end{remark}

Therefore, to sum up, the model is parametrised by the following \textit{control parameters}, which are held fixed in the thermodynamic limit:
\begin{itemize}
\item the \emph{load} of the network $\alpha$, defined in Eq. \eqref{eq:load};
\item the \emph{quality} $r\in(0,1]$ of the examples, introduced for the creation of the database in Eq. \eqref{eq:eta-def};
\item the per-channel \emph{dataset sizes} $M_L,M_U$. Equivalently, one can also consider the per-channel \emph{rescaled sizes}
\begin{equation}\label{eq:rho-def}
{\rho_X:=\frac{1-r^2}{\rho_X}=\frac{1-r^2}{M_X r^2},\qquad \Rcal_X:=r^2+\frac{1-r^2}{M_X}=r^2(1+\rho_X),}
\end{equation}
{for $X\in\{L,U\}$. The scaling $M_{X}\propto r^{-2}, \ X \in \{L,U\},$ keeps $\rho_X$ of order one\footnote{$\rho_X$ has been also denoted as \textit{entropy} of the dataset for its role in the description of the information contained in it. For more details, please see \cite{prlmiriam, super, unsup}.};}
\item the \emph{mixing parameter} $\lambda\in[0,1]$ introduced in the Hamiltonian in \eqref{eq:H-def} and focus of Remark \ref{rem:lambda}.
\item the \emph{inverse temperature} $\beta>0$ tuning the stochasticity of the spins. Large $\beta$ values, corresponding to low temperatures, strongly favor low-energy states, while small $\beta$ values, corresponding to high temperatures, make higher-energy states more accessible. We will exploit it in the definition of the thermodynamic variables of the model (i.e. Boltzmann Gibbs probability distribution and partition function);
\end{itemize}

Then, in order to describe the macroscopic behaviour of the system, one can introduce several \textit{order parameters}. In particular, 
the retrieval order parameter is the standard \emph{Mattis magnetization} which can be defined as the normalized dot product between $\bm \xi^\mu$ and $\bm \sigma$, namely for each $\mu=1, \hdots, K$ as 
\begin{equation}\label{eq:m-def}
m^\mu(\bm \sigma \vert \bm \xi^\mu):=\frac{1}{N}\sum_{i=1}^{N}\xi_i^\mu\sigma_i.
\end{equation}
We can even define, for each $\mu=1, \dots K$ and for each $a=1, \dots M_X$, with $X \in \{L,U\}$ the \textit{single example overlap}:
\begin{equation}\label{eq:n-mu-def}
n^{\mu,a}_X (\bm \sigma| \bm \eta^\mu) :=\frac{1}{N}\sum_{i=1}^{N}\eta_i^{\mu a}\sigma_i,
\end{equation}
which can be used for writing the Hamiltonian in the following form
\begin{equation}\label{eq:H-lambda-split}
-\beta\Hcal(\bm\sigma \vert \Scal, \lambda)=\lambda\,\beta N\,\Ecal_{L}(\bm\sigma \vert \Scal, \lambda)+(1-\lambda)\,\beta N\,\Ecal_{U}(\bm\sigma \vert \Scal, \lambda) + O(1),
\end{equation}
where
\begin{equation}\label{eq:ELU-def}
\Ecal_{L}(\bm\sigma):=\frac{1}{2\Gamma_{L}N}\sum_{\mu=1}^{K}\Bigl(\sum_{a=1}^{M_{L}}n_{L}^{\mu, a}\Bigr)^{\!2},
\qquad
\Ecal_{U}(\bm\sigma):=\frac{1}{2\Gamma_{U}N}\sum_{\mu=1}^{K}\sum_{c=1}^{M_{U}}(n_{U}^{\mu, c})^{2}.
\end{equation}
It is also useful to define the mean of the labeled and unlabeled examples:
\begin{equation}
    \bar{\eta}^\mu_{i,(X)} := \dfrac{1}{M_X}\sum_{a=1}^{M_X} \eta_i^{\mu a}
    \label{eq:etaavgX}
\end{equation}
and the corresponding \textit{mean example magnetizations} are defined respectively as
\begin{equation}\label{eq:nLU-def}
n_L^\mu(\bm \sigma \vert \bm {\bar \eta}^\mu):=\frac{r}{\Rcal_L}\,\frac{1}{N}\sum_{i=1}^{N}\bar \eta_{i,(L)}^{\mu}\sigma_i,\qquad
n_U^\mu(\bm \sigma \vert \bm {\bar \eta}^\mu):=\frac{r}{\Rcal_U}\,\frac{1}{N}\sum_{i=1}^{N} \bar \eta_{i, (U)}^{\mu}\sigma_i.
\end{equation}
The prefactor $r/\Rcal_X$ normalises each $n_X^\mu$ so that, at the retrieval saddle $\boldsymbol\sigma=\boldsymbol\xi^\mu$, one has $\E[n_X^\mu]=1=\E[m^\mu]$.

\begin{remark}
Using again \eqref{eq:n-mu-def}, it is possible to rewrite \eqref{eq:H-def} in a matrix form. Indeed, if we consider the vector
\begin{equation}\label{eq:n-mu-vec1}
\mathbf n_\mu:=\bigl(n^{\mu, 1}_L,\dots,n^{\mu, M_L}_L;\,
n^{\mu, 1}_U,\dots,n^{\mu, M_U}_U\bigr)^{\!\top}\in\R^{M},
\end{equation}
and we introduce the block matrix 
\begin{equation} \label{eq:W}
W:=\begin{pmatrix}
\dfrac{\beta\lambda}{\Gamma_L}\,\mathbf 1_{M_L}\mathbf 1_{M_L}^{\!\top}
& \mathbf 0\\[10pt]
\mathbf 0 &
\dfrac{\beta(1-\lambda)}{\Gamma_U}\,\mathbb I_{M_U}
\end{pmatrix}\in\R^{M\times M},
\end{equation}
where $\bm 1_{M_L}\in\R^{M_L}$ is the all-ones vector and $\mathbb I_{M_U}$ is the identity matrix $M_U \times M_U$, we can rewrite the Hamiltonian \eqref{eq:H-def} as 
\begin{equation} \label{eq:Hcompact}
-\beta\,\Hcal(\boldsymbol\sigma|\mathcal{S})
=\frac{N}{2}\sum_{\mu=1}^{K}\mathbf n_\mu^{\!\top}\,W\,\mathbf n_\mu
\;+\;O(1).
\end{equation}

\end{remark}

For the presence of quenched disorder, one introduce also the \textit{two-replicas overlap} (also known as \textit{Edwards-Anderson overlap}) of the Ising spins, reads as 
\begin{equation}\label{eq:q-def}
q_{ab}(\bm \sigma):=\frac{1}{N}\sum_{i=1}^{N}\sigma_i^{(a)}\sigma_i^{(b)},
\end{equation}
which measures the spin-glass character of the equilibrium measure (notice that, for $a=b$, $q_{aa}(\bm \sigma)$ is identically equal to $1$).

In order to develop the statistical mechanics techniques, we need to introduce also the \emph{partition function} and the \emph{quenched statistical pressure} at finite size $N$, defined as
\begin{align}
\Zcal_{N,\beta}(\Scal) &:=\sum_{\{\boldsymbol\sigma\}}\exp\!\bigl[-\beta\,\Hcal_N(\boldsymbol\sigma|\Scal)\bigr]\,,\label{eq:Z-def}\\[3pt]
\Acal_N(\beta,r,\rho_L,\rho_U,\lambda) &:=\frac{1}{N}\,\E\bigl[\ln\Zcal_{N,\beta}(\Scal)\bigr]\,,\label{eq:A-def}
\end{align}
where $\sum_{\{\boldsymbol\sigma\}}$ is intended for all the possible configurations of $\bm \sigma$, $\E:=\E_{\mathcal S}$ is the quenched expectation over the elements of the database and $\exp\!\bigl[-\beta\,\Hcal_N(\boldsymbol\sigma|\Scal)\bigr]$ is denoted as \textit{Boltzmann factor}. 

It is possible to define the thermodynamic quenched statistical pressure
\begin{align}
    \Acal(\alpha,\beta,r,\rho_L,\rho_U,\lambda) &:= \lim_{N \to +\infty} \Acal_N(\alpha,\beta,r,\rho_L,\rho_U,\lambda)=\lim_{N \to +\infty}\frac{1}{N}\,\E\bigl[\ln\Zcal_{N,\beta}(\Scal)\bigr],
\end{align}
supposing its existence and uniqueness\footnote{The existence and uniqueness of the thermodynamic quenched statistical pressure has already proven by Guerra and Toninelli and Talagrand in \cite{guerra2002thermodynamic, talagrand2001rigorous} for Sherrington-Kirkpatrick model \cite{sherrington1975solvable}. Moreover, recent work extends this result to Hopfield model \cite{agliari2024thermodynamic}.}. 

The Boltzmann-Gibbs measure on the spin configurations is
\begin{equation}\label{eq:BG-measure}
\Prob_{\beta, \lambda}(\boldsymbol\sigma|\Scal)=\frac{1}{\Zcal_{N,\beta}(\Scal)}\,e^{-\beta\Hcal(\boldsymbol\sigma|\Scal)},
\end{equation}
and the expectation of any observable $\mathcal{O}(\boldsymbol\sigma)$ is denoted by $\omega(\mathcal{O})=\sum_{\boldsymbol\sigma}\Prob_{\beta, \lambda}(\boldsymbol\sigma|\Scal)\mathcal{O}(\boldsymbol\sigma)$; the disorder-averaged Boltzmann expectation is $\avg{\mathcal{O}}:=\E[\omega(\mathcal{O})]$.

\section{Signal-to-noise and stability analysis}
\label{sec:S2N}

The signal-to-noise ratio (SNR) \cite{AGS} is a useful tool to understand some features of the model. In this section we will present the main results using this technique, for more details one can see Appendix \ref{app:S2N}.

In a nutshell, the SNR is based on the comparison between the role of the retrieval part, namely that one linked to the patterns we have chosen to be retrieved, from now on denoted as the \textit{signal}, and all the rest which act as an intrinsic \textit{noise} in the network. Without losing generality, we decide that the signal contribution is represented by the terms with $\mu=1$, instead that ones with $\mu \geq 2$ are the noise part: 
\begin{align}
     \Hcal_N(\boldsymbol\sigma|\Scal, \lambda)= - \dfrac{1}{2N} \sum_{i,j=1}^{N}\left[\;\frac{\lambda}{\Gamma_L}\sum_{a,b=1}^{M_L}\eta_i^{1, a}\eta_j^{1, b}\;+\;\frac{1-\lambda}{\Gamma_U}\sum_{c=1}^{M_U}\eta_i^{1, c}\eta_j^{1,c}\;\right] \sigma_i \sigma_j \notag \\
    -\dfrac{1}{2N} \sum_{\mu \ge 2}^{K} \sum_{i,j=1}^{N}\left[\;\frac{\lambda}{\Gamma_L}\sum_{a,b=1}^{M_L}\eta_i^{\mu a}\eta_j^{\mu b}\;+\;\frac{1-\lambda}{\Gamma_U}\sum_{c=1}^{M_U}\eta_i^{\mu c}\eta_j^{\mu c}\;\right] \sigma_i \sigma_j, 
    \label{eq:Hs2n}
\end{align}
therefore in terms of Boltzmann factor we get
\begin{equation}\label{eq:signal-noise-split}
e^{-\beta\Hcal(\boldsymbol\sigma)}=\underbrace{\exp\!\left[\,\Bcal_{\mathrm{sig}}\,\right]}_{\mu=1}\cdot\underbrace{\exp\!\left[\,\Bcal_{\mathrm{noise}}\,\right]}_{\mu\ge 2},
\end{equation}
with, using \eqref{eq:n-mu-def}:
\begin{align}
\Bcal_{\mathrm{sig}}&:=\frac{\beta\lambda N}{2\Gamma_L}\Bigl(\sum_{a=1}^{M_L}n_L^{1,a}\Bigr)^{\!2}+\frac{\beta(1-\lambda)N}{2\Gamma_U}\sum_{c=1}^{M_U}(n_U^{1,c})^2,\label{eq:Bsig}\\
\Bcal_{\mathrm{noise}}&:=\frac{\beta\lambda N}{2\Gamma_L}\sum_{\mu\ge 2}^{K}\Bigl(\sum_{a=1}^{M_L}n_L^{\mu ,a}\Bigr)^{\!2}+\frac{\beta(1-\lambda)N}{2\Gamma_U}\sum_{\mu\ge 2}^{K}\sum_{c=1}^{M_U}(n_U^{\mu ,c})^2.\label{eq:Bnoise}
\end{align}

We now define the model's dynamics in the $\beta \to \infty$ limit as described by the following update equation: the object is to assess the state of the neuron $\bm \sigma$ at time $t=1$, starting from the initial configuration $\bm \sigma^{(t=0)} = \bm \xi^1 $.
\begin{equation}
    \sigma_i^{(t+1)} = \mathrm{sign}(h_i \sigma^{(t)})
\end{equation}
where
\begin{align}
    h_i&:= h_i^{(L)} + h_i^{(U)} \notag \\
    &:=\frac{\lambda}{N\Gamma_L}\sum_{\mu=1}^{K}\sum_{a,b=1}^{M_L}\eta_i^{\mu a}\sum_{j\ne i}\eta_j^{\mu b}\,\xi_j^1 + \frac{1-\lambda}{N\Gamma_U}\sum_{\mu=1}^{K}\sum_{c=1}^{M_U}\eta_i^{\mu c}\sum_{j\ne i}\eta_j^{\mu c}\,\xi_j^1.
\end{align}
After deriving the state of the network at $t=1$,
we are interested in the behaviour of the one-step Mattis magnetization on the target archetype $\bm \xi^1$ in null-temperature and thermodynamic limits, namely
\begin{align}
    m^{(1)}(\bm \xi^1):=\frac{1}{N}\sum_{i=1}^{N}\mathrm{sgn}(h_i)\,\xi_i^1.
\end{align}
Since by the CLT the quantity $h_i \xi_i^1$ is asymptotically Gaussian as $N \to +\infty$, this allows us to compute the thermodynamic one-step Mattis magnetization on the target archetype $\bm \xi^1$ as 
\begin{align}
    m^{(1)}(\bm \xi^1) \xrightarrow{N\to\infty} \textnormal{erf} \left( \dfrac{S}{\sqrt{2V}}\right), 
    \label{eq:m1eq}
\end{align}
where $S:=\E\bigl[h_i\,\xi_i^1\bigr]=\E\bigl[(h_i^{(L)}+h_i^{(U)})\xi_i^1\bigr]$ and $V:=\Var\bigl(h_i\,\xi_i^1\bigr)$.
Let us start by the computation of the signal; for all the details we remind to Appendix \ref{app:S2N}. Focusing only on the results, the contributions for the signal of the supervised and unsupervised parts are 
\begin{align} &\E\bigl[h_i^{(L),\,\mu=1}\xi_i^1\bigr]=\frac{\lambda}{N\Gamma_L} M_L r(N-1)M_L r\;\xrightarrow{N\to\infty}\;\frac{\lambda M_L^2 r^2}{\Gamma_L}=\frac{\lambda}{1+\rho_L} \\
&\E\bigl[h_i^{(U),\,\mu=1}\xi_i^1\bigr]=\frac{1-\lambda}{N\Gamma_U} M_Ur (N-1)r\;\xrightarrow{N\to\infty}\;\frac{(1-\lambda)M_U r^2}{\Gamma_U}=1-\lambda,
\end{align}
therefore 
\begin{align}
    \label{eq:signal}S=\E\bigl[(h_i^{(L)}+h_i^{(U)})\xi_i^1\bigr]  \xrightarrow{N\to\infty} \dfrac{\lambda}{1+\rho_L} + (1-\lambda).
\end{align}

As far as the noise terms concerns, the calculations are more cumbersome: the variance of $\xi_i^1 h_i$ has several contributions from every cross product of $\left(h_i^{(L)} + h_i^{(U)}\right)^2$. As always shown in Appendix \ref{app:S2N}, one can find that the variance can be written as 
\begin{align}
V=\Var(h_i \xi_i^1) \xrightarrow{N\to\infty}C(\lambda,r,\rho_L,\rho_U)\,\alpha+R(\lambda,r,\rho_L,\rho_U), 
\end{align}
where we have put together the slow \emph{noise inference terms} in $C(\lambda,r,\rho_L,\rho_U)$ and the \emph{within-class noise} in $R(\lambda,r,\rho_L,\rho_U)$: 
\begin{align}
C(\lambda,r,\rho_L,\rho_U)&=\lambda^2+(1-\lambda)^2\Bigl[1+\frac{1-r^4}{M_Ur^4}\Bigr]+\frac{2\lambda(1-\lambda)}{1+\rho_L},\label{eq:C-final_main}\\[6pt]
R(\lambda,r,\rho_L,\rho_U)&=\frac{\lambda^2\rho_L}{(1+\rho_L)^2}+(1-\lambda)^2\rho_U.\label{eq:R-final_main}
\end{align}

Now, putting \eqref{eq:C-final_main} and \eqref{eq:R-final_main} in \eqref{eq:m1eq} we get
\begin{align}
\label{eq:m1snr}
    m^{(1)}(\bm \xi^1)=\textnormal{erf} \left( \dfrac{\dfrac{\lambda}{1+\rho_L} + (1-\lambda)}{\sqrt{2(C(\lambda,r,\rho_L,\rho_U) \alpha + R(\lambda,r,\rho_L,\rho_U))}}\right).
\end{align}

Some limiting behaviours are immediate:
\begin{itemize}[leftmargin=2em]
\item As $\alpha\to 0$ (\textit{low storage}), the variance $V$ tends only to $R(\lambda,r,\rho_L,\rho_U)$ and, so, $m^{(1)}\to\erf(S/\sqrt{2R})$. If, in addition, for $r=1$ (high-quality data), $\rho_L=\rho_U=0$ and, therefore, $R=0$. This gets to $m^{(1)} (\bm \xi^1)\to 1$, which means perfect retrieval.
\item As $r\to 0$ (\textit{low-quality data}), at fixed $\rho_U$, $C(\lambda,r \to 0,\rho_L,\rho_U)=O(r^2)$, and $m^{(1)}(\bm \xi^1)$ goes to $0$.
\item As $\lambda\to 1$ (\textit{supervised limit}), $S=1/(1+\rho_L)$, 
	$V=\alpha+\rho_L/(1+\rho_L)^2$, hence
	\begin{align}
	m^{(1)}_{\lambda=1}(\bm \xi^1)=\erf\!\Bigl(\frac{1}{\sqrt{2\alpha(1+\rho_L)^2+2\rho_L}}\Bigr),
	\end{align}
	reproducing the results obtained in~\cite{prlmiriam}.
	\item As $\lambda\to 0$ (\textit{unsupervised limit}), $S=1$, 
	$V=\alpha[1+(1-r^4)/(M_U r^4)]+\rho_U$, hence
	\begin{align}	m^{(1)}_{\lambda=0}=\erf\!\left(\frac{1}{\sqrt{2\alpha[1+(1-r^4)/(M_U r^4)]+2\rho_U}}\right),
	\end{align}
	reproducing~\cite{prlmiriam, AgliariDeMarzo}.
\end{itemize}

In order to understand the minimal number of example in dataset per archetype required to get stable retrieval phase, we use the stability condition $m^{(1)}(\bm \xi) > \textnormal{erf}(\theta)$ for $\theta$ a threshold typical of order unity (we choose $\theta=\dfrac{1}{\sqrt{2}}$ since it is the natural stability boundary when you want the signal at least large as the standard deviation). We get the condition 
\begin{align}
    C(\lambda,r,\rho_L,\rho_U)\alpha + R(\lambda,r,\rho_L,\rho_U) < \dfrac{\dfrac{\lambda}{1+\rho_L}+(1-\lambda)}{2\theta^2}. 
    \label{eq:stabcond}
\end{align}

\par\medskip
Our goal now would be to find the optimal number of examples in both unsupervised and supervised datasets using the condition \eqref{eq:stabcond}. However, this is not analytical tractable, since we would need of another condition to deal with both $M_L$ and $M_U$, so we can restrict to the condition of having the same example in each database, namely $M_L=M_U=M/2$.

\begin{figure}[t!]
    \centering
    \includegraphics[width=\linewidth]{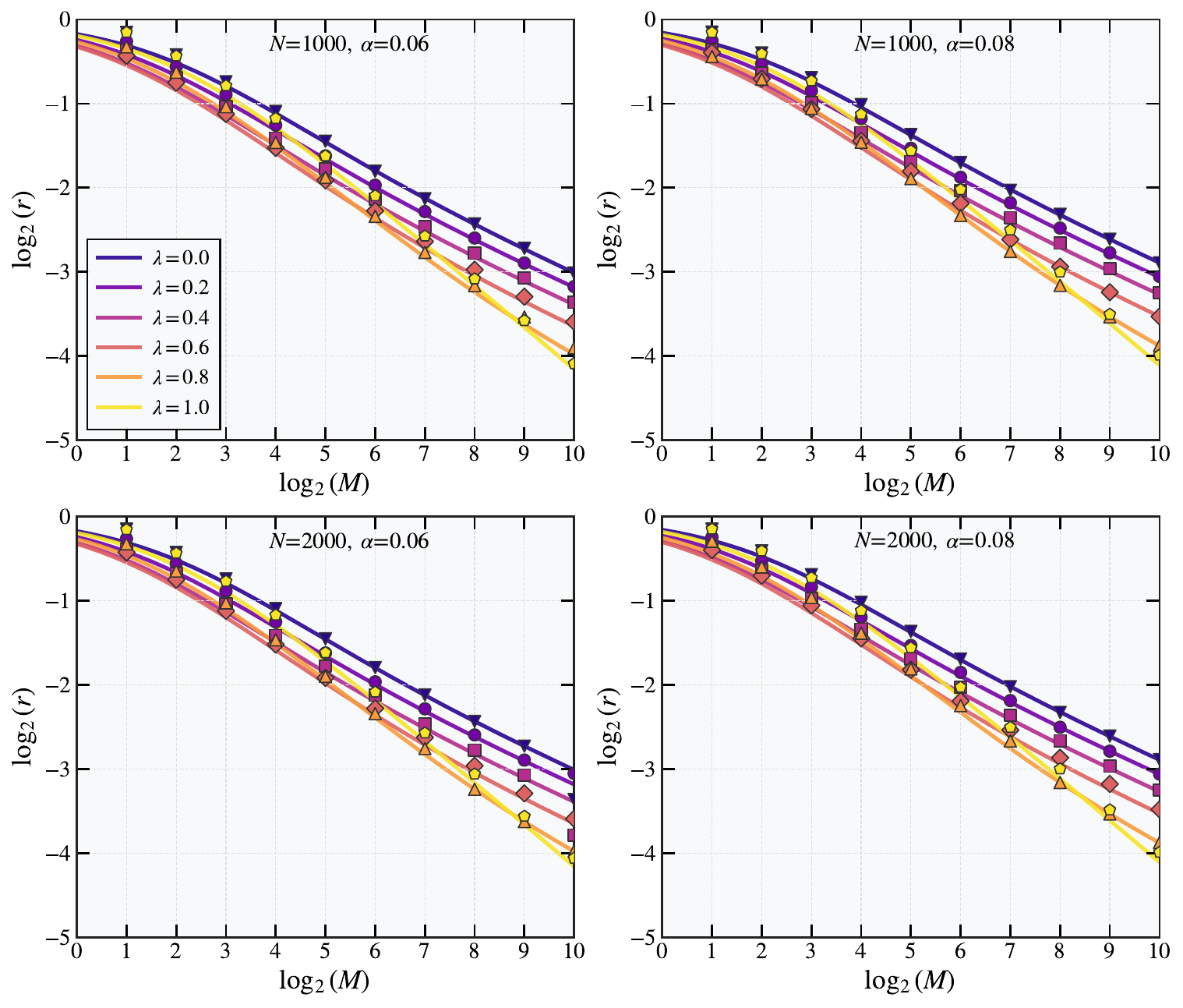}
    \caption{Comparison between SNR theoretical predictions (solid lines) and noiseless ($T=0$) MC simulations (symbols) for the semi-supervised Hopfield network, showing the phase boundaries in the $(\log_2 M, \log_2 r)$ plane for different values of the mixing parameter $\lambda \in \{0.0, 0.2, 0.4, 0.6, 0.8, 1.0\}$. The four panels display different system sizes $N$ and network loads $\alpha_N = K/N$: (a) $N = 1000,\  \alpha = 0.06$, (b) $N = 1000,\ \alpha = 0.08$, (c) $N = 2000,\ \alpha = 0.06$, and (d) $N = 2000,\ \alpha = 0.08$. The theoretical curves represent the learning threshold boundary derived from the stability condition $m^{(1)}(\bm \xi^1) > 0.9$, 
    using the analytical one-step Mattis magnetization \eqref{eq:m1snr}. The numerical MC data points are obtained by fixing the number of examples $M$ (ranging exponentially from $2$ to $2^{12}$) and determining via bisection the critical dataset quality $r$ such that the average one-step Mattis magnetization $ m^{(1)}$ reaches the retrieval threshold $0.9$. Note the excellent agreement between the theoretical predictions and numerical outcomes across all parameters.}
    \label{fig:s2nMC}
\end{figure}

In this case it is possible to find a closed-form of \eqref{eq:stabcond}, namely
\begin{equation}\label{eq:Mstar-final_main}
M^\star(\alpha,r,\lambda;\theta)\simeq\frac{\alpha(1-\lambda)^2\,\dfrac{1-r^4}{r^4}+\bigl[\lambda^2+(1-\lambda)^2\bigr]\,\dfrac{1-r^2}{r^2}}{\dfrac{1}{2\theta^2}-\alpha},\qquad \alpha<\frac{1}{2\theta^2}.
\end{equation}
The analytical behaviour of $M^\star$ is in complete agreement with Monte Carlo (MC) simulations, as shown in Fig. \ref{fig:s2nMC} for different values of the size of the network $N$ and number of patterns $K$.

\label{sec:s2n}

\section{Analytical results in RS assumption} \label{sec:RS}
The SNR analysis and the numerical simulations presented in Section \ref{sec:S2N} provide a solid heuristic foundation for understanding the learning thresholds of the semi-supervised Hopfield model. To place these findings on a rigorous mathematical footing, we now demonstrate how the same macroscopic behavior can be recovered analytically within the Replica Symmetry (RS) framework by means of \textit{Guerra's interpolation method} \cite{guerra_broken}.

Our main objective in this section is to derive the thermodynamic quenched statistical pressure $\Acal(\alpha,\beta,r,\rho_L,\rho_U,\lambda)$ in the high-storage regime. Since we are interested in the retrieval phase, we isolate, as in Sec. \ref{sec:s2n}, the first archetype $\bm \xi^1$ as the target pattern to be retrieved, separating its coherent contribution (the signal) from that of the remaining $K-1$ non-condensed archetypes. The latter act as a source of cross-talk noise, whose handling requires accounting for the structural differences between the supervised and unsupervised channels. While the detailed treatment of the signal component is relegated to Appendix \ref{app:signal}, the rigorous analysis of this structured noise is developed in the following subsections.

\subsection{Treating the noise with Gaussian approximation}
\label{ssec:gaussian}

Analyzing the Hamiltonian of the model \eqref{eq:H-def}, a structural difference between the two channels, supervised and unsupervised, appears. This makes the noise treatment subtle. Indeed, in a nutshell, the supervised channel produces a random field $J_i^\mu$ that is itself a {sum} of $M_L$ i.i.d.\ Rademacher random variables, hence amenable to a CLT argument. The unsupervised channel, conversely, exposes the individual random variable $\eta_i^{\mu , c}$, a single Rademacher entry, inside the quadratic form, and CLT does not apply to it. In this case one resorts to the {universality of the quenched noise} for spin-glass models established by Carmona and Hu~\cite{CarmonaWu} and extended to bipartite systems in~\cite{Genovese}. Let us see in details what this is faced.

Let us start by rewriting the supervised coupling $J_{ij}^{(L)}$ as 
\begin{align}
J_{ij}^{(L)} &= \dfrac{1}{N \Gamma_L} \sum_{\mu=2}^K \sum_{a,b=1}^{M_L} \eta_i^{\mu,a} \eta_j^{\mu,b}=\dfrac{1}{N\mathcal{R}_L} \sum_{\mu=2}^K \left( \dfrac{1}{M_L}\sum_{a=1}^{M_L}\eta_i^{\mu,a}\right)\left( \dfrac{1}{M_L}\sum_{b=1}^{M_L}\eta_j^{\mu,b}\right) \notag \\
&= \dfrac{1}{N\mathcal{R}_L} \sum_{\mu=2}^K  \bar \eta_{i, (L)}^\mu \bar \eta_{j, (L)}^\mu,
\label{eq:Jij_supG}
\end{align}
where in the last passage we have exploited that $\Gamma_L=\Rcal_L M_L^2$ and \eqref{eq:etaavgX}. \\
Now, focusing on $\eta_{i, (L)}^\mu$, we can rewrite it as 
\begin{equation}\label{eq:Ji-as-mean}
\bar{\eta}_{i, (L)}^\mu=\frac{1}{M_L}\sum_{a=1}^{M_L}\eta_i^{\mu, a}=\xi_i^\mu\frac{1}{M_L}\sum_{a=1}^{M_L}\chi_i^{\mu, a},
\end{equation}
where we have exploited the fact that our examples are now labeled, hence we know the corresponding archetype. For $\mu\ge 2$ the prefactor $\xi_i^\mu$ is a Rademacher independent of everything else (and independent across $i$), while the inner empirical mean of i.i.d.\ Bernoulli's converges, by CLT, to a Gaussian as $M_L\to\infty$. Specifically, by~\eqref{eq:chi-moments},
\begin{equation}\label{eq:chi-bar-moments}
\E\Bigl[\frac{1}{M_L}\sum_{a=1}^{M_L}\chi_i^{\mu ,a}\Bigr]=r,\qquad
\Var\Bigl[\frac{1}{M_L}\sum_{a=1}^{M_L}\chi_i^{\mu .a}\Bigr]=\frac{1-r^2}{M_L},
\end{equation}
hence, multiplied by the independent sign $\xi_i^\mu$ (whose effect is to recentre the mean to zero for $\mu\ge 2$ because $\E_\xi[\xi_i^\mu]=0$), one obtains
\begin{equation}
\bar{\eta}_i^\mu\xrightarrow{M_L\to\infty,\,\mu\ge 2}\;\sqrt{\Rcal_L}\,\tilde\lambda_i^{\mu, (L)},\qquad \tilde\lambda_i^{\mu, (L)}\stackrel{\text{i.i.d.}}{\sim}\mathcal{N}(0,1),
\label{eq:Ji-CLT}
\end{equation}
where $\Rcal_L=r^2+(1-r^2)/M_L$ is the second moment of $\bar{\eta}_i^\mu$ defined in \eqref{eq:rho-def}. 
Under the replacement, the supervised noise term \eqref{eq:Jij_supG} becomes
\begin{equation}\label{eq:Bnoise-L-Gaussian}
J_{ij}^{(L)} \to \dfrac{1}{N} \sum_{\mu=2}^K \lambda_i^{\mu, (L)}\lambda_j^{\mu,(L)}
\end{equation}

which can be replaced in the the Hamiltonian of the model \eqref{eq:Hs2n}. Now let us focus on the unsupervised term, namely
\begin{align}
    J_{ij}^{(U)}= \dfrac{1}{N \Gamma_U} \sum_{\mu=2}^K \sum_{c=1}^{M_U} \eta_i^{\mu,c} \eta_j^{\mu,c}. 
\end{align}

In the unsupervised channel the random field $\eta_i^{\mu , c}$ is a single Rademacher variable, not a sum, so CLT does not apply directly. However, what we ultimately need is not a distributional replacement of \emph{each} random variable, but a replacement that leaves the \emph{quenched pressure} of the model unchanged in the thermodynamic limit. This is exactly the content of the Carmona-Hu universality theorem for spin glass models~\cite{CarmonaWu}.

\par\medskip
The theorem, proved also to bipartite mean-field spin glasses in \cite{Genovese}, extends the setting closest to the two-species structure that emerges below. Its application to our unsupervised channel is direct: each $\eta_i^{\mu, c}$ can be considered as a random variable of zero mean (for $\mu\ge 2$) and unit variance, so the quenched pressure is invariant under the replacement
\begin{equation}\label{eq:CW-V}
{\;\eta_i^{\mu , c}\;\xrightarrow{\text{Carmona--Hu}}\;\tilde\lambda_i^{\mu ,c},\qquad \tilde\lambda_i^{\mu , c}\sim\mathcal{N}(0,1),\quad\mu\ge 2,\;}
\end{equation}
with covariance structure $\E[\tilde\lambda_i^{\mu, c}\tilde\lambda_i^{\mu, c'}]=\delta_{cc'}+(1-\delta_{cc'})r^2$. 

After the replacement, the unsupervised noise becomes
\begin{equation}\label{eq:Bnoise-U-Gaussian}
J_{ij}^{(U)}\longrightarrow\;\frac{\beta(1-\lambda)}{2N\Gamma_U}\sum_{\mu\ge 2}^K\sum_{c=1}^{M_U}\Bigl(\sum_i \tilde\lambda_i^{\mu, c}\sigma_i\Bigr)^{\!2}
\end{equation}
which can be also reinserted in \eqref{eq:Hs2n}. 

\subsection{Linearization of the partition function}
\label{ssec:linearization}

After the Gaussian replacements~\eqref{eq:Ji-CLT} and~\eqref{eq:CW-V} the noise contribution in \eqref{eq:Hs2n} is a sum of squares of linear combinations of $\sigma_i$. To convert each square into a linear form, we apply the standard Hubbard Stratonovich (HS) Gaussian identity
\begin{equation}\label{eq:HS-identity}
\exp\!\Bigl[\tfrac{1}{2}A\,x^2\Bigr]=\sqrt{\tfrac{1}{2\pi}}\int_{\R}dz\,\exp\!\Bigl[-\tfrac{1}{2}z^2+\sqrt{A}\,z\,x\Bigr]=\int\Drep z\,\exp\!\bigl[\sqrt{A}\,z\,x\bigr],
\end{equation}
where $\Drep z:=dz\,e^{-z^2/2}/\sqrt{2\pi}$ is the standard Gaussian measure on $\R$.
Then, we apply~\eqref{eq:HS-identity} to each $\mu\ge 2$ in the supervised term of~\eqref{eq:Bnoise-L-Gaussian} with $A=\beta\lambda/N$ and $x=\sum_i\tilde\lambda_i^{\mu, (L)}\sigma_i$, introducing one Gaussian variable $z_\mu^{(L)}$ per non-condensed archetype, and we get
\begin{equation}\label{eq:HS-L}
\exp\!\Bigl[\frac{\beta\lambda}{2N}\Bigl(\sum_i\tilde\lambda_i^{\mu, (L)}\sigma_i\Bigr)^{\!2}\Bigr]=\int\Drep z_\mu^{(L)}\,\exp\!\Bigl[\sqrt{\frac{\beta\lambda}{N}}\,z_\mu^{(L)}\sum_i\tilde\lambda_i^{\mu, (L)}\sigma_i\Bigr].
\end{equation}
Instead, for the unsupervised part, we apply~\eqref{eq:HS-identity} to each $(\mu,c)$ with $\mu\ge 2$ and $c=1,\dots,M_U$ in \eqref{eq:Hs2n}, with $A=\beta(1-\lambda)/(N\,r^2)$ (using $\Gamma_U=M_U r^2$, the prefactor reads $\beta(1-\lambda)/(N M_U r^2)$, that is shared by the $M_U$ squares; the $M_U$ factor is absorbed in the per-example HS variable). Introducing one Gaussian variable $z_\mu^c$ per pair $(\mu,c)$,
\begin{equation}\label{eq:HS-U}
\exp\!\Bigl[\frac{\beta(1-\lambda)}{2N\Gamma_U}\Bigl(\sum_i\tilde\lambda_i^{\mu, c}\sigma_i\Bigr)^{\!2}\Bigr]=\int\Drep z_\mu^c\,\exp\!\Bigl[\sqrt{\frac{\beta(1-\lambda)}{N\Gamma_U}}\,z_\mu^c\sum_i\tilde\lambda_i^{\mu, c}\sigma_i\Bigr].
\end{equation}
Replacing \eqref{eq:HS-L} and~\eqref{eq:HS-U} into \eqref{eq:Hs2n}, the partition function \eqref{eq:Z-HS} can be rewritten as 
\begin{equation}\label{eq:Z-HS}
\begin{aligned}
\Zcal_{N,\beta}(\Scal) =\sum_{\{\boldsymbol\sigma\}}\int\prod_{\mu\ge 2}\Drep z_\mu^{(L)}\,\prod_{c=1}^{M_U}\Drep z_\mu^c\;\exp\Biggl[\;&\frac{\beta N}{2}\bigl[\lambda(1+\rho_L)\,n_L^2+(1-\lambda)(1+\rho_U)\,n_U^2\bigr]\\
&+\sqrt{\frac{\beta\lambda}{N}}\sum_{\mu\ge 2}z_\mu^{(L)}\sum_i\tilde\lambda_i^{\mu, (L)}\sigma_i\\
&+\sqrt{\frac{\beta(1-\lambda)}{N\Gamma_U}}\sum_{\mu\ge 2}\sum_c z_\mu^c\sum_i\tilde\lambda_i^{\mu, c}\sigma_i\Biggr].
\end{aligned}
\end{equation}
where the first term in the exponential is simply an algebraic manipulation of the function using the order parameters and it is computed in Appendix \ref{app:signal}. This is our starting point for the recovery of the expression of the quenched statistical pressure \eqref{eq:A-def} via Guerra's interpolation of the next section. 

\begin{remark}
{(The relation between databases)}

We highlight that the two databases are made from the same archetypes: this means that there is a strong covariance between them. Therefore, we need to analyze also their relationship. 

Collecting the cross-covariances we have established, the Gaussian random fields acting on each spin $\sigma_i$ from a non-condensed archetype $\mu\ge 2$ form an $M_U+1$-dimensional centred Gaussian vector
\begin{equation}\label{eq:lambda-vec}
\widetilde{\boldsymbol\lambda}_i^\mu:=\bigl(\tilde\lambda_i^{\mu, (L)},\;\tilde\lambda_i^{\mu, 1},\;\dots,\;\tilde\lambda_i^{\mu, M_U}\bigr)\in\R^{1+M_U}, i=1, \hdots, N, \mu=1, \hdots, K
\end{equation}
with covariance matrix structured by the labeled vs unlabeled split. 
The slow noise in \eqref{eq:Hs2n} is governed by the $M\times M$ matrix $SW$, where $W$ is the block-diagonal weight matrix, already defined in \eqref{eq:W}, but rewritten here for simplicity\footnote{One can see in the Hubbard-Stratonovich transformation the rising of the coefficients.}
\begin{equation}\label{eq:Wmatrix}
W=\begin{pmatrix}W_L & 0\\ 0 & W_U\end{pmatrix},\qquad W_L:=\frac{\beta\lambda}{\Gamma_L}\mathbf{1}_{M_L}\mathbf{1}_{M_L}^{\!\top},\quad W_U:=\frac{\beta(1-\lambda)}{\Gamma_U}\,\mathbb{I}_{M_U}, 
\end{equation}
 and $S$ is the Gram matrix in the example space $\R^M=\R^{M_L}\oplus\R^{M_U}$ governed by\footnote{The vanishing of the prefactors $\sqrt{\beta\lambda/\Gamma_L}$ and $\sqrt{\beta(1-\lambda)/\Gamma_U}$ in front of the supervised and unsupervised contributions is justified from the Hubbard-Stratonovich.} 
\begin{align}\label{eq:Smatrix}
S_{ab}=r^2+(1-r^2)\delta_{ab},\qquad a,b=1,\dots,M.    
\end{align}
Although $W$ is block-diagonal, $S$ is not: the rank-one piece $r^2\mathbf 1_M\mathbf 1_M^{\!\top}$ of~\eqref{eq:Smatrix} couples {every} labeled example to {every} unlabeled one, so $SW$ acquires non-vanishing off-diagonal blocks and the two channels do \emph{not} decouple. Using $\Gamma_L = M_L^2 r^2(1+\rho_L)$ and $\Gamma_U=M_U r^2$ the four blocks read
\begin{equation}\label{eq:SW-blocks}
\begin{aligned}
(SW)_{LL}&=\frac{\beta\lambda}{M_L}\,\mathbf 1_{M_L}\mathbf 1_{M_L}^{\!\top}, & (SW)_{LU}&=\frac{\beta(1-\lambda)}{M_U}\,\mathbf 1_{M_L}\mathbf 1_{M_U}^{\!\top},\\[2pt]
(SW)_{UL}&=\frac{\beta\lambda}{M_L(1+\rho_L)}\,\mathbf 1_{M_U}\mathbf 1_{M_L}^{\!\top}, & (SW)_{UU}&=\frac{\beta(1-\lambda)}{\Gamma_U}\bigl[r^2\mathbf 1_{M_U}\mathbf 1_{M_U}^{\!\top}+(1-r^2)\mathbb I_{M_U}\bigr],
\end{aligned}
\end{equation}
The cross-block {entries} are $O(1/M)$, but since $\|\mathbf 1_{M_X}\|=\sqrt{M_X}$ their {spectral weight} is $O(1)$ and they cannot be dropped. The (non-symmetric) matrix $SW$ leaves three mutually orthogonal subspaces of $\R^M$ invariant.
\begin{itemize}[leftmargin=2em]
\item \emph{Unsupervised bulk} (multiplicity $M_U-1$). For $\mathbf v\in\R^{M_U}$ with $\mathbf 1_{M_U}^{\!\top}\mathbf v=0$, the vector $(\mathbf 0_{M_L},\mathbf v)$ is annihilated by $(SW)_{LL}$ and $(SW)_{LU}$, while $(SW)_{UU}(\mathbf 0_{M_L},\mathbf v)=\tfrac{\beta(1-\lambda)(1-r^2)}{\Gamma_U}\mathbf v$, giving the eigenvalue, with multiplicity $d=M_U-1$
\begin{equation}\label{eq:mubulk}
\mu_U^{\mathrm{bulk}}=\frac{\beta(1-\lambda)(1-r^2)}{M_Ur^2}=\beta(1-\lambda)\,\rho_U.
\end{equation}
\item \emph{Supervised null space} (multiplicity $M_L-1$). For $\mathbf u\in\R^{M_L}$ with $\mathbf 1_{M_L}^{\!\top}\mathbf u=0$ one has $(SW)(\mathbf u,\mathbf 0_{M_U})^{\!\top}=\mathbf 0$, therefore the value $0$ is the eigenvalue with multiplicity $d= M_L-1$.
\item \emph{Condensed sector} (the remaining $2$ dimensions). The plane spanned by $\bm e_L=(\mathbf 1_{M_L},\mathbf 0_{M_U})$ and $\bm e_U=(\mathbf 0_{M_L},\mathbf 1_{M_U})$ is invariant, $SW\,\bm e_L=\beta\lambda\,\bm e_L+\tfrac{\beta\lambda}{1+\rho_L}\,\bm e_U$ and $SW\,\bm e_U=\beta(1-\lambda)\,\bm e_L+\beta(1-\lambda)\kappa\,\bm e_U$, where $\kappa := 1 + \rho_U$, so in the basis $\{\bm e_L,\bm e_U\}$ the action is
\begin{equation}\label{eq:M2}
\mathbb M_2=\beta\begin{pmatrix}\lambda & (1-\lambda)\\[2pt] \dfrac{\lambda}{1+\rho_L} & (1-\lambda)\kappa\end{pmatrix},
\end{equation}
with $\det\mathbb M_2=\beta^2\lambda(1-\lambda)\bigl(\kappa-\tfrac{1}{1+\rho_L}\bigr)=\beta^2\lambda(1-\lambda)\bigl(\rho_U+\tfrac{\rho_L}{1+\rho_L}\bigr)$ and eigenvalues 
\begin{equation}\label{eq:mupm}
\mu_\pm=\frac\beta2\Bigl[\lambda+(1-\lambda)\kappa\pm\sqrt{\bigl(\lambda-(1-\lambda)\kappa\bigr)^2+\dfrac{4\lambda(1-\lambda)}{1+\rho_L}}\,\Bigr].
\end{equation}
\end{itemize}
The dimensions add up, $2+(M_U-1)+(M_L-1)=M$. Writing $\tilde\mu_k:=\mu_k/\beta$, the spectrum of $SW$ is
\begin{equation}\label{eq:spectrum}
{\;\{(\tilde\mu_k,d_k)\}_{k=1, \hdots, M}=\bigl\{(\tilde\mu_+,1),\;(\tilde\mu_-,1),\;(\tilde\mu_U^{\mathrm{bulk}},\,M_U-1),\;(0,\,M_L-1)\bigr\}.\;}
\end{equation}
 Being similar to the symmetric $S^{1/2}WS^{1/2}$, $SW$ has a real non-negative spectrum, which is what permits the channel diagonalization used in the interpolation.
\end{remark}
Coming back to the definition of the partition function, the problem is that ~\eqref{eq:Z-HS} contains $1+M_U$ analog
variables $(z_\mu^{(L)};\,z_\mu^{1},\dots,z_\mu^{M_U})$ per non-condensed
archetype $\mu\ge 2$, coupled to the spins through the quenched random
fields $(\lambda_i^{\mu, (L)}, \lambda_i^{\mu, 1}, \hdots, \lambda_i^{\mu, M_U} )$, with the two channel-specific
prefactors $\sqrt{\beta\lambda/N}$ and
$\sqrt{\beta(1-\lambda)/(N\Gamma_U)}$. It is possible to show now that, exploiting an orthogonal rotation
into the eigenbasis of $SW$ as in \eqref{eq:Smatrix} and \eqref{eq:Wmatrix}, this noise
contribution becomes a sum of independent analog variables, with the
$\lambda$-dependence absorbed entirely into the spectrum. 

Therefore, thanks to this analysis, one can write the noise terms of the semi-supervised model \eqref{eq:Hs2n} in a compact form similar to the construction of the supervised classic case \cite{prlmiriam}. 

Indeed, extending the $\bm z_\mu$ vector to $\R^M$ by appending $M_L-1$ phantom
components in the null space of $W_L$ (which carry no coupling to the
spins and will play no role) we can write the noise part of ~\eqref{eq:Z-HS} as
\begin{equation}\label{eq:HS-matrix}
\Bcal_\mu^{\rm noise}=\sqrt{\frac{1}{N}}\sum_{i=1}^{N}\sigma_i\,
\bigl(\tilde{\mathbf z}_\mu\bigr)^{\!\top}\,W^{1/2}\,
\boldsymbol{\tilde\Lambda}_i^\mu,
\end{equation}
where $W^{1/2}$ is the positive square root of the weight matrix
$W$ defined in~\eqref{eq:Wmatrix}, and
$\boldsymbol{\tilde\Lambda}_i^\mu\in\R^M$ is the vector of (Gaussian)
random fields at site $i$ for archetype $\mu$, with covariance
\begin{equation}\label{eq:Lambda-cov}
\E\bigl[\tilde\Lambda^{\,a}_{\mu i}\,\tilde\Lambda^{\,b}_{\mu j}\bigr]
=\delta_{ij}\,S_{ab},
\end{equation}
with $S$ the Gram matrix of~\eqref{eq:Smatrix}. Equation~\eqref{eq:HS-matrix}
reduces to~\eqref{eq:Z-HS} block-wise: the rank-one piece of $W^{1/2}$
in the $L$-block projects the supervised coupling onto the symmetric
direction $\mathbf 1_L/\sqrt{M_L}$, while the diagonal $U$-block of
$W^{1/2}$ is proportional to $\sqrt{\beta(1-\lambda)/\Gamma_U}\,
\mathbb I_{M_U}$.

Now, let $O$ be the orthogonal $M\times M$ matrix that diagonalises
$W^{1/2}SW^{1/2}$,
\begin{equation}\label{eq:O-diag}
O^{\!\top}\,(W^{1/2}SW^{1/2})\,O=
\operatorname{diag}(\beta\tilde\mu_1,\dots,\beta\tilde\mu_M),
\end{equation}
with $\{\tilde\mu_k\}$ the spectrum in \eqref{eq:spectrum}
(the non-zero spectrum of $W^{1/2}SW^{1/2}$ coincides with that of $SW$).
Introducing rotated variables
\begin{equation}\label{eq:rotated-vars}
\mathbf z_\mu^{\rm rot}:= O^{\!\top}\tilde{\mathbf z}_\mu,\qquad
\boldsymbol\Lambda_i^{{\rm rot},\mu}:= O^{\!\top}W^{1/2}\,
\boldsymbol{\tilde\Lambda}_i^\mu,
\end{equation}
the orthogonal invariance of the Gaussian measure ensures
$\Drep\tilde{\mathbf z}_\mu=\Drep\mathbf z_\mu^{\rm rot}$, while the rotated
random fields decouple across $k$ with variance $\beta\tilde\mu_k$,
\begin{equation}\label{eq:rotated-Lambda-cov}
\E\bigl[\Lambda^{{\rm rot},\mu k}_i\,\Lambda^{{\rm rot},\mu k'}_j\bigr]
=\delta_{ij}\,\delta_{kk'}\,\beta\tilde\mu_k.
\end{equation}
Writing $\Lambda^{{\rm rot},\mu k}_i=\sqrt{\beta\tilde\mu_k}\,
\lambda_i^{\mu, k}$ with $\lambda_i^{\mu, k}\stackrel{\rm i.i.d.}{\sim}
\mathcal N(0,1)$ standard, inserting $\mathbb I=OO^{\!\top}$ and renaming $z_\mu^{{\rm rot},k}$ as $z_\mu^k$ the noise becomes
in~\eqref{eq:HS-matrix},
\begin{equation}\label{eq:noise-eigenchannel}
\Bcal_\mu^{\rm noise}=\sqrt{\frac{\beta}{N}}\sum_{k:\,\tilde\mu_k\ne 0}
\sqrt{\tilde\mu_k}\,z_\mu^k\sum_{i=1}^{N}\lambda_i^{\mu k}\sigma_i,\qquad
\lambda_i^{\mu k}\stackrel{\rm i.i.d.}{\sim}\mathcal N(0,1).
\end{equation}
The $k$-sum runs only over the $M_U+1$ eigen-channels with non-zero
$\tilde\mu_k$: the $M_L-1$ phantom components corresponding to the
supervised null space carry
$\sqrt{\tilde\mu_k}=0$ and decouple identically.
\\Thus, after the rotation, we have independent eigen-modes $z_\mu^k$, one per eigenvalue $\tilde\mu_k$ of $SW$ with multiplicity $d_k$, and the natural slow-noise order parameters are the two-replica overlaps
\begin{equation}\label{eq:pk-def-intro}
p^k_{ab}:=\frac{1}{K-1}\sum_{\mu\ge 2}^{K}z_\mu^{k,(a)}z_\mu^{k,(b)} \quad k = 1, \dots, M_U + 1 \ \textnormal{such that} \ \tilde \mu_k \neq 0.
\end{equation} 

\subsection{Guerra's interpolation}
\label{ssec:guerra}

After the discussion in Subsecs. \ref{ssec:gaussian} and \ref{ssec:linearization}, the final Hamiltonian taken under consideration is the following
\begin{align}
    -\beta \Hcal_N(\boldsymbol\sigma|\Scal, \lambda, \bm z) =&  \frac{\beta N}{2}\bigl[\lambda(1+\rho_L)\,n_L^2+(1-\lambda)(1+\rho_U)\,n_U^2\bigr]\notag \\
    &+ \sqrt{\frac{\beta}{N}}\sum_{k:\,\tilde\mu_k\ne 0}
\sqrt{\tilde\mu_k}\,z_\mu^k\sum_{i=1}^{N}\lambda_i^{\mu k}\sigma_i.
\label{eq:Hguerra}
\end{align}

Our goal now is to find the quenched statistical pressure \eqref{eq:A-def} of the model in \eqref{eq:Hguerra} using Guerra's interpolation. The strategy is standard~\cite{guerra_broken,GuerraNN, Fachechi1,prlmiriam, Albanese2021}: we introduce a one-parameter family of partition functions $\Zcal_N(t)$, $t\in[0,1]$, interpolating between the original model at $t=1$ and a one-body solvable model at $t=0$. Then, we solve the simpler $t=0$ model exactly, compute the streaming $\partial_t\Acal(t)$, and recover the original quenched pressure via the fundamental theorem of calculus,
\begin{equation}\label{eq:sumrule}
\Acal(t=1)=\Acal(t=0)+\int_0^1\partial_t\Acal(t)\,dt.
\end{equation}
We are interested in Replica Symmetry (RS) approximation, namely the relevant order parameters self-average around deterministic values in the thermodynamic limit. More precisely, for any order parameter $X$ with limiting value $\langle X \rangle$, we assume
\begin{equation}
\lim_{N\to\infty}\mathbb{P}_N(X)=\delta(X-\langle X \rangle).
\label{eq:RS-ansatz}
\end{equation}
We will see that under the RS ansatz the integrand in \eqref{eq:sumrule} becomes $t$-independent so the integral becomes trivial.

For all the detailed computation of each term presented before, we remind to Appendix \ref{app:guerra}; in this section we report the main tools and results. 

\par\medskip
Let $\boldsymbol\theta=\{\theta_i\}_{i=1,\dots,N}$ and $\boldsymbol\psi=\{\psi_\mu^k\}_{\mu=1, \hdots, K}^{k=1, \hdots, M}$, where $k$ counts all the eigenvalues of $SW$ and carry the multiplicities $d_k$ of~\eqref{eq:spectrum}, be auxiliary standard Gaussian fields, independent of the quenched disorder and of each other, and let $J_m\in\R$ be an external field conjugate to the magnetization $m$. The Guerra interpolating partition function is
\begin{equation}\label{eq:Z-interp}
\Zcal_N(t,J_m)=\lim_{J_m \to 0}\sum_{\{\boldsymbol\sigma\}}\int\prod_{\mu\ge 2}\prod_{k:\tilde \mu_k \neq 0}\Drep z_\mu^k\;\Bcal(t,J_m),
\end{equation}
with the interpolating Boltzmann factor $\Bcal(t,J_m)$ is written, after the diagonalization in the eigen-channel basis \eqref{eq:noise-eigenchannel} as \begin{equation}\label{eq:Bt-def}
\begin{aligned}
\Bcal(t,J_m)=\;
&\exp \left\{ -\dfrac{1}{2}\sum_{\mu\ge 2}\sum_{k:\,\tilde\mu_k\neq0}\bigl[1-(1-t)\beta\tilde\mu_k(1-\avg{q_{12}})\bigr](z_\mu^k)^2\right.\\
&+\sqrt{t}\,\sqrt{\dfrac{\beta}{N}}\sum_{\mu\ge 2}\sum_{k:\,\tilde\mu_k\neq0}\sqrt{\tilde\mu_k}\,z_\mu^k\sum_i\lambda_i^{\mu, k}\sigma_i\\
&+t\,\dfrac{\beta N}{2}\bigl[\lambda(1+\rho_L)\,n_L^2+(1-\lambda)(1+\rho_U)\,n_U^2\bigr]\\
&+(1-t)\,\beta\bigl[\lambda(1+\rho_L)\avg{n_L}\,N n_L+(1-\lambda)(1+\rho_U)\avg{n_U}\,N n_U\bigr]\\
&+\sqrt{(1-t)\,\alpha\beta\,\sum_{k:\,\tilde\mu_k\neq0} d_k\,\tilde\mu_k\,\avg{p_{12}^k}}\;\sum_i\theta_i\sigma_i\\
&\left.+\sqrt{(1-t)\,\beta\avg{q_{12}}}\sum_{\mu\ge 2}\sum_{k:\,\tilde\mu_k\neq0}\sqrt{\tilde\mu_k}\,\psi_\mu^k z_\mu^k-\,J_m\,\beta N m \right\},
\end{aligned}
\end{equation}
where $J_m$ is a conjugate field $J_m$ which allows us to recover the magnetization through
\begin{equation}\label{eq:m-from-Jm}
\avg{m}=-\beta^{-1}\partial_{J_m}\Acal(t=1,J_m)\Big|_{J_m=0}.
\end{equation}

The interpolating quenched statistical pressure is defined as
\begin{equation}\label{eq:Acal-interp}
\Acal_N(t,J_m):=\frac{1}{N}\,\E\ln\Zcal_N(t,J_m),\qquad \Acal_N(t):=\Acal_N(t,0),
\end{equation}
and the corresponding generalised Boltzmann expectation is
\begin{equation}\label{eq:omega-t}
\omega_t(\,\cdot\,):=\frac{1}{\Zcal_N(t,J_m)}\sum_{\{\boldsymbol\sigma\}}\int\!\!\prod\Drep z\,(\,\cdot\,)\,{\Bcal(t,J_m)},\qquad \avg{\,\cdot\,}_t:=\E[\omega_t(\,\cdot\,)].
\end{equation}

We stress that for $t=1$ all the one body terms vanish and we recover the original model in \eqref{eq:Hguerra}. Instead, for $t=0$ we get a new solvable model whose quenched statistical pressure is easier to find (see Appendix \ref{app:guerra} for details): 
\begin{equation}\label{eq:cauchy-final}
\begin{aligned}
&\Acal_{N, \lambda}(t{=}0,J_m)=\;
-\frac{K}{2N}\sum_k d_k\Bigl[\ln\!\bigl(1-\beta\tilde\mu_k(1-\avg{q_{12}})\bigr)-\frac{\beta\tilde\mu_k\avg{q_{12}} }{1-\beta\tilde\mu_k(1-\avg{q_{12}})}\Bigr]+\ln 2\\
&+\E_{\chi,\theta}\ln\cosh\!\left[\beta\avg{m}\left(\,\dfrac{\lambda}{M_L r}\sum_{a=1}^{M_L}\chi^{1,a}+\,\dfrac{1-\lambda}{M_U r}\sum_{c=1}^{M_U}\chi^{1,c}\right)-\beta J_m+\theta\sqrt{\dfrac{K}{N}\beta \sum_k d_k\,\tilde\mu_k\,\avg{p_{12}^k}}\right].
\end{aligned}
\end{equation}

All we need to apply the Fundamental Theorem of Calculus is the computation of the integral in \eqref{eq:sumrule} which gets
\begin{equation}\label{eq:dt-Acal}
\begin{aligned}
\partial_t\Acal_N(t,J_m)=\;
&\frac{\beta}{2}\sum_{X\in\{L,U\}}c_X(1+\rho_X)\bigl(\Delta[n_X^2]-\avg{n_X}^2\bigr)\\
&-\frac{K\beta}{2N}\sum_k d_k\,\tilde\mu_k\bigl(\Delta[p_{12}^k q_{12}]+\avg{p_{12}^k}(1-\avg{q_{12}})\bigr),
\end{aligned}
\end{equation}
with $c_L=\lambda,\ c_U=1-\lambda$ and we have exploited the definition of the variances of $n_X$ and $p_{12}^k q_{12}$ as 
\begin{equation}\label{eq:Deltas}
\Delta[n_X^2]:=\E\avg{(n_X-\avg{n_X})^2}_t,\qquad \Delta[p_{12}^k q_{12}]:=\E\avg{(p^k_{12}-\avg{p_{12}^k})(q_{12}-\avg{q_{12}})}_t.
\end{equation} 

Under the RS ansatz~\eqref{eq:RS-ansatz} in the thermodynamic limit every centred fluctuation in~\eqref{eq:Deltas} vanishes, so $\Delta[n_X^2]\to 0$ and $\Delta[p_{12}^k q_{12}]\to 0$. The streaming becomes {$t$-independent}:
\begin{equation}\label{eq:dt-RS}
\partial_t\Acal(t,J_m)=-\frac{\beta}{2}\bigl[\lambda(1+\hat\rho_L)\avg{n_L}^2+(1-\lambda)(1+\hat\rho_U)\avg{n_U}^2\bigr]-\frac{\alpha\beta}{2}(1-\avg{q_{12}})\sum_k d_k\,\tilde\mu_k\,\avg{p_{12}^k}, 
\end{equation}
where we have used also the definition of the load of the network in \eqref{eq:load}. Since $\partial_t\Acal$ is $t$-independent, the integral $\int_0^1\partial_t\Acal\,dt$ in~\eqref{eq:sumrule} equals~\eqref{eq:dt-RS} directly. Also in this case, for all the computation we remind to Appendix \ref{app:guerra}. 
\begin{figure}
    \centering
\includegraphics[width=1\linewidth]{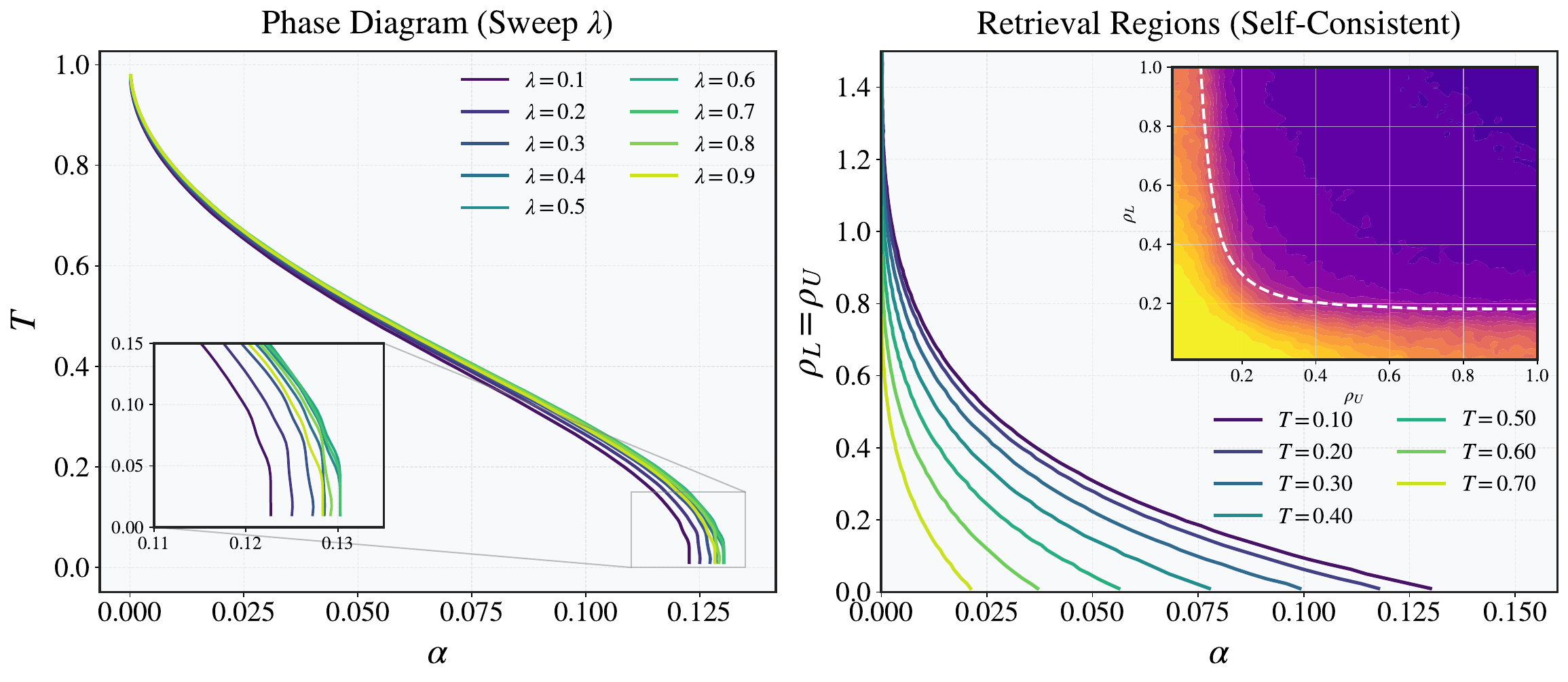}
    \caption{%
Sensitivity of the retrieval phase to the mixing parameter $\lambda$ and to
the dataset noise level $\rho=\rho_L=\rho_U$.
\emph{Left panel:} RS retrieval boundary $\alpha_c(T)$ in the $(\alpha,T)$
plane for $r=0.8$, $M_L=M_U=30$, and nine values of the mixing parameter
$\lambda\in\{0.1,0.2,\ldots,0.9\}$ (colour-coded from dark blue to yellow).
The maximum critical load $\alpha_c(T\to 0)$ is not attained at the extremal
values $\lambda=0$ or $\lambda=1$, but at an intermediate value
$\lambda\approx 0.5$--$0.6$, indicating that a mixed semi-supervised coupling
genuinely outperforms any pure learning strategy.
The inset magnifies the low-temperature region
($T\lesssim 0.15$, $\alpha\in[0.11,0.135]$), resolving the non-monotone
ordering of the curves with respect to $\lambda$.
\emph{Right panel:} RS retrieval boundary in the $(\alpha,\rho)$ plane
for $r=0.7$ and fixed mixing parameter $\lambda=0.5$, at temperatures
$T\in\{0.10,0.20,\ldots,0.70\}$.
Each curve separates the retrieval phase (below, high magnetization) from the
paramagnetic phase (above): lower temperature and smaller $\rho$ (better
dataset quality) expand the retrieval basin.
The inset shows a MC heat map of the Mattis magnetization in the
$(\rho_U,\rho_L)$ plane, with the white dashed line marking the analytical
Signal To Noise stability boundary \eqref{eq:stabcond}; the colour encodes the Mattis magnetization $m(\bm \xi^1|\bm \sigma)\in[0,1]$.}
\label{fig:phase_diagrams}
\end{figure}
We can apply the definition of the load \eqref{eq:load} also for the one body terms in order to finally get the following expression of the quenched statistical pressure of the model whose Hamiltonian in \eqref{eq:H-def}:

\begin{align}
\label{eq:Acal-RS-fullmain}
&\Acal_{\mathrm{RS}}=\lim_{J_m \to 0}\Acal_{\mathrm{RS}}(J_m)\notag \\
&= \lim_{J_m \to 0}\Biggl\{-\frac{\alpha}{2}\sum_k d_k\Bigl[\ln\!\bigl(1-\beta\tilde\mu_k(1-\avg q)\bigr)-\frac{\beta\tilde\mu_k\avg q}{1-\beta\tilde\mu_k(1-\avg q)}\Bigr] \notag \\
&-\frac{\beta}{2}\bigl[\lambda(1+\rho_L)\avg{n_L}^2+(1-\lambda)(1+\rho_U)\avg{n_U}^2\bigr] -\frac{\alpha\beta}{2}(1-\avg q)\sum_k d_k\tilde\mu_k\avg{p_{12}^k}+\ln 2 \notag
\\
&+\E_{\chi,\theta}\,\ln\cosh\!\Bigl[\beta\bigl(\lambda\avg{n_L}\tfrac{1}{M_Lr}\!\sum_a\chi^{1a}+(1-\lambda)\avg{n_U}\tfrac{1}{M_Ur}\!\sum_c\chi^{1c}\bigr)-\beta J_m+\theta\sqrt{\alpha\beta\sum_k d_k\tilde\mu_k\,\avg{p_{12}^k}}\Bigr]\Biggr\}. 
\end{align}
\normalsize

The order parameters $\avg{m},\avg{n_L},\avg{n_U},\avg{q_{12}},\avg{p_{12}^k}$ are determined by extremising the RS statistical pressure~\eqref{eq:Acal-RS-fullmain} with respect to each order parameters. We compute the derivatives one by one in Appendix \ref{app:guerra}. Here, we report the self-consistency equations in the large dataset limit $M_X \to \infty$, for $ X \in \{L,U\}$ (see Appendix \ref{ssec:largeM}):

\begin{align}
\begin{split}\label{eq:m-largeM}
\avg m &=\;\E_z\Bigl[\tanh\Bigl(\beta\avg m\\
&\hspace{1cm}+\beta z\sqrt{\alpha\sum_k d_k\,\frac{\tilde\mu_k^{\,2}\,\avg{q_{12} }}{\bigl[1-\beta\tilde\mu_k(1-\avg{q_{12} })\bigr]^2}+\avg m^2\bigl(\lambda^2\rho_L+(1-\lambda)^2\rho_U\bigr)}\,\Bigr)\Bigr],
\end{split} \\ 
\begin{split}\label{eq:q-largeM}
\avg{q_{12} }&=\;\E_z\Bigl[\tanh^2\Bigl(\beta\avg m\\
&\hspace{1cm}+\beta z\sqrt{\alpha\sum_k d_k\,\frac{\tilde\mu_k^{\,2}\,\avg{q_{12} }}{\bigl[1-\beta\tilde\mu_k(1-\avg{q_{12} })\bigr]^2}+\avg m^2\bigl(\lambda^2\rho_L+(1-\lambda)^2\rho_U\bigr)}\,\Bigr)\Bigr],
\end{split} \\
\avg{p_{12}^k} &= \frac{\beta\tilde\mu_k\,\avg{q_{12} }}{\bigl[1-\beta\tilde\mu_k(1-\avg{q_{12} })\bigr]^2},\label{eq:pk-largeM}\\[3pt]
\avg{n_L} &= \avg{n_U}=\avg m.\label{eq:nLU-largeM}
\end{align}

\begin{figure}[t!]
    \centering
    \includegraphics[width=1\linewidth]{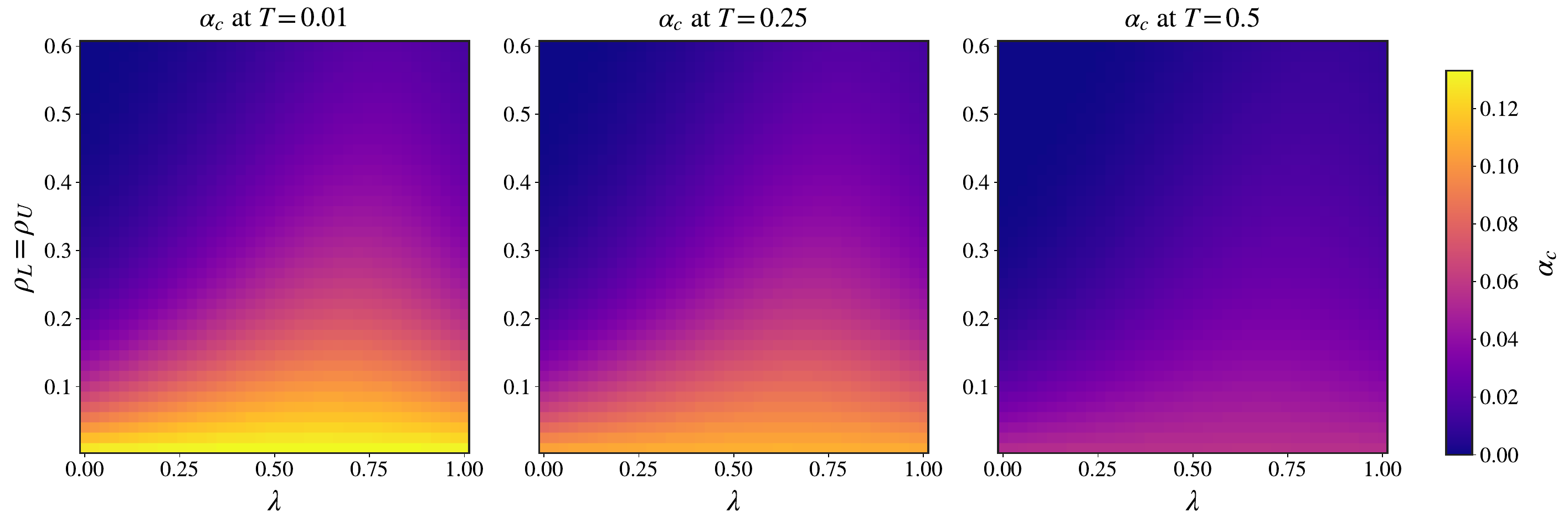}
    \caption{%
{Heat map of the critical retrieval load $\alpha_c$ as a function of the
mixing parameter $\lambda\in[0,1]$ and the rescaled noise level
$\rho=\rho_L=\rho_U\in[0.01,0.60]$, computed from the RS
self-consistency equations for $M_L=M_U=30$ at three temperatures
$T\in\{0.01, 0.25, 0.50\}$.
In all panels the colour encodes the maximal load $\alpha_c$:
high-capacity configurations appear bright (yellow), low-capacity ones dark
(blue).
As expected, increasing $\rho$ (deteriorating dataset quality) reduces
$\alpha_c$ uniformly, while increasing temperature contracts the retrieval
region.
The key observation, visible across all temperatures, is that for any fixed
$\rho > 0$ the maximum of $\alpha_c$ over $\lambda$ is attained at an
\emph{interior} value $\lambda\in(0,1)$, strictly away from both the purely
unsupervised ($\lambda=0$) and purely supervised ($\lambda=1$) extremes.
Quantitatively, at $T=0.01$ and $\rho=0.01$, one finds
$\alpha_c(\lambda=0)\approx 0.127$, $\alpha_c(\lambda=1)\approx 0.131$, and
$\alpha_c(\lambda\approx 0.59)\approx 0.133$: a balanced mixed strategy
outperforms both pure channels.
}}
\label{fig:lambda_hatrho}
\end{figure}

To extract the retrieval boundary and the structure of the phase diagram, we specialise
the large-dataset RS self-consistency system~\eqref{eq:m-largeM}--\eqref{eq:pk-largeM}
(where $\avg{n_L}=\avg{n_U}=\avg m$) to the zero-temperature regime $\beta\to\infty$,
following Amit's approach~\cite{Amit}. In this limit the spin-glass overlap saturates,
 $\avg q\to 1$, while the combination $\mathcal C:=\beta(1-\avg q)$ stays finite.
The two surviving order parameters $\avg m$ and $\mathcal C$ are then fixed by the closed
zero-temperature self-consistency system (derived in detail in Appendix~\ref{app:T0})
\begin{empheq}{align}
\avg m &=\erf\!\left(\frac{\avg m}{\sqrt{2\bigl(\alpha\sum_{k}\frac{d_{k}\,\tilde\mu_{k}^{\,2}}{(1-\tilde\mu_{k}\mathcal{C})^{2}}+\avg m^{2}\,\lambda^2\rho_L+(1-\lambda)^2\rho_U\bigr)}}\right),
\label{eq:m-T0}\\[4pt]
\begin{split}\label{eq:C-T0}
\mathcal{C} &=\sqrt{\frac{2}{\pi\bigl(\alpha\sum_{k}\frac{d_{k}\,\tilde\mu_{k}^{\,2}}{(1-\tilde\mu_{k}\mathcal{C})^{2}}+\avg m^{2}(\lambda^2\rho_L+(1-\lambda)^2\rho_U)\bigr)}}\;
\\ &\cdot \exp\!\Bigl[-\frac{\avg m^{2}}{2\bigl(\alpha\sum_{k}\frac{d_{k}\,\tilde\mu_{k}^{\,2}}{(1-\tilde\mu_{k}\mathcal{C})^{2}}^{2}+\avg m^{2}(\lambda^2\rho_L+(1-\lambda)^2\rho_U)\bigr)}\Bigr].
\end{split}
\end{empheq}
Equation~\eqref{eq:m-T0} is the zero-temperature retrieval equation, whose non-trivial
solution $\avg m>0$ delimits the retrieval phase, while~\eqref{eq:C-T0} closes the system
by fixing $\mathcal C=\beta(1-\avg q)$. We solve this system numerically to obtain the phase diagram of
Figure~\ref{fig:phase_diagrams}; evaluating the resulting critical retrieval load over the
$(\lambda,\rho)$ plane, with $\rho:=\rho_L=\rho_U$, then yields the capacity map of
Figure~\ref{fig:lambda_hatrho}.

\section{Optimal mixing parameter}\label{ssec:opt-lambda}
\begin{figure}[t!]
    \centering
    \includegraphics[width=1\linewidth]{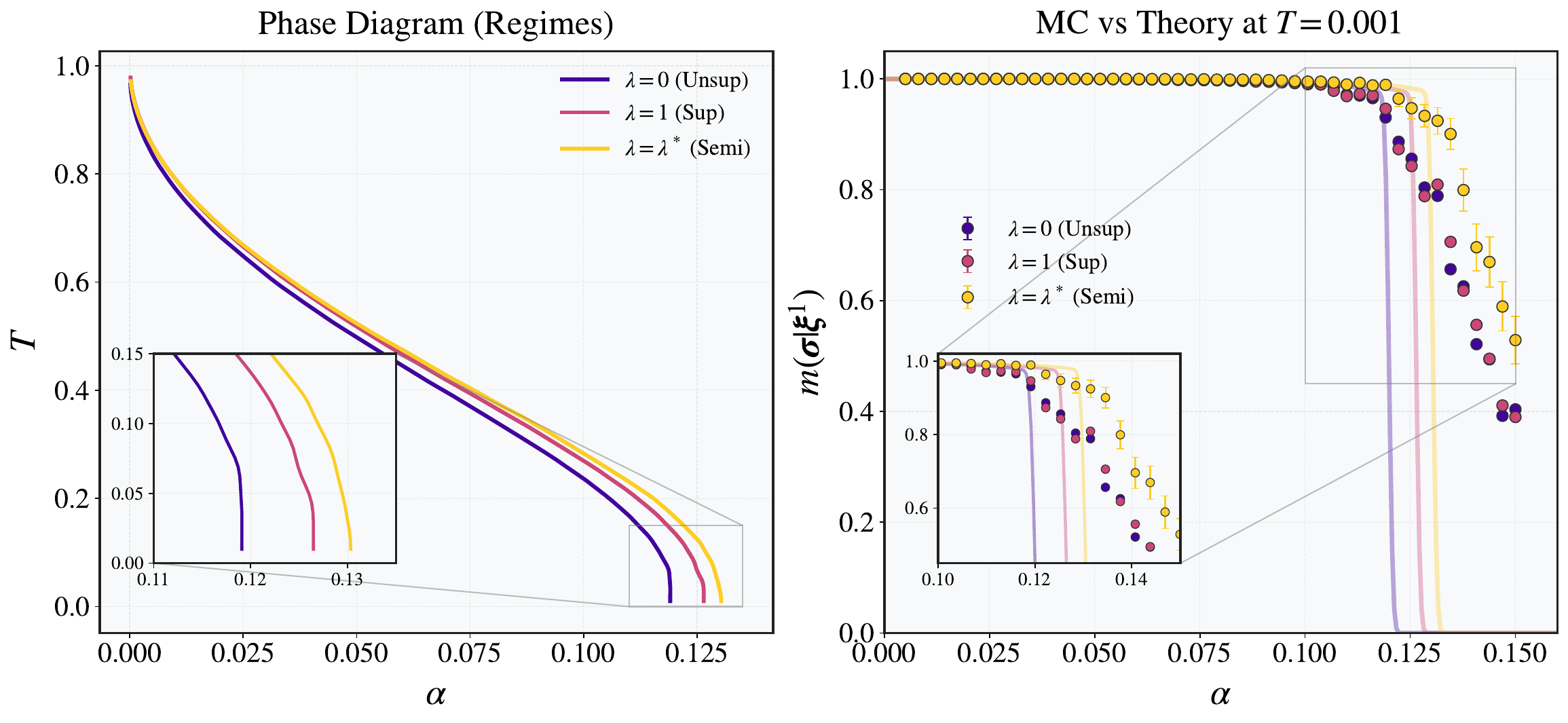}
    \caption{%
{Phase diagrams and retrieval performance of the semi-supervised Hopfield
model for $r=0.8$, $M_L=M_U=30$.
\emph{Left panel:} retrieval boundary $\alpha_c(T)$ in the $(\alpha,T)$ plane
for the three representative learning regimes: fully unsupervised
($\lambda=0$), fully supervised ($\lambda=1$), and semi-supervised with the
pointwise-optimal mixing $\lambda=\lambda^*$
from~\eqref{eq:lambdastar}.
The optimal semi-supervised boundary lies strictly to the right of both
limiting cases, demonstrating that the enlarged retrieval region is a genuine
effect of the mixed Hebbian learning rule.
The inset magnifies the low-temperature region ($T\lesssim 0.15$,
$\alpha\in[0.11,0.135]$), resolving the separation between the three curves.
\emph{Right panel:} Mattis magnetization $m= m(\boldsymbol{\sigma}|\boldsymbol{\xi}^1)$
as a function of the storage load $\alpha$ at near-zero temperature
($T=0.001$), for the same three regimes.
Solid lines are replica-symmetric (RS) theoretical predictions; symbols with
error bars are Monte Carlo estimates ($N=2000$, $50$ disorder realizations).
Both the analytical curves and the numerical data display a sharp, collective
drop of $m$ within a narrow window of $\alpha$, signalling a thermodynamic
phase transition.
The inset zooms into the transition region
($\alpha\in[0.10,0.15]$, $m\in[0.45,1.0]$), highlighting the quantitative
agreement between theory and simulation.
}}
\label{fig:entropy}
\end{figure}

\begin{figure}[t!]
    \centering
    \includegraphics[width=1\linewidth]{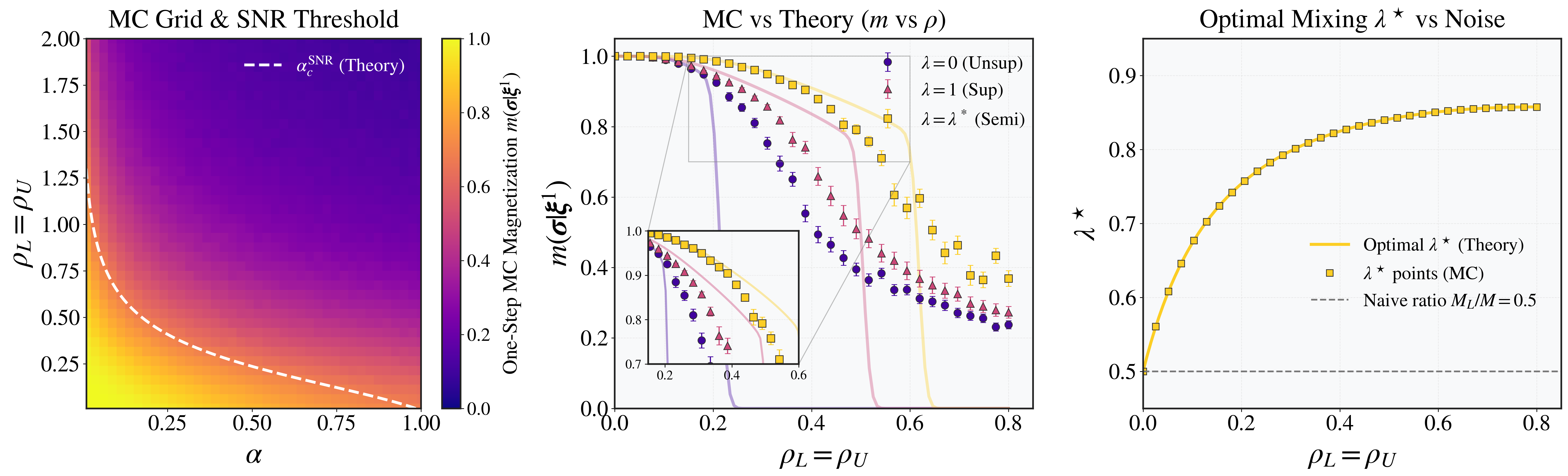}
    \caption{%
{MC validation of the retrieval phase boundaries in the
$(\alpha,\rho)$ plane.
\emph{Left panel:} heat map of the one-step Mattis magnetization
$m(\boldsymbol{\sigma}|\boldsymbol{\xi}^1)$ on a $40\times 40$ grid of
$(\alpha,\rho)$ values ($N=2000$, $M=30$, $\lambda=0.5$,
zero-temperature single-step dynamics, $32$ disorder realizations).
The colour encodes $m\in[0,1]$ (plasma scale); the white dashed curve is the
analytical SNR threshold $\alpha_c^{\rm SNR}(\rho)$
from~\eqref{eq:alpha_c}, which faithfully tracks the numerically observed
boundary between high-$m$ and low-$m$ regions across the entire parameter
space.
\emph{Center panel:} magnetization
$m(\boldsymbol{\sigma}|\boldsymbol{\xi}^1)$ as a function of the
rescaled noise parameter $\rho$ at fixed load $\alpha=0.02$ and near-zero
temperature ($T=0.001$), for the three regimes: fully unsupervised
($\lambda=0$, circles), fully supervised ($\lambda=1$, triangles), and
semi-supervised with the pointwise-optimal mixing
$\lambda=\lambda^*(\alpha,r)$.
Solid lines are RS theoretical predictions; symbols with error bars are Monte
Carlo data ($N=2000$, $M_L=M_U=30$, $50$ disorder realizations).
As $\rho$ increases (dataset quality deteriorates), all curves decay, but
the optimal semi-supervised strategy maintains a higher magnetization over the
entire range.
The inset magnifies the high-magnetization region
($\rho\in[0.15,0.6]$, $m\in[0.7,1.0]$).
\emph{Right panel:} closed-form optimal mixing parameter $\lambda^\star$ as a function of
$\rho_L=\rho_U$.
The solid curve denotes the exact stationary root $\lambda^\star=1-x^\star$
from~\eqref{eq:lambdastar}; square markers indicate the discrete values evaluated in the MC simulations.
Notably, at perfect dataset quality $\rho_L=\rho_U=0$ (corresponding to $r=1$), within-class
noise vanishes and the system enters the degenerate regime where the optimal mixing reduces to the
empirical fraction of labeled examples, $\lambda^\star\big|_{r\to 1} = M_L/(M_L+M_U) = 0.5$ (see Remark~\ref{rem:r1-limit}).
For any $r<1$ ($\rho_L=\rho_U>0$), noise corrections detach $\lambda^\star$ from the naive ratio $0.5$,
smoothly increasing it towards $\sim 0.86$ to adaptively suppress unlabeled interference.
}}
\label{fig:phase}
\end{figure}
Asking which mixing $\lambda\in[0,1]$ is \textit{optimal} is ambiguous until one
declares by which principle the selection is made, and the two natural
principles pull in opposite directions. On the one hand one may treat
$\lambda$ on the same footing as the other macroscopic variables and let
equilibrium thermodynamics select it, i.e.\ look for the stationary point of
the quenched pressure along the mixing family: this is a \emph{global}
criterion, blind to which minimum the system is sitting in. On the other hand
one may keep $\lambda$ for what it operationally is, a hyperparameter of the
synaptic rule~\eqref{eq:J-def}, fixed by the experimenter before the dynamics
starts, and tune it so as to maximise the \emph{local} stability of the
target pattern, hence the critical load: this is the criterion implicit in the
signal-to-noise analysis of Section~\ref{sec:S2N}. The two routes do not
select the same value, and the discrepancy is not a technical accident: as we
show below, thermodynamics alone \emph{refuses} the mixture altogether. We
denote by $\lambda_{\rm eq}$ the thermodynamic stationarity point and by
$\lambda^{\star}$ the retrieval-optimal mixing, and we address them in turn.
Throughout this section we work in the large dataset limit
$M_{L},M_{U}\to\infty$, in which the self-consistency system reduces
to~(\ref{eq:m-largeM}--\ref{eq:pk-largeM}).

\subsection*{Thermodynamic stationarity }

Let $\Omega:=\{\avg m,\avg{ q_{12}},\avg{p_{12}^{1}}, \avg{p^{2}_{12}}, \hdots, \avg{p^{M_U+1}_{12}}\}$ denote the set of RS
order parameters. The pressure $\mathcal A=\mathcal A(\lambda,\Omega)$
depends on $\lambda$ both explicitly and implicitly through the elements of $\Omega$,
which are fixed by the self-consistency
equations~(\ref{eq:m-largeM}--\ref{eq:pk-largeM}). Its total derivative
with respect to $\lambda$ is
\begin{equation}\label{eq:df-dlambda}
\frac{d\mathcal A}{d\lambda}=\frac{\partial \mathcal A}{\partial\lambda}
+\sum_{O\in\Omega}\frac{\partial \mathcal A}{\partial O}\,\frac{\partial O}{\partial\lambda}.
\end{equation}
The self-consistency equations are precisely the stationarity conditions
$\partial \mathcal A/\partial O=0$, so every term in the implicit sum vanishes and only the
explicit $\lambda$-dependence survives:%
\footnote{This stationarity) property of the RS
variational free energy is standard in the statistical mechanics of disordered
systems~\cite{MPV,nishimori2001statistical,MezardMontanari}: the total derivative of the
quenched pressure with respect to a control parameter reduces to its explicit part
because the order parameters sit at a saddle point. It is the finite-temperature,
classical counterpart of the Hellmann-Feynman theorem of quantum mechanics \cite{Hellmann1937, Feynman1939, Guttinger1932}, where the
derivative of an eigenvalue with respect to a parameter of the Hamiltonian $\mathcal H$ equals the
expectation of $\partial_\lambda\mathcal H$ on the corresponding eigenstate, the
first-order variation of the state itself dropping out by normalization and
stationarity.}
\begin{equation}\label{eq:HF}
\frac{d\mathcal A}{d\lambda}=\frac{\partial \mathcal A}{\partial\lambda}\bigg|_{\Omega\ \text{fixed}}.
\end{equation}
The $\lambda-$gradient of the pressure is therefore controlled by the
{explicit} $\lambda-$dependence of the quenched statistical pressure $\mathcal A$ alone, with the order
parameters frozen at their self-consistent values.

We remind that we can write the Boltzmann
weight as in Eq. \eqref{eq:H-lambda-split}:
\begin{equation}\label{eq:Ham_sec5}
-\beta\Hcal(\bm\sigma)=\lambda\,\beta N\,\Ecal_{L}(\bm\sigma)+(1-\lambda)\,\beta N\,\Ecal_{U}(\bm\sigma),
\end{equation}
and $\Ecal_X$, for $X \in {L,U}$ were already defined in Eq. \eqref{eq:ELU-def}.

Differentiating~\eqref{eq:Ham_sec5} at fixed $\bm\sigma$ gives
$\partial_{\lambda}(-\beta\Hcal)=\beta N[\Ecal_{L}-\Ecal_{U}]$, so by
the identity~\eqref{eq:HF},
\begin{equation}\label{eq:HF-explicit}
\frac{\partial \mathcal A}{\partial\lambda}=\beta\lim_{N\to\infty}\E\,\omega\bigl(\Ecal_{L}-\Ecal_{U}\bigr).
\end{equation}
The stationarity condition $\partial \mathcal A/\partial\lambda=0$ therefore
acquires a transparent physical content: the thermodynamically stationary
mixing $\lambda_{\rm eq}$ {equipartitions the macroscopic energy}
extracted from the labeled and unlabeled datasets,
\begin{empheq}[]{equation}\label{eq:energy-balance}
\E\,\omega(\Ecal_{L})=\E\,\omega(\Ecal_{U}).
\end{empheq}

The variational status of this condition, whether $\lambda_{\rm eq}$ is a
maximum or a minimum of the pressure, and whether it is unique, is completely
fixed by an elementary but consequential structural fact: the
Hamiltonian~\eqref{eq:H-lambda-split} is {convex} in $\lambda$, so that
$\lambda\mapsto N^{-1}\ln\Zcal_{N,\beta}$ has second derivative
\begin{equation}\label{eq:d2A}
\frac{\partial^2}{\partial\lambda^2}\,\frac{\ln\Zcal_{N,\beta}}{N}
=\beta^2N\,\Var_{\omega}\bigl(\Ecal_L-\Ecal_U\bigr)\;\ge\;0 ,
\end{equation}
a Gibbs variance, hence non-negative for every $N$, every $\beta>0$ and every
realization of the dataset $\Scal$. Convexity is preserved by the quenched
average and by pointwise limits, so both $\Acal_N$ and $\Acal$ are convex
in the mixing parameter\footnote{Here and in Appendix~\ref{app:convexity} the
energy densities of \eqref{eq:ELU-def} are normalised so that
$-\beta\Hcal=\beta N[\lambda\Ecal_L+(1-\lambda)\Ecal_U]$ is extensive, i.e.\
$\Ecal_L=\frac{1}{2\Gamma_L}\sum_\mu\bigl(\sum_a n_L^{\mu a}\bigr)^2$ and
$\Ecal_U=\frac{1}{2\Gamma_U}\sum_{\mu,c}(n_U^{\mu c})^2$.}. The consequences we
need are collected, with their proofs, in Appendix~\ref{app:convexity}; here we
only recall them and comment on their physical content.

\par\medskip
First, the derivative $d\Acal/d\lambda=\beta\,\E\,\omega(\Ecal_L-\Ecal_U)$ is
non-decreasing, so the equipartition condition~\eqref{eq:energy-balance} has a
solution set which is a closed interval, reduced to the single point
$\lambda_{\rm eq}$ whenever the convexity is strict: the energy balance
identifies a well-defined mixing, and no spurious multiplicity of thermodynamic
optima can arise. Second, and decisively, an interior solution
of~\eqref{eq:energy-balance} is a global \emph{minimum} of
$\lambda\mapsto\Acal(\beta,\lambda)$, i.e.\ a global \emph{maximum} of the
quenched free energy; equivalently, by the inequality
\begin{equation}\label{eq:chord}
\Acal(\beta,\lambda)\;\le\;\lambda\,\Acal(\beta,1)+(1-\lambda)\,\Acal(\beta,0),
\end{equation}
the pressure of the mixture never exceeds the interpolation of the two pure
channels. The mixed protocol is thus, thermodynamically speaking, the
\emph{least} favourable one: a selection principle based on relaxation to the
lowest free energy at fixed temperature would push the synaptic rule towards the
pure channels $\lambda\in\{0,1\}$ and would never stop at an interior point.
Equilibrium thermodynamics, by itself, does not select semi-supervised learning, it rejects it. This is the precise sense in which the convexity
of $\Acal$ plays a ``no-go'' role: it forbids reading $\lambda$ as an order
parameter that the system chooses spontaneously, and it thereby provides the
formal justification for treating it as what it operationally is, an externally
imposed hyperparameter of the learning rule.

Once $\lambda$ is recognised as a hyperparameter, the objective changes
accordingly. One is no longer interested in the global free-energy landscape,
but in the stability of one specific minimum, the one aligned with the target
pattern. The natural figure of merit is then the local signal-to-noise ratio of
Section~\ref{sec:S2N}: the signal $S(\lambda)$ against the total noise variance
$V(\lambda)=C(\lambda)\alpha+R(\lambda)$, where $C$ collects the
cross-pattern interference and $R$ the within-class fluctuations of the
examples. These two contributions compete. The interference term alone is
minimised by pure supervision, $C'(\lambda)<0$ on $(0,1)$, and would drive
$\lambda\to1$; but $R(\lambda)$ is convex and penalises both extremes of the
mixing range, since discarding either channel throws away the averaging that
suppresses the within-class noise. It is exactly this trade-off, absent from
the pressure, which only sees the total energy balance, that places the noise
minimum, hence the capacity maximum, at an interior $\lambda^\star\in(0,1)$.
Far from diminishing the role of $\lambda$, the dichotomy confirms it as a
genuine hyperparameter: its optimal tuning encodes information on the dataset
(quality $r$ and per-channel sizes $M_L,M_U$) that the free energy alone cannot
see. Note also that, although exact, the equipartition
characterization~\eqref{eq:energy-balance} is not usable in practice to locate
any mixing in closed form, which makes the passage to the SNR observables
necessary as well as conceptually mandatory.

\subsection*{Retrieval capacity maximization}

We remind that the one-step magnetization has this form:
\begin{align}
    m^{(1)}(\bm \xi^1)=\textnormal{erf} \left( \dfrac{\dfrac{\lambda}{1+\rho_L} + (1-\lambda)}{\sqrt{2(C(\lambda,r,\rho_L,\rho_U) \alpha + R(\lambda,r,\rho_L,\rho_U))}}\right).
\end{align}
asking the usual  condition $m^{(1)}(\bm \xi^1)>\erf(\theta)$, with the canonical choice $\theta=1/\sqrt 2$ we can get an estimate for the load:
\begin{equation}
\label{eq:alpha_c}
    \alpha_c^{\rm SNR}=\frac{(S(\lambda,r,\rho_L,\rho_U))^2-R(\lambda,r,\rho_L,\rho_U)}{C(\lambda,r,\rho_L,\rho_U)}.
\end{equation}
with $S(\lambda,r,\rho_L,\rho_U),\ R(\lambda,r,\rho_L,\rho_U),\ C(\lambda,r,\rho_L,\rho_U)$ of equations \eqref{eq:signal}, \eqref{eq:C-final_main} and \eqref{eq:R-final_main}. Since $\alpha_c^{\rm SNR}$ is a physical
capacity only when non-negative, we determine the optimal mixing as the constrained
maximiser
\begin{equation}\label{eq:constrained-program}
\lambda^\star=\arg\max_{\lambda\in[0,1]}\alpha_c^{\rm SNR}(\lambda)\qquad
\text{subject to}\qquad \alpha_c^{\rm SNR}(\lambda)\ge 0 .
\end{equation}
The stationarity $\partial_{\lambda}\alpha_{c}=0$, contains cubic terms that cancels identically. Reducing to the quadratic (we have set $x:=1-\lambda$):
\begin{equation}\label{eq:lambdastar-quadratic}
\mathfrak{A}\,x^{2}-\mathfrak{B}\,x+\mathfrak{C}=0,
\end{equation}
with coefficients:
\begin{empheq}[]{align}
\mathfrak{A}&=\eta\bigl[\rho_{U}+\eta(1-2\eta)-2(1-\eta)(B+2\eta)\bigr],\label{eq:coef-A}\\[3pt]
\mathfrak{B}&=\rho_{U}+\eta(1-2\eta)+(1-\eta)(1-2\eta)(B+2\eta),\label{eq:coef-B}\\[3pt]
\mathfrak{C}&=\eta(1-\eta)(3-2\eta),\label{eq:coef-C} \\
\eta &=\frac{\rho_L}{1+\rho_L}, \\
B &= \dfrac{1 - r^4}{M_Ur^4}.
\end{empheq}
The physical root, i.e.\ the one lying in $[0,1]$, is
\begin{empheq}{equation}\label{eq:lambdastar}
\lambda^{\star}=1-x^{\star},\qquad
x^{\star}=\frac{2\,\mathfrak{C}}{\mathfrak{B}+\sqrt{\mathfrak{B}^{2}-4\mathfrak{A}\mathfrak{C}}}.
\end{empheq}
Since $C(\lambda)>0$, the constraint in~\eqref{eq:constrained-program} reduces to
$\alpha_c^{\rm SNR}\ge 0\Leftrightarrow S^2\ge R$ and never binds at an interior optimum:
either $\alpha_c^{\rm SNR}(\lambda^\star)>0$, and~\eqref{eq:lambdastar} is the maximiser,
or $\alpha_c^{\rm SNR}(\lambda^\star)\le 0$, in which case $\alpha_c^{\rm SNR}(\lambda)\le 0$
for all $\lambda$ and no retrieval solution exists. The equality
$S^2(\lambda^\star)=R(\lambda^\star)$ therefore fixes a critical quality
$r_{\min}(M_L,M_U)$.

An explicit expression for the threshold follows from the fact that, at the boundary of
the retrieval region, the constrained maximum vanishes, $\alpha_c^{\rm SNR}(\lambda^\star)=0$
together with $\partial_\lambda\alpha_c^{\rm SNR}=0$. Since $\alpha_c^{\rm SNR}=(S^2-R)/C$
with $C>0$, this forces the numerator $N(\lambda)=S^2-R$ to develop a double root. Being
quadratic in $\lambda$,
\begin{equation}\label{eq:Nnum}
N(\lambda)=n_0+n_1\lambda+n_2\lambda^2,\qquad
n_0=1-\rho_U,\quad n_1=2(\rho_U-\eta),\quad n_2=\eta(2\eta-1)-\rho_U ,
\end{equation}
its discriminant reads $\tfrac14\Delta_\lambda=(1-\eta)\bigl[\eta-\rho_U(2\eta-1)\bigr]$
(note that $N$ is independent of $B$, so the threshold does not depend on the load or on
the interference structure). Setting $\Delta_\lambda=0$ and using $\eta=\rho_L/(1+\rho_L)$,
the boundary reduces to the symmetric condition
\begin{equation}\label{eq:threshold-rho}
\frac{1}{\rho_L}+\frac{1}{\rho_U}=1
\qquad\Longleftrightarrow\qquad
\rho_L\rho_U=\rho_L+\rho_U .
\end{equation}
Inserting $\rho_{L,U}=(1-r^2)/(M_{L,U}\,r^2)$ gives $(M_L+M_U)\,r^2=1-r^2$, i.e.\ the
closed-form retrieval threshold
\begin{empheq}{equation}\label{eq:rmin}
r_{\min}=\frac{1}{\sqrt{M_L+M_U+1}} .
\end{empheq}
For $r<r_{\min}$ the within-class fluctuations $\rho_{L,U}\sim 1/r^2$ overcome the signal
and no retrieval solution exists; the optimal mixing $\lambda^\star$ is therefore defined
only for $r>r_{\min}$.

\begin{remark}\label{rem:r1-limit}
At perfect dataset quality $r\to 1$ the two channels become noiseless,
$\rho_L,\rho_U\to 0$ (hence $\eta\to 0$) and $B\to 0$, so that all the coefficients of
the stationarity quadratic~\eqref{eq:lambdastar-quadratic} vanish simultaneously,
$\mathfrak A,\mathfrak B,\mathfrak C\to 0$, and the closed form~\eqref{eq:lambdastar}
becomes indeterminate. This is not a divergence but a
{degeneracy}: at $r=1$ one has $S=1$, $C=1$ and $R=0$ for every $\lambda$, so the
capacity is flat,
\begin{equation}\label{eq:alphac-flat}
\alpha_c^{\rm SNR}(\lambda)=\frac{S^2-R}{C}= 1\qquad\text{for all }\lambda\in[0,1],
\end{equation}
and the maximiser is not unique,  every mixing is optimal.

The degeneracy is lifted at first order in $\varepsilon:=1-r^2$. Using
$\eta\simeq\varepsilon/M_L$, $\rho_U\simeq\varepsilon/M_U$ and $B\simeq 2\varepsilon/M_U$,
one finds $\mathfrak A=O(\varepsilon^{2})$ while $\mathfrak B,\mathfrak C=O(\varepsilon)$,
so that
\begin{equation}\label{eq:xstar-r1}
x^\star=\frac{2\mathfrak C}{\mathfrak B+\sqrt{\mathfrak B^{2}-4\mathfrak A\mathfrak C}}
\;\xrightarrow{r\to 1}\;\frac{\mathfrak C}{\mathfrak B}
=\frac{3\eta}{\rho_U+B+3\eta}\;\longrightarrow\;\frac{M_U}{M_L+M_U},
\end{equation}
and therefore
\begin{empheq}{equation}\label{eq:lambdastar-r1}
\lambda^\star\big\vert_{r\to 1}=\frac{M_L}{M}.
\end{empheq}
Thus $\lambda^\star(r)$ is continuous and finite up to $r=1$, where it reduces to the
{empirical fraction of labeled examples}. Remarkably, this is the {only} regime
in which the optimal synaptic weight coincides with the naive proportion $M_L/M$ that,
as stressed in the Remark \ref{rem:lambda}, does
{not} in general control the learning rule: when the dataset is ideal the
within-class noise balance vanishes and only the counting of examples survives, whereas
for any $r<1$ the $O(1-r^2)$ corrections detach $\lambda^\star$ from $M_L/M$.

\par\medskip
The retrieval consequences of the optimal mixing $\lambda^\star$ are illustrated in
Figures~\ref{fig:entropy} and~\ref{fig:phase}. In the $(\alpha,T)$ plane
(Fig.~\ref{fig:entropy}, left) the retrieval boundary evaluated at $\lambda=\lambda^\star$
lies strictly to the right of both pure channels at every temperature, so the mixed rule
genuinely enlarges the retrieval region; the near-zero-temperature magnetization curves
(right) exhibit a sharp retrieval-to-spin-glass transition, with the RS
prediction and the MC data collapsing onto the same critical load.
The robustness of this picture against dataset quality is summarised in
Figure~\ref{fig:phase}: the one-step magnetization heat map (left) shows that the
closed-form SNR threshold $\alpha_c^{\rm SNR}(\rho)$ of~\eqref{eq:alpha_c} faithfully
tracks the numerically observed boundary across the whole $(\alpha,\rho)$ plane, while at
fixed load (center) the optimal mixture sustains a higher magnetization than either pure
channel as the noise $\rho$ grows. Finally (right), $\lambda^\star(\rho)$ detaches from the
naive labeled fraction $M_L/M=1/2$ as soon as $\rho>0$ and increases monotonically towards
$\simeq 0.86$: the network adaptively shifts weight onto the supervised channel to suppress
the diverging unlabeled interference, confirming that $\lambda^\star$ is a genuine,
dataset-dependent hyperparameter rather than a proxy for the proportion of labels.
 \end{remark}

\section{Conclusions and outlooks} \label{sec:conclusions}
In this paper we introduced and solved a semi-supervised extension of the Hopfield model, conceived to mirror the operationally relevant scenario of Machine Learning practice, where labeled examples are scarce and unlabeled ones abundant. The network is supplied with a dataset made of $M_L$ labeled and $M_U$ unlabeled noisy examples of $K$ unknown archetypes, and its synaptic coupling is the convex combination $J^{\lambda}=\lambda J^{(L)}+(1-\lambda)J^{(U)}$ of the supervised and unsupervised Hebbian kernels, with the mixing parameter $\lambda\in[0,1]$ tuning the relative weight assigned by the network to the two channels of information.

Hereafter we summarize the main outcomes of our work.

First, by a signal-to-noise analysis of the zero-temperature one-step dynamics we obtained the Mattis magnetization on the target archetype in closed form, whence we derived the threshold for learning $M^\star(\alpha,r,\lambda)$, namely the minimal dataset size guaranteeing a stable retrieval state at given storage load $\alpha$ and dataset quality $r$: the resulting learning boundaries in the $(M,r)$ plane are in complete agreement with Monte Carlo simulations, at all the inspected values of $\lambda$ and of the network size.

Second, by adapting Guerra's interpolation to the present setting, we derived the RS quenched statistical pressure in the high-storage regime, together with the self-consistency equations for the order parameters and their zero-temperature limit. The main technical hurdle lies in the quenched noise: since labeled and unlabeled examples stem from the same archetypes, the two channels do not decouple and the slow noise is governed by a non-trivial matrix whose exact spectral decomposition, two hybridized condensed modes, an unsupervised bulk and a supervised null space, allows to rewrite the disorder as a collection of independent eigen-channels. The resulting phase diagrams show that the retrieval region achieved at intermediate mixing is strictly larger than those of both the pure protocols: semi-supervised Hebbian learning genuinely outperforms its supervised and unsupervised limits.

Finally, we proved that the quenched pressure is convex in the mixing parameter, so that its interior stationary point, characterized by the exact equipartition of the macroscopic energies extracted from the two subsets of the dataset, is a maximum of the free energy: equilibrium thermodynamics, by itself, rejects the mixture and would drive the synaptic rule toward the pure channels. Far from being a drawback, this no-go result provides the formal justification for treating $\lambda$ as what it operationally is, an externally imposed hyperparameter of the learning rule, rather than an order parameter spontaneously selected by the system.

 Once the optimality criterion is shifted from the free energy to the retrieval capacity, the trade-off between cross-archetype interference (minimized by pure supervision) and within-class fluctuations (which penalize both extremes of the mixing range) places the capacity maximum at an interior value $\lambda^\star\in(0,1)$, for which we provided a closed-form expression, together with the minimal dataset quality $r_{\min}=1/\sqrt{M_L+M_U+1}$ below which no retrieval is possible, whatever the mixing. Remarkably, $\lambda^\star$ coincides with the empirical fraction of labeled examples $M_L/(M_L+M_U)$ only for noiseless datasets, while, as the dataset quality deteriorates, it progressively detaches from it, shifting weight onto the supervised channel so as to suppress the diverging unlabeled interference: the optimal mixing is a genuine, dataset-dependent hyperparameter and not a proxy for the proportion of labels.

As a technical remark, we stress that the strategy pursued here, the gaussianization of the two channels via CLT and universality of the quenched noise \cite{CarmonaWu, Genovese}, followed by the exact diagonalization of the correlated disorder into independent eigen-channels, is of broad applicability and can be exported to other multi-channel Hebbian settings beyond the present one.

\par\medskip
As for future perspectives, several directions naturally stem from this work. First, the whole construction can be extended to dense architectures, where neurons interact in groups of $P>2$ units \cite{super, unsup}: there, the interplay between density and semi-supervision is expected to be particularly rich, as the supervised and unsupervised channels are known to respond very differently to the interaction order. Second, our analysis is confined to the RS level of description: relaxing this assumption, e.g.\ by inspecting the instability line of the RS solution \cite{albanese2023almeida} and by moving to broken replica-symmetry schemes\cite{guerra_broken}, would sharpen the low-temperature portion of the phase diagram. Third, the present framework is naturally suited to be exported to hetero-associative architectures \cite{agliari2025generalized,AGLIARI2026108181}, where multiple layers of neurons learn, store and retrieve structured tuples of patterns and whose supervised and unsupervised Hebbian protocols have been recently worked out \cite{alessandrelli2025supervised, alessandrelli2025beyond, AGLIARI2026131134}: there, semi-supervision would intertwine with the cooperative interplay among layers, possibly enlarging the retrieval region along both directions at once. Finally, having recognized $\lambda$ as a hyperparameter, it is natural to ask whether the network itself could adaptively adjust it during training, namely to promote the mixing parameter to a slow dynamical variable: we plan to report soon on this topic.

\acknowledgments
L.A. acknowledges funding from the project “Patto Territoriale del Sistema Universitario Pugliese” (CUP F61B23000370006). \\
L.A., A.Ladiana and A.Lepre are members of the GNFM group within INdAM which is acknowledged. 
\noindent
\\
The authors acknowledge the use of the Lagrange Multi-GPU Server at the Department of Mathematics, Sapienza University of Rome, for computational resources supporting this work.

\appendix

\section{Evaluation of the momenta of the effective post synaptic potential}
\label{app:S2N}

In this section we compute in details the two main quantities of section \ref{sec:S2N}: $S:=\E\bigl[h_i\,\xi_i^1\bigr]=\E\bigl[(h_i^{(L)}+h_i^{(U)})\xi_i^1\bigr]$ and $V:=\Var\bigl(h_i\,\xi_i^1\bigr)$, where we remind that:
\begin{align}
    h_i&:= h_i^{(L)} + h_i^{(U)} \notag \\
    &=\frac{\lambda}{N\Gamma_L}\sum_{\mu=1}^{K}\sum_{a,b=1}^{M_L}\eta_i^{\mu a}\sum_{j\ne i}\eta_j^{\mu b}\,\xi_j^1 + \frac{1-\lambda}{N\Gamma_U}\sum_{\mu=1}^{K}\sum_{c=1}^{M_U}\eta_i^{\mu c}\sum_{j\ne i}\eta_j^{\mu c}\,\xi_j^1.
\end{align}
 We split each contribution into a \emph{signal} part (terms with $\mu=1$, the archetype to be retrieved) and a \emph{noise} part (terms with $\mu\ge 2$, the other archetypes that act as quenched interference):
\begin{align}
h_i^{(L)}&=h_i^{(L),\,\mu=1}+h_i^{(L),\,\mu\ge 2},\\[2pt]
h_i^{(U)}&=h_i^{(U),\,\mu=1}+h_i^{(U),\,\mu\ge 2},
\end{align}
where, using $\eta_j^{1b}\xi_j^1=\chi_j^{1b}$,
\begin{align}
h_i^{(L),\,\mu=1}&=\frac{\lambda}{N\Gamma_L}\Bigl(\sum_{a=1}^{M_L}\chi_i^{1a}\Bigr)\Bigl(\sum_{b=1}^{M_L}\sum_{j\ne i}\chi_j^{1b}\Bigr)\,\xi_i^1,\label{eq:hiL-mu1}\\[4pt]
h_i^{(U),\,\mu=1}&=\frac{1-\lambda}{N\Gamma_U}\sum_{c=1}^{M_U}\chi_i^{1c}\sum_{j\ne i}\chi_j^{1c}\,\xi_i^1,\label{eq:hiU-mu1}\\[4pt]
h_i^{(L),\,\mu\ge 2}&=\frac{\lambda}{N\Gamma_L}\sum_{\mu\ge 2}\Bigl(\sum_a\eta_i^{\mu a}\Bigr)\Bigl(\sum_b\sum_{j\ne i}\eta_j^{\mu b}\xi_j^1\Bigr),\label{eq:hiL-noise}\\[4pt]
h_i^{(U),\,\mu\ge 2}&=\frac{1-\lambda}{N\Gamma_U}\sum_{\mu\ge 2}\sum_c\eta_i^{\mu c}\sum_{j\ne i}\eta_j^{\mu c}\xi_j^1.\label{eq:hiU-noise}
\end{align}
The factor $\xi_i^1$ pulled out of the signal terms makes manifest that, in expectation, these terms produce a contribution \emph{aligned} with $\xi_i^1$,precisely the signal we want to control.
\vspace{2mm}
We start computing the signal, and we notice that
for $\mu\ge 2$, the inner sum $\sum_{j\ne i}\eta_j^{\mu b}\xi_j^1$ involves the product $\xi_j^\mu\xi_j^1$ of two independent Rademacher variables, hence its expectation vanishes: $\E\bigl[\sum_{j\ne i}\eta_j^{\mu b}\xi_j^1\bigr]=0$. The only non-zero contribution to $S$ comes from the $\mu=1$ pieces~\eqref{eq:hiL-mu1} and \eqref{eq:hiU-mu1}.

Starting from~\eqref{eq:hiL-mu1} and multiplying by $\xi_i^1$ (which cancels the $\xi_i^1$ already in the formula since $(\xi_i^1)^2=1$):
\begin{equation}\label{eq:hiL-mu1-times-xi}
h_i^{(L),\,\mu=1}\xi_i^1=\frac{\lambda}{N\Gamma_L}\Bigl(\sum_{a=1}^{M_L}\chi_i^{1a}\Bigr)\Bigl(\sum_{b=1}^{M_L}\sum_{j\ne i}\chi_j^{1b}\Bigr).
\end{equation}
The two factors involve disjoint sets of site indices, hence by independence of the $\chi$'s across sites,
\begin{equation}\label{eq:signal-L-step1}
\E\bigl[h_i^{(L),\,\mu=1}\xi_i^1\bigr]=\frac{\lambda}{N\Gamma_L}\Bigl(\sum_a\E[\chi_i^{1a}]\Bigr)\Bigl(\sum_b\sum_{j\ne i}\E[\chi_j^{1b}]\Bigr)=\frac{\lambda}{N\Gamma_L} M_L r(N-1)M_L r,
\end{equation}
having used $\E[\chi_i^{1a}]=r$. Taking $N\to\infty$ and substituting $\Gamma_L=M_L^2\Rcal_L=M_L^2 r^2(1+\rho_L)$,
\begin{equation}\label{eq:signal-L-final}
\E\bigl[h_i^{(L),\,\mu=1}\xi_i^1\bigr]\;\xrightarrow{N\to\infty}\;\frac{\lambda M_L^2 r^2}{\Gamma_L}=\frac{\lambda}{1+\rho_L}.
\end{equation}
The $1/(1+\rho_L)$ comes from the within-class fluctuation correction $1-r^2$ in $\Rcal_L$. In the large dataset limit $M_L\to\infty$ one has $\rho_L\to 0$ and the signal reduces to $\lambda$ exactly.

\vspace{2mm}
Starting from~\eqref{eq:hiU-mu1} and multiplying by $\xi_i^1$:
\begin{equation}\label{eq:hiU-mu1-times-xi}
h_i^{(U),\,\mu=1}\xi_i^1=\frac{1-\lambda}{N\Gamma_U}\sum_{c=1}^{M_U}\chi_i^{1c}\sum_{j\ne i}\chi_j^{1c}.
\end{equation}
By the same disjoint-site argument as before, but now with each example $c$ contributing independently,
\begin{equation}\label{eq:signal-U-step1}
\E\bigl[h_i^{(U),\,\mu=1}\xi_i^1\bigr]=\frac{1-\lambda}{N\Gamma_U}\sum_{c=1}^{M_U}\E[\chi_i^{1c}]\sum_{j\ne i}\E[\chi_j^{1c}]=\frac{1-\lambda}{N\Gamma_U} M_Ur (N-1)r.
\end{equation}
Taking $N\to\infty$ and using $\Gamma_U=M_U r^2$,
\begin{equation}\label{eq:signal-U-final}
\E\bigl[h_i^{(U),\,\mu=1}\xi_i^1\bigr]\;\xrightarrow{N\to\infty}\;\frac{(1-\lambda)M_U r^2}{\Gamma_U}=1-\lambda.
\end{equation}
Notice that no $\rho_U$ correction appears here, because $\Gamma_U$ has been chosen to be exactly $M_U r^2$.
\vspace{2mm}
The variance of $h_i$ receives contributions from \emph{every} cross-product in $\left(h_i^{(L)}+h_i^{(U)}\right)^2$. There are five qualitatively distinct contributions, that we label from (a) to (e):
\begin{itemize}[leftmargin=2.5em]
\item[(a)] supervised \emph{interference} (terms with $\mu\ne 1$ in $(h_i^{(L)})^2$);
\item[(b)] supervised \emph{within-class fluctuations} (terms with $\mu=1$ in $(h_i^{(L)})^2$);
\item[(c)] unsupervised \emph{interference} (terms with $\mu\ne 1$ in $(h_i^{(U)})^2$);
\item[(d)] unsupervised \emph{within-class fluctuations} (terms with $\mu=1$ in $(h_i^{(U)})^2$);
\item[(e)] supervised-unsupervised \emph{cross-channel} covariance ($\mu\ne 1$ in $2h_i^{(L)}h_i^{(U)}$).
\end{itemize}
The cross-channel covariance at $\mu=1$ vanishes once we recombine the two channels with the signal $\xi_i^1$ already subtracted, and is therefore not listed.

\vspace{2mm}
Starting from (a), we define for simplicity:
\begin{equation}
A_\mu^{(L)}:=\sum_{a=1}^{M_L}\eta_i^{\mu a},\qquad B_\mu^{(L)}:=\sum_{b=1}^{M_L}\sum_{j\ne i}\eta_j^{\mu b}\xi_j^1,
\end{equation}
so that $h_i^{(L),\mu\ge 2}=(\lambda/(N\Gamma_L))\sum_{\mu\ge 2}A_\mu^{(L)}B_\mu^{(L)}$. The two factors involve disjoint site indices, hence they are independent. Using $\E[\chi^a\chi^{a'}]=\delta_{aa'}+(1-\delta_{aa'})r^2$,
\begin{align}
\E\bigl[(A_\mu^{(L)})^2\bigr]&=\sum_{a,a'}\E[\chi^a\chi^{a'}]=M_L+M_L(M_L-1)r^2=M_L^2 r^2+M_L(1-r^2)=M_L^2\Rcal_L=\Gamma_L,\label{eq:A2}\\[3pt]
\E\bigl[(B_\mu^{(L)})^2\bigr]&=\sum_{b,b'}\E[\chi^b\chi^{b'}]\sum_{j\ne i}1\;=\;(N-1)\,M_L^2\Rcal_L\;\simeq\;N\,\Gamma_L.\label{eq:B2}
\end{align}
The means $\E[A_\mu^{(L)}]$ and $\E[B_\mu^{(L)}]$ vanish for $\mu\ne 1$ (by $\E_\xi[\xi^\mu]=0$), so the variance of $A_\mu^{(L)}B_\mu^{(L)}$ is just $\E[A^2]\E[B^2]=N\Gamma_L^2$. Summing over the $K-1\simeq K$ independent contributions,
\begin{equation}\label{eq:Var-a}
\Var\bigl(h_i^{(L),\,\mu\ge 2}\bigr)=\frac{\lambda^2}{N^2\Gamma_L^2}(K-1)N\Gamma_L^2\;\xrightarrow{N\to\infty}\;{\;\lambda^2\,\alpha.\;}
\end{equation}
\vspace{2mm}
Instead, for the case (b), and so $\mu=1$, $A_1^{(L)}=\sum_a\chi_i^{1a}$ and $B_1^{(L)}=\sum_b\sum_{j\ne i}\chi_j^{1b}$ have non-zero means $\E[A_1]=M_L r$, $\E[B_1]=(N-1)M_L r$, and variances $\Var(A_1)=M_L(1-r^2)$, $\Var(B_1)=(N-1)M_L(1-r^2)$. The full variance of the product $A_1 B_1$ (with $A_1,B_1$ independent) is given by the standard formula
\begin{equation}\label{eq:Var-AB}
\Var(A_1 B_1)=\Var(A_1)\Var(B_1)+\Var(A_1)\E[B_1]^2+\E[A_1]^2\Var(B_1).
\end{equation}
The three contributions scale, respectively, as $N M_L^2(1-r^2)^2$, $N^2 M_L^3 r^2(1-r^2)$, and $N M_L^3 r^2(1-r^2)$. The dominant one in $N$ is the middle term $N^2 M_L^3 r^2(1-r^2)$. Using $\Gamma_L^2=M_L^4\Rcal_L^2=M_L^4 r^4(1+\rho_L)^2$,
\begin{equation}\label{eq:Var-b}
\Var\bigl(h_i^{(L),\,\mu=1}\bigr)=\frac{\lambda^2 M_L^3 r^2(1-r^2)}{\Gamma_L^2}=\frac{\lambda^2(1-r^2)}{M_L r^2(1+\rho_L)^2}\;\xrightarrow{N\to\infty}\;{\;\frac{\lambda^2\rho_L}{(1+\rho_L)^2}.\;}
\end{equation}
The factor $1/(1+\rho_L)^2$ is the same factor that appears (squared in the signal) in~\eqref{eq:signal-L-final}: in the SNR ratio $S/\sqrt V$ the two cancel cleanly.
\vspace{2mm}
For the case (c) and (d), we define $Y_\mu^c\equiv\eta_i^{\mu c}\sum_{j\ne i}\eta_j^{\mu c}\xi_j^1$. Then $\E[Y_\mu^c]=0$ (because $\E_\xi[\xi^\mu]=0$) and
\begin{equation}\label{eq:Yc-square}
\E[(Y_\mu^c)^2]=\E[(\eta_i^{\mu c})^2]\sum_{j\ne i}\E\bigl[(\xi_j^\mu)^2(\xi_j^1)^2(\chi_j^{\mu c})^2\bigr]=1\cdot(N-1)\simeq N.
\end{equation}
For $c\ne c'$, both $\eta_i^{\mu c},\eta_i^{\mu c'}$ share the same archetype $\xi_i^\mu$ but the two $\chi$'s are independent, so $\E[\eta_i^{\mu c}\eta_i^{\mu c'}]=r^2$ and (by the same site-by-site analysis) $\E[Y_\mu^c Y_\mu^{c'}]=(N-1)r^4\simeq Nr^4$. Summing:
\begin{equation}\label{eq:sum-Yc}
\sum_{c,c'=1}^{M_U}\E[Y_\mu^c Y_\mu^{c'}]=M_U N+M_U(M_U-1) Nr^4\simeq NM_U(1+(M_U-1)r^4)=N M_U^2 r^4\left[1+\dfrac{1-r^4}{M_U r^4}\right].
\end{equation}
Multiplying by $(K-1)\simeq\alpha N$ and $(1-\lambda)^2/(N^2\Gamma_U^2)$ with $\Gamma_U=M_U r^2$,
\begin{equation}\label{eq:Var-c}
\Var\bigl(h_i^{(U),\,\mu\ge 2}\bigr)=\frac{(1-\lambda)^2}{N^2\Gamma_U^2}(\alpha N)(N M_U^2 r^4)\Bigl[1+\frac{1-r^4}{M_U r^4}\Bigr]\;\xrightarrow{N\to\infty}\;{\;(1-\lambda)^2\alpha\Bigl[1+\frac{1-r^4}{M_Ur^4}\Bigr].\;}
\end{equation}

The unsupervised within-class fluctuations have the structure $\sum_{c=1}^{M_U} X_c$ (sum over examples), not $(\sum_{c=1}^{M_U} X_c)^2$, so each example contributes independently to the variance. With $X_c=\chi_i^{1,c}\sum_{j\ne i}\chi_j^{1,c}$ (after using $\eta^{1,c}=\xi^1\chi^{1,c}$ and pulling out $\xi_i^1\xi_j^1$),
\begin{equation}\label{eq:Var-Xc}
\Var(X_c)=\Var(\chi_i^{1c})\Var\bigl(\sum_{j\ne i}\chi_j^{1c}\bigr)+\E[\chi_i^{1c}]^2\Var\bigl(\sum_{j\ne i}\chi_j^{1c}\bigr)+\Var(\chi_i^{1c})\E\bigl[\sum_{j\ne i}\chi_j^{1c}\bigr]^2.
\end{equation}
The dominant term is
$\Var(\chi_i^{1c})\,\E\bigl[\sum_{j\ne i}\chi_j^{1c}\bigr]^2
=(1-r^2)(N-1)^2 r^2\simeq N^2 r^2(1-r^2)$
(it dominates the other two cross-products, which are only $O(N)$ each).
Summing over $c$ (independent examples),
$\sum_c\Var(X_c)\simeq M_U N^2 r^2(1-r^2)$.
Multiplying by $(1-\lambda)^2/(N^2\Gamma_U^2)$ with $\Gamma_U^2=M_U^2 r^4$,
\begin{equation}\label{eq:Var-d}
\Var\!\bigl(h_i^{(U),\mu=1}\bigr)
=\frac{(1-\lambda)^2}{N^2 M_U^2 r^4} M_U N^2 r^2(1-r^2)
=\frac{(1-\lambda)^2(1-r^2)}{M_U r^2}
\;\xrightarrow{N\to\infty}\;{\;(1-\lambda)^2\rho_U.\;}
\end{equation}

The last term (e) is unique to the semi-supervised model. Contributions (a) and (c) both involve sums over $\mu\ne 1$, sharing the \emph{same} non-condensed archetype $\boldsymbol\xi^\mu$ (only the example index $a$ vs.\ $c$ differs). They are therefore \emph{not} independent, and their covariance must be added to the variance with a factor $2$. Writing $h_i^{(L),\,\mu\ge 2}=(\lambda/(N\Gamma_L))\sum_{\mu\ge 2}A_\mu^{(L)} B_\mu^{(L)}$ and $h_i^{(U),\,\mu\ge 2}=((1-\lambda)/(N\Gamma_U))\sum_{\mu\ge 2}\sum_{c=1}^{M_U} Y_\mu^c$,
\begin{equation}\label{eq:Cov-e-def}
\Cov\bigl(h_i^{(L),\,\mu\ge 2},h_i^{(U),\,\mu\ge 2}\bigr)=\frac{\lambda(1-\lambda)}{N^2\Gamma_L\Gamma_U}\,(K-1)\sum_{c=1}^{M_U}\E\bigl[A_\mu^{(L)}\,Y_\mu^c\bigr].
\end{equation}
By disjoint-site factorization,
\begin{equation}\label{eq:E-AY}
\E[A_\mu^{(L)} Y_\mu^c]=\E\bigl[A_\mu^{(L)}\eta_i^{\mu c}\bigr]\cdot\E\Bigl[B_\mu^{(L)\,'}\sum_{j\ne i}\eta_j^{\mu c}\xi_j^1\Bigr],
\end{equation}
where $B_\mu^{(L)\,'}$ denotes the $j$-sum part of $B_\mu^{(L)}$ (the $b$-sum factors out cleanly). Now
\begin{align}
\E\bigl[A_\mu^{(L)}\eta_i^{\mu c}\bigr]&=\sum_{a=1}^{M_L}\E\bigl[\eta_i^{\mu a}\eta_i^{\mu c}\bigr]=\sum_{a=1}^{M_L} r^2=M_L r^2,\\[3pt]
\E\Bigl[B_\mu^{(L)\,'}\sum_{j\ne i}\eta_j^{\mu c}\xi_j^1\Bigr]&=\sum_{b=1}^{M_L}\sum_{j\ne i}\E\bigl[\eta_j^{\mu b}\eta_j^{\mu c}\bigr]=(N-1)M_L r^2,
\end{align}
where we used $\E[\eta_i^{\mu a}\eta_i^{\mu c}]=\E[(\xi_i^\mu)^2\chi_i^{\mu a}\chi_i^{\mu c}]=\E[\chi]^2=r^2$ (different examples $a\ne c$ have independent $\chi$'s). Inserting:
\begin{align}
\Cov\bigl(h_i^{(L),\,\mu\ge 2},h_i^{(U),\,\mu\ge 2}\bigr)&=\frac{\lambda(1-\lambda)}{N^2\Gamma_L\Gamma_U}\cdot(K-1)\cdot M_U\cdot M_L r^2\cdot(N-1)M_L r^2\notag\\
&\xrightarrow{N\to\infty}\;\frac{\lambda(1-\lambda)\,\alpha\,M_U M_L^2 r^4}{\Gamma_L\Gamma_U}=\frac{\lambda(1-\lambda)\,\alpha\,M_U M_L^2 r^4}{M_L^2\Rcal_L\cdot M_U r^2}\notag\\
&=\frac{\lambda(1-\lambda)\,\alpha\,r^2}{\Rcal_L}={\;\frac{\lambda(1-\lambda)\,\alpha}{1+\rho_L}.\;}\label{eq:Cov-e-final}
\end{align}
This contribution enters the variance with the factor $2$ from $\Var(h^L+h^U)=\Var(h^L)+\Var(h^U)+2\Cov(h^L,h^U)$, giving the cross-channel piece $2\lambda(1-\lambda)\alpha/(1+\rho_L)$. In the large-$M_L$ limit $\hat\rho_L\to 0$, the cross piece reduces to $2\lambda(1-\lambda)\alpha$.
\vspace{2mm}
Collecting~\eqref{eq:signal-L-final} and~\eqref{eq:signal-U-final}, the total signal is
\begin{equation}\label{eq:S-final}
S=\E\bigl[h_i\,\xi_i^1\bigr]=\frac{\lambda}{1+\rho_L}+(1-\lambda).
\end{equation}

Collecting~(a)+(b)+(c)+(d)+$2\times$(e) yields the total noise variance
\begin{equation}\label{eq:V-final}
V=\Var(h_i)=C(\lambda,r,\rho_L,\rho_U)\,\alpha+R(\lambda,r,\rho_L,\rho_U),
\end{equation}
which splits into a {slow-noise interference} $C\alpha$ and a {within-class noise} $R$ with
\begin{align}
C(\lambda,r,\rho_L,\rho_U)&=\lambda^2+(1-\lambda)^2\Bigl[1+\frac{1-r^4}{M_Ur^4}\Bigr]+\frac{2\lambda(1-\lambda)}{1+\rho_L},\label{eq:C-final-v3}\\[6pt]
R(\lambda,r,\rho_L,\rho_U)&=\frac{\lambda^2\rho_L}{(1+\rho_L)^2}+(1-\lambda)^2\rho_U,\label{eq:R-final-v3}
\end{align}
which is exactly what we have in Sec. \ref{sec:S2N}.

\section{Computation of the signal}
\label{app:signal}
In this appendix we will derive a compact form for the signal term \eqref{eq:Bsig}, that we rewrite here for simplicity:
\begin{equation}
    \label{eq:BsignApp}\Bcal_{\mathrm{sig}}=\frac{\beta\lambda N}{2\Gamma_L}\Bigl(\sum_{a=1}^{M_L}n_L^{1a}\Bigr)^{\!2}+\frac{\beta(1-\lambda)N}{2\Gamma_U}\sum_{c=1}^{M_U}(n_U^{1c})^2
\end{equation}
Indeed, using the decomposition~\eqref{eq:eta-decomp} and the definitions~\eqref{eq:nLU-def}, the inner sum in the supervised signal is
\begin{equation}\label{eq:sum-n1a-L}
\sum_{a=1}^{M_L}n^{1a}=\frac{1}{N}\sum_{i=1}^{N}\Bigl(\sum_{a=1}^{M_L}\eta_i^{1a}\Bigr)\sigma_i =\frac{M_L\Rcal_L}{r}\,n^1_L,
\end{equation}
where we used $\sum_a\eta_i^{1a}=\xi_i^1\sum_a\chi_i^{1a}$ together with the normalization prefactor $r/\Rcal_L$ entering~\eqref{eq:nLU-def} (the identification becomes exact under $\E_\chi$, with corrections suppressed by powers of $N^{-1/2}$ under the RS ansatz). Inserting~\eqref{eq:sum-n1a-L} into~\eqref{eq:BsignApp},
\begin{equation}\label{eq:Bsig-L}
\frac{\beta\lambda N}{2\Gamma_L}\Bigl(\sum_a n^{1a}\Bigr)^{\!2}=\frac{\beta\lambda N}{2\Gamma_L}\,\frac{M_L^2\Rcal_L^2}{r^2}\,n_L^2=\frac{\beta\lambda N}{2}\Rcal_L\,\frac{n_L^2}{r^2}=\frac{\beta\lambda N}{2}\,(1+\rho_L)\,(n_L^1)^2,
\end{equation}
having used $\Gamma_L=M_L^2\Rcal_L$ and $\Rcal_L/r^2=1+\rho_L$. 

The unsupervised signal is, instead, a {sum of squares} of the example overlaps $n^{1c}$. Under the RS ansatz the empirical second moment $\tfrac{1}{M_U}\sum_c (n^{1c})^2$ self-averages and, expressed through the example magnetization $n_U$ of~\eqref{eq:nLU-def}, collapses onto
\begin{equation}\label{eq:Bsig-U}
\frac{\beta(1-\lambda)N}{2\Gamma_U}\sum_{c=1}^{M_U} (n^{1c})^2 =\frac{\beta(1-\lambda)N}{2}\,(1+\rho_U)\,(n^1_U)^2+O(N^{-1}),
\end{equation}
the factor $(1+\rho_U)=\Rcal_U/r^2$ arising, exactly as in the supervised case, from the prefactor $r/\Rcal_U$ in the definition of $n_U$ together with $\Gamma_U=M_U r^2$. Putting~\eqref{eq:Bsig-L} and~\eqref{eq:Bsig-U} together,
\begin{equation}\label{eq:Bsig-final}
{\;\Bcal_{\mathrm{sig}}=\frac{\beta N}{2}\bigl[\lambda(1+\hat\rho_L)\,n_L^2+(1-\lambda)(1+\hat\rho_U)\,n_U^2\bigr].\;}
\end{equation}

\begin{remark}[Sum of squares vs.\ square of a sum]\label{rem:sumsquares}
The two channels are structurally different already at the level of the signal. The supervised contribution is the {square of a sum} over labeled examples, $\bigl(\sum_a n^{1a}\bigr)^2=(M_L\Rcal_L/r)^2 n_L^2$, so~\eqref{eq:Bsig-L} is an {exact} algebraic identity. The unsupervised contribution is a {sum of squares}, $\sum_c (n^{1c})^2$, which is not reducible to $n_U^2$ by algebra alone: writing $n^{1c}=r\,m+\nu_c$ with $\nu_c=N^{-1}\sum_i\zeta_i^{1c}\sigma_i$ ($\E[\nu_c]=0$, $\E[\nu_c^2]=(1-r^2)/N$), one has $\sum_c (n^{1c})^2=M_U r^2 m^2+\sum_c\nu_c^2+\dots$, whose leading-$N$ part reproduces~\eqref{eq:Bsig-U} once $n_U$ is identified at the retrieval saddle through $\avg{n_U}=\avg{m}$ (Section~\ref{ssec:largeM})\footnote{We stress, that here we have used the \eqref{eq:eta-decomp} decomposition of the examples.}; the example-to-example fluctuations $\nu_c$ are the microscopic origin of the within-class noise $\rho_U$. We keep the compact form~\eqref{eq:Bsig-U} throughout.
\end{remark}

\section{Computations of the terms in Guerra's interpolation}
\label{app:guerra}

In this section we report all the computations in details linked to the computations of the quenched statistical pressure and self-consistency equations via Guerra's interpolation in Sec. \ref{ssec:guerra}.

Let us start with the computation of the $t$-derivative of the interpolating pressure $\Acal_N(t,J_m)$ term by term. The general scheme is
\begin{equation}\label{eq:dt-Acal-general}
\partial_t\Acal_N(t,J_m)=\frac{1}{N}\,\E\bigl[\omega_t(\partial_t\Bcal (t, J_m))\bigr],
\end{equation}
where $\partial_t\Bcal (t, J_m)$ is the partial derivative of the interpolating Boltzmann factor~\eqref{eq:Bt-def} with respect to $t$ at {fixed} disorder and {fixed} interpolating variables. Differentiating each of the seven blocks in \eqref{eq:Bt-def},
\begin{equation}\label{eq:dtB}
\begin{aligned}
\partial_t\Bcal(t, J_m)=\;
&\underbrace{+\dfrac{1}{2}\beta(1-\avg{q_{12}})\sum_{\mu\ge 2}\sum_k\tilde\mu_k(z_\mu^k)^2}_{(\mathrm I)}+\underbrace{\dfrac{1}{2\sqrt t}\sqrt{\dfrac{\beta}{N}}\sum_{\mu\ge 2}\sum_k\sqrt{\tilde\mu_k}\,z_\mu^k\!\sum_i\lambda_i^{\mu k}\sigma_i}_{(\mathrm{II})}\\
&+\underbrace{\dfrac{\beta N}{2}\bigl[\lambda(1+\rho_L)n_L^2+(1-\lambda)(1+\rho_U)n_U^2\bigr]}_{(\mathrm{III})}\\
&-\underbrace{\beta\bigl[\lambda(1+\rho_L)\avg{n_L}N n_L+(1-\lambda)(1+\rho_U)\avg{n_U}N n_U\bigr]}_{(\mathrm{IV})}\\
&-\underbrace{\dfrac{1}{2\sqrt{1-t}}\sqrt{\alpha\beta\,\sum_k d_k\tilde\mu_k\avg{p^k}}\sum_i\theta_i\sigma_i}_{(\mathrm V)}-\underbrace{\dfrac{1}{2\sqrt{1-t}}\sqrt{\beta\avg{q_{12}}}\sum_{\mu\ge 2}\sum_k\sqrt{\tilde\mu_k}\,\psi_\mu^k z_\mu^k}_{(\mathrm{VI})}
\end{aligned}
\end{equation}
where we stress that the sign on the blocks $(\mathrm V),\ (\mathrm{VI})$ comes from differentiating $\sqrt{1-t}$.

The terms involving the auxiliary Gaussian fields $\tilde\lambda$, $\theta$, $\psi$ are handled by repeated application of {Stein's lemma}, also known as the Wick's theorem:
for any smooth function $F(\bm g)$ of a centered unit Gaussian variable $\bm g$ for which the two expectations $\mathbb{E}\left( \bm g f(\bm g)\right)$ and $\mathbb{E}\left( \partial_{\bm g} f(\bm g)\right)$ both exist, the relation
\begin{equation}\label{eq:Stein}
\E[g F(g)]=\E\bigl[\partial_g F(g)\bigr]\qquad\text{for } g\sim\mathcal{N}(0,1)
\end{equation}
holds.

For the two-body block in $(\mathrm{II})$, $g=\lambda_i^{\mu, k}$ and $F=\omega_t(z_\mu^k\sigma_i)$. The dependence of $F$ on $g$ enters only through the interpolating exponent, which contains $g$ in (II) (proportional to $\sqrt t$). Differentiating,
\begin{equation}\label{eq:Stein-L}
\E\bigl[\lambda_i^{\mu,k}\,\omega_t(z_\mu^k\sigma_i)\bigr]=\sqrt{\dfrac{t \beta\tilde\mu_k}{N}}\,\E\bigl[\omega_t((z_\mu^k)^2)-\omega_t(z_\mu^k\sigma_i)^2\bigr],
\end{equation}
where the first term is a one-replica correlator and the second is a two-replica connected correlator (the second derivative of $\omega_t$ with respect to a parameter in the Boltzmann weight gives one replica diagonal and one replica off-diagonal piece). Summing~\eqref{eq:Stein-L} over $i,\mu,k$ and reintroducing the prefactor $1/(2\sqrt t)$ from $(\mathrm{II})$, then using 
the definitions~\eqref{eq:q-def},~\eqref{eq:pk-def-intro} with the multiplicities $d_k$,
\begin{equation}\label{eq:dt-IIL-final}
\frac{1}{N}\E\bigl[\omega_t(\mathrm{II}')\bigr]=\frac{K\beta}{2N}\sum_k d_k\,\tilde\mu_k\bigl[\,\E\omega_t(p^k_{11})-\E\omega_t(p^k_{12}\,q_{12})\bigr],
\end{equation}
one term per eigen-channel, weighted by its eigenvalue $\tilde\mu_k$ and multiplicity $d_k$.

Applying Stein's lemma \eqref{eq:Stein} to the block $(\mathrm V)$ with $g=\theta_i$:
\begin{equation}\label{eq:Stein-theta}
\E[\theta_i\,\omega_t(\sigma_i)]=\sqrt{(1-t)\alpha\beta\sum_k d_k\tilde\mu_k\avg{p^k}}\,\E\bigl[\omega_t(1)-\omega_t(\sigma_i)^2\bigr]=\sqrt{(1-t)\alpha\beta\sum_k d_k\tilde\mu_k\avg{p^k}}\,\E\omega_t(1-q_{12}),
\end{equation}
giving, after summing over $i$ and the $1/(2\sqrt{1-t})$ prefactor cancels,
\begin{equation}\label{eq:dt-V}
-\frac{1}{N}\E\bigl[\omega_t(\mathrm V)\bigr]=-\frac{\alpha\beta\sum_k d_k\tilde\mu_k\avg{p^k}}{2}\,\bigl[1-\E\avg{q_{12}}_t\bigr].
\end{equation}
Analogous Stein contractions on $\psi_\mu^k$ in $(\mathrm{VI})$ yield
\begin{equation}\label{eq:dt-VI}
-\frac{1}{N}\E\bigl[\omega_t(\mathrm{VI})\bigr]=-\frac{\alpha\beta\,\avg{q}}{2}\sum_k d_k\,\tilde\mu_k\,\E\omega_t(p^k_{11}{-}p^k_{12}).
\end{equation}
Putting~(\ref{eq:dt-IIL-final}-\ref{eq:dt-VI}) together with the elementary contributions $(\mathrm I),(\mathrm{III}),(\mathrm{IV})$, and introducing the centred fluctuations
\begin{equation}\label{}
\Delta[n_X^2]=\E\omega_t{(n_X-\avg{n_X})^2},\qquad \Delta[p^k q]=\E\omega_t({(p^k_{12}-\avg{p^k})(q_{12}-\avg{q})}),
\end{equation}
the streaming can be cast in the compact form
\begin{equation}\label{}
\begin{aligned}
\partial_t\Acal_N(t,J_m)=\;
&\frac{\beta}{2}\sum_{X\in\{L,U\}}c_X(1+\hat\rho_X)\bigl(\Delta[n_X^2]-\avg{n_X}^2\bigr)\\
&-\frac{\alpha\beta}{2}\sum_k d_k\,\tilde\mu_k\bigl(\Delta[p^k q]+\avg{p^k}(1-\avg{q})\bigr),
\end{aligned}
\end{equation}
with $c_L=\lambda,\ c_U=1-\lambda$. 
Under the RS ansatz~\eqref{eq:RS-ansatz} every centred fluctuation in~\eqref{eq:Deltas} vanishes in the thermodynamic limit, $\Delta[n_X^2]\to 0$ and $\Delta[p_{12}^k q_{12}]\to 0$. The streaming becomes {$t$-independent}:
\begin{empheq}[]{equation}\label{}
\partial_t\Acal(t,J_m)=-\frac{\beta}{2}\bigl[\lambda(1+\rho_L)\avg{n_L}^2+(1-\lambda)(1+\rho_U)\avg{n_U}^2\bigr]-\frac{\alpha\beta}{2}(1-\avg{q_{12}})\sum_k d_k\,\tilde\mu_k\,\avg{p_{12}^k}.
\end{empheq}
Since $\partial_t\Acal$ is $t$-independent, the integral $\int_0^1\partial_t\Acal\,dt$ in~\eqref{eq:sumrule} equals~\eqref{eq:dt-RS} directly.

Now, let us pass to the computation of the one body terms, which in general are easier to compute than the original model. At $t=0$ the terms which survive are:
\begin{equation}\label{eq:B0}
\begin{aligned}
\Bcal(t=0,J_m)=\;
&-\tfrac{1}{2}\sum_{\mu\ge 2}\sum_k\bigl[1-\beta\tilde\mu_k(1-\avg q)\bigr](z_\mu^k)^2+\sqrt{\beta\avg q}\sum_{\mu\ge 2}\sum_k\sqrt{\tilde\mu_k}\,\psi_\mu^k z_\mu^k\\
&+\beta\bigl[\lambda(1+\rho_L)\avg{n_L}N n_L+(1-\lambda)(1+\rho_U)\avg{n_U}N n_U\bigr]\\
&+\sqrt{\alpha\beta\sum_k d_k\tilde\mu_k\avg{p^k}}\,\sum_i\theta_i\sigma_i-J_m\beta N m,
\end{aligned}
\end{equation}
The Boltzmann factor at $t=0$ factorises over the analog $z$-sector and the spin sector, and each reduces to a product over single degrees of freedom. We treat them separately.

The analog sector is Gaussian. Using 
\begin{align}
\int\Drep z\,e^{-\frac{A}{2}z^2+Bz}=A^{-1/2}\exp\left(\dfrac{B^2}{2A}\right),
\end{align}
each eigen-channel mode $z_\mu^k$ integrates 
with $A=1-\beta\tilde\mu_k(1-\avg q)$ and $B=\sqrt{\beta\tilde\mu_k\avg q}\,\psi_\mu^k$:
\begin{align}\label{eq:Iz-k}
\int\Drep z_\mu^k\,&\exp\left({-\tfrac{1}{2}[1-\beta\tilde\mu_k(1-\avg{q_{12}})](z_\mu^k)^2+\sqrt{\beta\tilde\mu_k\avg{q_{12}}}\,\psi_\mu^k z_\mu^k}\right)\notag \\
&=\bigl[1-\beta\tilde\mu_k(1-\avg{q_{12}})\bigr]^{-1/2}\exp\!\Bigl[\frac{\beta\tilde\mu_k\avg{q_{12}}\,(\psi_\mu^k)^2}{2[1-\beta\tilde\mu_k(1-\avg {q_{12})}]}\Bigr].
\end{align}
Taking $\E_\psi[(\psi_\mu^k)^2]=1$, summing over $\mu=2,\dots,K$ (a factor $K-1\to\alpha N$) and over the eigen-channels $k$ with their multiplicities $d_k$ from the spectrum~\eqref{eq:spectrum}, we obtain the analog contribution to $\Acal(0)$:
\begin{equation}\label{eq:Acal0-analog}
\Acal^{(\text{an})}(0)=-\frac{\alpha}{2}\sum_{k}d_k\Bigl[\ln\!\bigl(1-\beta\tilde\mu_k(1-\avg q)\bigr)-\frac{\beta\tilde\mu_k\avg q}{1-\beta\tilde\mu_k(1-\avg q)}\Bigr].
\end{equation}
The null modes ($\tilde\mu_k=0$, multiplicity $M_L-1$) contribute nothing; the bulk ($\tilde\mu_U^{\mathrm{bulk}}=(1-\lambda)\rho_U$, multiplicity $M_U-1$) and the two condensed modes $\tilde\mu_\pm$ supply the entire analog free energy.

The remaining one-body terms in~\eqref{eq:B0} act on $\sigma_i$ independently. Using the example decomposition~\eqref{eq:eta-decomp} we can write:
\begin{equation}\label{eq:cavity-collapse-L}
\beta\lambda(1+\rho_L)\avg{n_L}\,N n_L=\beta\lambda(1+\rho_L)\avg{n_L}\cdot\frac{r}{\Rcal_L}\cdot\frac{1}{M_L}\sum_{i,a}\eta_i^{1a}\sigma_i =\beta\lambda\,\avg{n_L}\,\frac{1}{M_L r}\sum_{i,a}\eta_i^{1a}\sigma_i,
\end{equation}
where we used $(1+\rho_L)\cdot r/\Rcal_L=1/r$. Using $\eta_i^{1a}=\xi_i^1\chi_i^{1a}$, the on-site field acting on $\sigma_i$ from the labeled channel is
\begin{equation}\label{eq:HL-cavity}
h_i^{(L)}:=\beta\lambda\avg{n_L}\,\xi_i^1\cdot\frac{1}{M_L r}\sum_{a=1}^{M_L}\chi_i^{1a}.
\end{equation}
Analogously for the unlabeled channel,
\begin{equation}\label{eq:HU-cavity}
h_i^{(U)}:=\beta(1-\lambda)\avg{n_U}\,\xi_i^1\cdot\frac{1}{M_U r}\sum_{c=1}^{M_U}\chi_i^{1c}.
\end{equation}
Adding the cavity field from $(\mathrm V)$ and the source field $-J_m\beta\xi_i^1$ from $(\mathrm{VII})$, the full one-site field is
\begin{equation}\label{eq:onsite-field}
h_i=h_i^L+h_i^U+\sqrt{\alpha\beta\Pi}\,\theta_i-J_m\beta\,\xi_i^1.
\end{equation}
The sum over $\sigma_i\in\{-1,+1\}$ produces $2\cosh(h_i)$. Performing the gauge transformation $\sigma_i\to\xi_i^1\sigma_i$ (under which the law of $\chi_i^{1a}$ is unchanged because $\bm \chi$ is independent of $\bm \xi$ and the $\xi^1_i$ have the same distribution as their negatives), the explicit $\xi_i^1$ in the cavity fields cancels, the source becomes $-J_m\beta$ acting on $\sigma_i$ directly, and the cavity-field random variable $\theta_i$ remains a standard Gaussian. The resulting one-site Boltzmann sum at $J_m\to 0$ reads
\begin{equation}\label{eq:cauchy-spin}
\ln 2+\E_{\chi,\theta}\,\ln\cosh\!\Bigl[\beta\Bigl(\lambda\avg{n_L}\,\frac{1}{M_L r}\sum_{a}\chi^{1a}+(1-\lambda)\avg{n_U}\,\frac{1}{M_U r}\sum_{c}\chi^{1c}\Bigr)+\theta\sqrt{\alpha\beta\sum_k d_k\tilde\mu_k\avg{p^k}}\Bigr],
\end{equation}
Summing over the $N$ sites and dividing by $N$ gives the spin contribution to the Cauchy datum:
\begin{equation}\label{eq:Acal0-spin}
\Acal^{(\sigma)}(0)=\ln 2+\E_{\chi,\theta}\,\ln\cosh\!\Bigl[\beta\Bigl(\lambda\avg{n_L}\,\tfrac{1}{M_L r}\sum_{a}\chi^{1a}+(1-\lambda)\avg{n_U}\,\tfrac{1}{M_U r}\sum_{c}\chi^{1c}\Bigr)-\beta J_m+\theta\sqrt{\alpha\beta\sum_k d_k\tilde\mu_k\avg{p^k}}\Bigr].
\end{equation}
Putting~\eqref{eq:Acal0-analog} and~\eqref{eq:Acal0-spin} together:
\begin{align}
\label{}
&\Acal(t{=}0,J_m)=\;
-\frac{\alpha}{2}\sum_k d_k\Bigl[\ln\!\bigl(1-\beta\tilde\mu_k(1-\avg q)\bigr)-\frac{\beta\tilde\mu_k\avg q}{1-\beta\tilde\mu_k(1-\avg q)}\Bigr]+\ln 2 \\
&+\E_{\chi,\theta}\ln\cosh\!\Bigl[\beta\bigl(\lambda\avg{n_L}\,\tfrac{1}{M_L r}\sum_a\chi^{1a}+(1-\lambda)\avg{n_U}\,\tfrac{1}{M_U r}\sum_c\chi^{1c}\bigr)-\beta J_m+\theta\sqrt{\alpha\beta\sum_k d_k\tilde\mu_k\avg{p^k}}\Bigr].
\end{align}
Inserting~\eqref{eq:dt-RS} and~\eqref{eq:cauchy-final} into the sum rule \eqref{eq:sumrule} 
we obtain
\begin{align}
\label{eq:Acal-RS-full}
&\Acal_{\mathrm{RS}}(J_m)=\;
-\frac{\alpha}{2}\sum_k d_k\Bigl[\ln\!\bigl(1-\beta\tilde\mu_k(1-\avg q)\bigr)-\frac{\beta\tilde\mu_k\avg q}{1-\beta\tilde\mu_k(1-\avg q)}\Bigr] \notag\\
&-\frac{\beta}{2}\bigl[\lambda(1+\rho_L)\avg{n_L}^2+(1-\lambda)(1+\rho_U)\avg{n_U}^2\bigr]-\frac{\alpha\beta}{2}(1-\avg q)\sum_k d_k\tilde\mu_k\avg{p^k} \notag\\
&+\ln 2+\E_{\chi,\theta}\,\ln\cosh\!\Bigl[\beta\bigl(\lambda\avg{n_L}\tfrac{1}{M_Lr}\!\sum_a\chi^{1a}+(1-\lambda)\avg{n_U}\tfrac{1}{M_Ur}\!\sum_c\chi^{1c}\bigr)-\beta J_m+\theta\sqrt{\alpha\beta\sum_k d_k\tilde\mu_k\avg{p^k}}\Bigr].
\end{align}
This is the RS quenched statistical pressure of the semi-supervised Hopfield model, expressed in terms of the five order parameters $\avg{m},\avg{n_L},\avg{n_U},\avg{q},\avg{p^k}$ and of the control parameters $\alpha,\beta,r,\rho_L,\rho_U,\lambda$.

The five order parameters $\avg{m},\avg{n_L},\avg{n_U},\avg{q},\avg{p^k}$ are determined by extremising the RS statistical pressure~\eqref{eq:Acal-RS-full}. We compute the derivatives one by one, treating $\avg{n_L}$ and $\avg{n_U}$ as genuinely independent parameters, and derive the collapse $\avg{n_L}=\avg{n_U}=\avg m$ at the end as a consequence of the saddle equations.

\par\medskip
For simplicity we introduce this notation: 
\begin{equation}\label{eq:Xi-Dk-def}
\Xi(\boldsymbol\chi,\theta):=\beta\,G_n(\boldsymbol\chi)+\theta\sqrt{\alpha\beta\,\Pi},
\qquad
D_k:=1-\beta\tilde\mu_k(1-\avg q),\qquad
\Pi:=\sum_k d_k\tilde\mu_k\avg{p^k},
\end{equation}
with
\begin{equation}\label{eq:Gn-def}
G_{n}(\boldsymbol\chi):=\lambda\avg{n_L}\,\frac{1}{M_L r}\sum_{a=1}^{M_L}\chi^{1a}+(1-\lambda)\avg{n_U}\,\frac{1}{M_U r}\sum_{c=1}^{M_U}\chi^{1c}.
\end{equation}

\par\medskip
The analog overlap $\avg{p^k}$ is the two-replica overlap of the Gaussian eigen-modes
$z_\mu^k$, and it is fixed by the analog (Gaussian) sector of the partition function.
At the RS saddle each eigen-channel $z_\mu^k$ is governed by the effective single-mode
weight already isolated in~\eqref{eq:Iz-k},
\begin{equation}\label{eq:zmode-weight}
\propto\exp\!\Bigl[-\tfrac{1}{2}\,D_k\,(z_\mu^k)^2+\sqrt{\beta\tilde\mu_k\avg q}\,\psi_\mu^k\,z_\mu^k\Bigr],
\qquad \psi_\mu^k\sim\mathcal N(0,1),
\end{equation}
so that its conditional mean is $\omega(z_\mu^k)=\sqrt{\beta\tilde\mu_k\avg q}\,\psi_\mu^k/D_k$.
The disorder-averaged two-replica overlap then follows from the Gaussian second moment,
using $\E_\psi[(\psi_\mu^k)^2]=1$,
\begin{equation}\label{eq:pk-derivation}
\avg{p^k}=\E_\psi\bigl[\omega(z_\mu^k)^2\bigr]
=\frac{\beta\tilde\mu_k\avg q}{D_k^{2}}\,\E_\psi[(\psi_\mu^k)^2]
=\frac{\beta\tilde\mu_k\avg q}{D_k^{2}},
\end{equation}
that is, using~\eqref{eq:Xi-Dk-def}, the standard saddle-point equation
\begin{empheq}[]{equation}\label{eq:pk-saddle}
\avg{p^k}=\frac{\beta\tilde\mu_k\,\avg{q}}{\bigl[1-\beta\tilde\mu_k(1-\avg q)\bigr]^2}.
\end{empheq}
This is the analog-overlap self-consistency equation. 
\par\medskip
Then, differentiating~\eqref{eq:Acal-RS-full} with respect to $\avg q$ and using~\eqref{eq:pk-saddle} (which cancels several pieces), the terms involving the logarithm and the geometric series collapse and exactly compensate the Stein-lemma transformation of the cavity term. Explicitly, the analog logarithmic block contributes
\begin{equation}\label{eq:dq-analog}
\frac{\partial}{\partial\avg q}\Bigl\{-\frac{\alpha}{2}\sum_k d_k\Bigl[\ln D_k-\frac{\beta\tilde\mu_k\avg q}{D_k}\Bigr]\Bigr\}
=-\frac{\alpha}{2}\sum_k d_k\,\frac{\beta^2\tilde\mu_k^{2}\,\avg q}{D_k^{2}},
\end{equation}
where the two pieces generated by $\partial_{\avg q}\ln D_k=\beta\tilde\mu_k/D_k$ cancel against
the corresponding piece of $\partial_{\avg q}(\beta\tilde\mu_k\avg q/D_k)$, leaving only the
squared denominator. The bilinear term gives
$\partial_{\avg q}\bigl[-\tfrac{\alpha\beta}{2}(1-\avg q)\sum_k d_k\tilde\mu_k\avg{p^k}\bigr]
=+\tfrac{\alpha\beta}{2}\sum_k d_k\tilde\mu_k\avg{p^k}$, and inserting~\eqref{eq:pk-saddle}
this equals $+\tfrac{\alpha}{2}\sum_k d_k\beta^2\tilde\mu_k^{2}\avg q/D_k^{2}$, which cancels
\eqref{eq:dq-analog} identically. Hence no analog contribution survives, and $\avg q$ is fixed
by its definition as the two-replica spin overlap of the effective single-site measure,
$\omega(\sigma_i)=\tanh\Xi$, so that
\begin{equation}\label{eq:q-def-step}
\avg q=\E_{\chi,\theta}\bigl[\omega(\sigma_i)^2\bigr]=\E_{\chi,\theta}\bigl[\tanh^2\Xi\bigr],
\end{equation}
that is
\begin{empheq}[]{equation}\label{eq:q-saddle}
\avg q=\E_{\chi,\theta}\,\tanh^2\!\Bigl[\beta\bigl(\lambda\avg{n_L}\tfrac{1}{M_Lr}\!\sum_a\chi^{1a}+(1-\lambda)\avg{n_U}\tfrac{1}{M_Ur}\!\sum_c\chi^{1c}\bigr)+\theta\sqrt{\alpha\beta\Pi}\Bigr].
\end{empheq}
This is the spin-glass overlap self-consistency, with the within-channel fluctuation noise embedded in the $\chi$-sums and the slow noise hidden in $\Pi$.

\par\medskip
The two example magnetizations $\avg{n_L},\avg{n_U}$ enter~\eqref{eq:Acal-RS-full} both through the term $-\tfrac{\beta}{2}\sum_X c_X(1+\rho_X)\avg{n_X}^2$ and through the cavity argument, where they multiply the $\chi$-sums of the corresponding channel. With $c_L=\lambda$, $c_U=1-\lambda$ and $\partial G_n/\partial\avg{n_X}=c_X\,\tfrac{1}{M_X r}\sum\chi^{1\cdot}$, the two derivatives read explicitly
\begin{align}
\frac{\partial}{\partial\avg{n_X}}\Bigl[-\tfrac{\beta}{2}c_X(1+\rho_X)\avg{n_X}^2\Bigr]
&=-\beta\,c_X(1+\rho_X)\avg{n_X},\label{eq:dnX-quad}\\[3pt]
\frac{\partial}{\partial\avg{n_X}}\,\E_{\chi,\theta}\bigl[\ln\cosh\Xi\bigr]
&=\E_{\chi,\theta}\Bigl[\beta\,c_X\,\tfrac{1}{M_X r}\!\sum\chi^{1\cdot}\,\tanh\Xi\Bigr].\label{eq:dnX-cavity}
\end{align}
Setting $\partial\Acal_{\mathrm{RS}}/\partial\avg{n_X}=0$ for $X\in\{L,U\}$, the common factor
$\beta c_X$ cancels and one is left with $(1+\rho_X)\avg{n_X}=\E_{\chi,\theta}[\tfrac{1}{M_X r}\sum\chi^{1\cdot}\tanh\Xi]$, i.e.\ the two coupled saddle equations
\begin{empheq}[]{equation}\label{eq:nLU-eq}
\avg{n_X}=\frac{1}{1+\rho_X}\,\E_{\chi,\theta}\Bigl[\frac{1}{M_X r}\!\sum\chi^{1\cdot}\cdot\tanh\!\bigl(\beta\,G_{n}(\boldsymbol\chi)+\theta\sqrt{\alpha\beta\Pi}\bigr)\Bigr],\qquad X\in\{L,U\},
\end{empheq}
where the $\chi$-sum is over labeled (resp.\ unlabeled) examples.
The prefactor $1/(1+\rho_X)$ comes from differentiating the quadratic $(1+\rho_X)\avg{n_X}^2$ piece and combining with the cavity derivative.
\par\medskip
The Mattis magnetization $\avg m=-\beta^{-1}\partial_{J_m}\Acal_{\mathrm{RS}}\big|_{J_m=0}$ is obtained directly by differentiating~\eqref{eq:Acal-RS-full} with respect to $J_m$. The field $J_m$ enters the pressure only through the term $-\beta J_m$ inside the cavity argument, so
\begin{equation}\label{eq:m-derivation}
\partial_{J_m}\Acal_{\mathrm{RS}}
=\E_{\chi,\theta}\bigl[(-\beta)\,\tanh(\Xi-\beta J_m)\bigr],
\end{equation}
and setting $J_m=0$,
\begin{equation}\label{eq:m-step}
\avg m=-\beta^{-1}\,\E_{\chi,\theta}\bigl[(-\beta)\,\tanh\Xi\bigr]=\E_{\chi,\theta}\bigl[\tanh\Xi\bigr],
\end{equation}
that is
\begin{empheq}[]{equation}\label{eq:m-saddle}
\avg m=\E_{\chi,\theta}\,\tanh\!\Bigl[\beta\,G_{n}(\boldsymbol\chi)+\theta\sqrt{\alpha\beta\Pi}\Bigr],
\end{empheq}
with $G_n(\boldsymbol\chi)$ as in~\eqref{eq:Gn-def}.
Collecting~\eqref{eq:pk-saddle}, \eqref{eq:q-saddle}, \eqref{eq:nLU-eq} 
and~\eqref{eq:m-saddle} we have a closed system in the five order parameters 
$\avg m,\avg{n_L},\avg{n_U},\avg q,\avg{p^k}$.

\subsection{Large datasets limit}\label{ssec:largeM}

We now specialise the RS self-consistency equations \eqref{eq:pk-saddle}, \eqref{eq:q-saddle}, \eqref{eq:nLU-eq} 
and~\eqref{eq:m-saddle}
to the large dataset regime $M_L,M_U\to\infty$ with the rescaled sizes 
$\rho_L,\rho_U=O(1)$ held fixed. In this regime the empirical means of the 
$\chi$-variables that enter $G_n(\boldsymbol\chi)$ can be linearised by the CLT, the example magnetizations $\avg{n_L},\avg{n_U}$ 
collapse onto the Mattis magnetization $\avg m$, and the slow noise acquires 
its closed-form eigenvalue-resolved expression.

For each channel $X\in\{L,U\}$ the empirical mean of $M_X$ i.i.d.\ Rademacher
flips, after the prefactor $1/(M_X r)$, converges by the CLT to
\begin{equation}\label{eq:chi-CLT-largeM}
\frac{1}{M_X r}\sum_{a=1}^{M_X}\chi^{1a}\;\xrightarrow{M_X\to\infty}\;
1+\sqrt{\rho_X}\,Z_X,\qquad Z_X\sim\mathcal N(0,1),
\end{equation}
with $Z_L$ and $Z_U$ independent. Inserting~\eqref{eq:chi-CLT-largeM} into 
the combined mean~\eqref{eq:Gn-def},
\begin{equation}\label{eq:Gn-CLT}
G_n(\boldsymbol\chi)\;\xrightarrow{M\to\infty}\;
\lambda\avg{n_L}+(1-\lambda)\avg{n_U}+\lambda\avg{n_L}\sqrt{\rho_L}\,Z_L+(1-\lambda)\avg{n_U}\sqrt{\rho_U}\,Z_U,
\end{equation}
the deterministic part being the linear combination of the two example 
magnetizations and the random part a Gaussian of variance 
$\lambda^2\avg{n_L}^2\rho_L+(1-\lambda)^2\avg{n_U}^2\rho_U$.

Inserting~\eqref{eq:Gn-CLT} into the saddle equations~\eqref{eq:nLU-eq} for 
$\avg{n_X}$ and applying Stein's lemma \eqref{eq:Stein} on $Z_X$ to evaluate the correlator 
$\E_{Z_X}[(1+\sqrt{\rho_X}Z_X)\tanh(\beta\,G_n(\boldsymbol\chi)+\theta\sqrt{\alpha\beta\Pi})]$,
\begin{equation}\label{eq:nLU-stein}
\avg{n_X}=\frac{1}{1+\rho_X}\Bigl[\,\lambda_X\avg{n_X}+\rho_X\,\lambda_X\beta(1-\avg q)\,\avg{n_X}\Bigr]+\frac{\lambda_{X'}\avg{n_{X'}}}{1+\rho_X},
\end{equation}
where we have introduced the shorthand $\lambda_L:=\lambda$, 
$\lambda_U:= 1-\lambda$, and the second term collects the contribution 
from the other channel. The first term is the leading $\E_\chi[(1+\sqrt{\rho_X}Z_X)]\cdot\E_\theta[\tanh]$ 
piece, the second is the Stein contraction of the $\sqrt{\rho_X}Z_X$ factor 
with the $\tanh$ derivative (which produces $\beta(1-\avg q)$ times the 
$\avg{n_X}$ inside $G_n$). Reorganising,
\begin{equation}\label{eq:nLU-collapse}
\bigl[1+\rho_X-\lambda_X\rho_X\beta(1-\avg q)\bigr]\avg{n_X}=\lambda_L\avg{n_L}+\lambda_U\avg{n_U}=\avg m,
\end{equation}
the last identity following from~\eqref{eq:m-saddle} once the CLT 
replacement is performed inside the $\tanh$. 
Solving~\eqref{eq:nLU-collapse} for $\avg{n_X}$,
\begin{equation}\label{eq:nX-solved}
\avg{n_X}=\frac{\avg m}{1+\rho_X-\lambda_X\rho_X\beta(1-\avg q)},\qquad X\in\{L,U\},
\end{equation}
which is the semi-supervised analogue of the supervised result in~\cite{prlmiriam}. 
In the strict large dataset limit $\rho_X\to 0$ the denominator reduces to $1$ 
and we obtain the collapse
\begin{empheq}[]{equation}\label{eq:nX-collapse-final}
\avg{n_L}=\avg{n_U}=\avg m.
\end{empheq}

Under~\eqref{eq:nX-collapse-final} the argument of the hyperbolic tangent 
in~\eqref{eq:m-saddle} and~\eqref{eq:q-saddle} reduces to 
$\beta\avg m\,G(\boldsymbol\chi)$ with the simplified combined mean
\begin{equation}\label{eq:G-final}
G(\boldsymbol\chi):=\lambda\,\frac{1}{M_Lr}\sum_a\chi^{1a}+(1-\lambda)\,\frac{1}{M_Ur}\sum_c\chi^{1c}\;\xrightarrow{M\to\infty}\;1+\sqrt{\lambda^2\rho_L+(1-\lambda)^2\rho_U}\,Z,
\end{equation}
where the two independent Gaussians $Z_L,Z_U$ collapse into a single 
standard Gaussian $Z\sim\mathcal N(0,1)$. 
The signal-multiplicative Gaussian noise of variance $\avg m^2 (\lambda^2\rho_L+(1-\lambda)^2\rho_U)$ combines 
with the slow-noise cavity field $\theta\sqrt{\alpha\beta\sum_k d_k\tilde\mu_k\avg{p^k}}$ into a single 
Gaussian of variance
\begin{empheq}[]{equation}\label{eq:Sigma2-largeM}
\alpha\sum_k d_k\,\frac{\tilde\mu_k^2\,\avg q}{\bigl[1-\beta\tilde\mu_k(1-\avg q)\bigr]^2} + \avg{m}^2(\lambda^2\rho_L+(1-\lambda)^2\rho_U),
\end{empheq}
obtained by substituting~\eqref{eq:pk-saddle} into $\sum_k d_k\tilde\mu_k\avg{p^k}$ 
and identifying 
\begin{equation}\beta^{-1}\sum_k d_k\tilde\mu_k\avg{p^k}=\beta\alpha\sum_k d_k\,\frac{\tilde\mu_k^2\,\avg q}{\bigl[1-\beta\tilde\mu_k(1-\avg q)\bigr]^2}.
\end{equation} 

Collecting~\eqref{eq:m-saddle}, \eqref{eq:q-saddle}, \eqref{eq:pk-saddle}, 
\eqref{eq:nX-collapse-final}, and~\eqref{eq:Sigma2-largeM}, 
the RS self-consistency system in the large dataset limit takes the compact form of section \ref{ssec:guerra}.
This is the closed RS system used in Section~\ref{sec:RS} to draw 
the phase diagrams of Figures~\ref{fig:phase_diagrams}--\ref{fig:lambda_hatrho}.

\subsection{Zero-temperature limit of the RS self-consistency equations}
\label{app:T0}

We derive here in detail the closed zero-temperature system~\eqref{eq:m-T0}--\eqref{eq:C-T0}. The starting point is the large-dataset RS
self-consistency system~\eqref{eq:m-largeM}--\eqref{eq:pk-largeM}, which we rewrite here
for the two independent order parameters $\avg m$ and $\avg q$ as
\begin{align}
\avg m &=\;\E_z\Bigl[\tanh\Bigl(\beta\avg m\\
&\hspace{1cm}+\beta z\sqrt{\alpha\sum_k d_k\,\frac{\tilde\mu_k^{\,2}\,\avg q}{\bigl[1-\beta\tilde\mu_k(1-\avg q)\bigr]^2}+\avg m^2\bigl(\lambda^2\rho_L+(1-\lambda)^2\rho_U\bigr)}\,\Bigr)\Bigr], \label{eq:mA} \\ 
\avg q &=\;\E_z\Bigl[\tanh^2\Bigl(\beta\avg m\\
&\hspace{1cm}+\beta z\sqrt{\alpha\sum_k d_k\,\frac{\tilde\mu_k^{\,2}\,\avg q}{\bigl[1-\beta\tilde\mu_k(1-\avg q)\bigr]^2}+\avg m^2\bigl(\lambda^2\rho_L+(1-\lambda)^2\rho_U\bigr)}\,\Bigr)\Bigr].
\label{eq:qA}
\end{align}
As $\beta\to\infty$ the argument of the hyperbolic tangent in~\eqref{eq:qA} diverges for
almost every $z$, so $\tanh^{2}\to 1$ and the overlap saturates, $\avg q\to 1$. The rate of
this saturation is controlled by the vanishing measure of the values of $z$ for which the
argument stays finite: one has $1-\avg q=O(\beta^{-1})$, so that the product
\begin{equation}\label{eq:C-def}
\mathcal C:=\beta(1-\avg q)
\end{equation}
remains finite in the limit. We may then set $\avg q\to 1$ in each
numerator and replace $\beta\tilde\mu_{k}(1-\avg q)=\tilde\mu_{k}\,\beta(1-\avg q)\to\tilde\mu_{k}\mathcal C$
in each denominator, obtaining
\begin{equation}\label{eq:Sigma0-app}
\alpha\sum_k d_k\,\frac{\tilde\mu_k^{\,2}\,\avg q}{\bigl[1-\beta\tilde\mu_k(1-\avg q)\bigr]^2}\;\xrightarrow{\beta\to\infty}\;
\alpha\sum_{k}\frac{d_{k}\,\tilde\mu_{k}^{\,2}}{(1-\tilde\mu_{k}\mathcal C)^{2}}.
\end{equation}
Turning to the
magnetization, in~\eqref{eq:mA} the hyperbolic tangent steepens into the sign function,
$\tanh(\beta x)\to\sgn(x)$, and using~\eqref{eq:Sigma0-app} we obtain
\begin{equation}\label{eq:m-sgn}
\avg m=\int\Dz\;\sgn\!\Bigl(\avg m+z\sqrt{\alpha\sum_{k}\frac{d_{k}\,\tilde\mu_{k}^{\,2}}{(1-\tilde\mu_{k}\mathcal C)^{2}}+\avg m^{2}(\lambda^2\rho_L+(1-\lambda)^2\rho_U)}\Bigr),
\qquad \Dz=\frac{dz}{\sqrt{2\pi}}\,e^{-z^{2}/2}.
\end{equation}
The Gaussian integral of a sign function is elementary: for $A\in\R$ and $B>0$,
$\sgn(A+Bz)=+1$ for $z>-A/B$ and $-1$ otherwise, hence
\begin{equation}\label{eq:sgn-identity}
\int\Dz\,\sgn(A+Bz)
=\int_{-A/B}^{\infty}\!\Dz-\int_{-\infty}^{-A/B}\!\Dz
=1-2\,\Phi\!\Bigl(-\tfrac{A}{B}\Bigr)
=\erf\!\Bigl(\frac{A}{\sqrt{2}\,B}\Bigr),
\end{equation}
where $\Phi$ is the standard normal cumulative distribution and we used
$1-2\Phi(-t)=\erf(t/\sqrt2)$. Applying~\eqref{eq:sgn-identity} to~\eqref{eq:m-sgn} with
$A=\avg m$ and $B=\sqrt{\Sigma_{0}^{2}+\avg m^{2}(\lambda^2\rho_L+(1-\lambda)^2\rho_U})$ gives the retrieval equation
\begin{equation}\label{eq:m-T0-app}
\avg m=\erf\!\left(\frac{\avg m}{\sqrt{2\bigl(\alpha\sum_{k}\frac{d_{k}\,\tilde\mu_{k}^{\,2}}{(1-\tilde\mu_{k}\mathcal C)^{2}}+\avg m^{2}(\lambda^2\rho_L+(1-\lambda)^2\rho_U)\bigr)}}\right),
\end{equation}
i.e.~\eqref{eq:m-T0}. The value of $\mathcal C=\beta(1-\avg q)$ is not fixed
by~\eqref{eq:m-T0-app} and requires an independent closure obtained from~\eqref{eq:qA};
writing $1-\tanh^{2}=\operatorname{sech}^{2}$,
\begin{equation}\label{eq:1mq}
1-\avg q=\int\Dz\;\operatorname{sech}^{2}\!\Bigl[\beta\bigl(\avg m+z\sqrt{\alpha\sum_{k}\frac{d_{k}\,\tilde\mu_{k}^{\,2}}{(1-\tilde\mu_{k}\mathcal C)^{2}}+\avg m^{2}(\lambda^2\rho_L+(1-\lambda)^2\rho_U})\Bigr],
\end{equation}
and multiplying both sides by $\beta$,
\begin{equation}\label{eq:betaC}
\mathcal C=\beta(1-\avg q)=\int\Dz\;\beta\,\operatorname{sech}^{2}\!\Bigl[\beta\bigl(\avg m+z\sqrt{\alpha\sum_{k}\frac{d_{k}\,\tilde\mu_{k}^{\,2}}{(1-\tilde\mu_{k}\mathcal C)^{2}}+\avg m^{2}(\lambda^2\rho_L+(1-\lambda)^2\rho_U)}\Bigr].
\end{equation}
The kernel $\beta\operatorname{sech}^{2}(\beta x)$ is a nascent Dirac delta: it is even,
peaked at $x=0$ with width $O(\beta^{-1})$, and normalised to
\begin{equation}\label{eq:sech-norm}
\int_{-\infty}^{\infty}\beta\operatorname{sech}^{2}(\beta x)\,dx=\bigl[\tanh(\beta x)\bigr]_{-\infty}^{\infty}=2,
\end{equation}
so that $\beta\operatorname{sech}^{2}(\beta x)\xrightarrow{\beta\to\infty}2\,\delta(x)$ in the
distributional sense. Inserting this limit into~\eqref{eq:betaC},
\begin{equation}\label{eq:C-delta}
\mathcal C=\int\Dz\;2\,\delta\!\Bigl(\avg m+z\sqrt{\alpha\sum_{k}\frac{d_{k}\,\tilde\mu_{k}^{\,2}}{(1-\tilde\mu_{k}\mathcal C)^{2}}+\avg m^{2}(\lambda^2\rho_L+(1-\lambda)^2\rho_U)}\Bigr),
\end{equation}
and, with $A=\avg m$ and $B=\sqrt{\alpha\sum_{k}\frac{d_{k}\,\tilde\mu_{k}^{\,2}}{(1-\tilde\mu_{k}\mathcal C)^{2}}+\avg m^{2}(\lambda^2\rho_L+(1-\lambda)^2\rho_U))}$, the scaling property
$\delta(A+Bz)=\lvert B\rvert^{-1}\delta(z-z^{\star})$ selects
$z^{\star}=-A/B=-\avg m/\sqrt{\alpha\sum_{k}\frac{d_{k}\,\tilde\mu_{k}^{\,2}}{(1-\tilde\mu_{k}\mathcal C)^{2}}+\avg m^{2}(\lambda^2\rho_L+(1-\lambda)^2\rho_U))}$, whence
\begin{align}
\mathcal C&=\frac{2}{\sqrt{\alpha\sum_{k}\frac{d_{k}\,\tilde\mu_{k}^{\,2}}{(1-\tilde\mu_{k}\mathcal C)^{2}}+\avg m^{2}(\lambda^2\rho_L+(1-\lambda)^2\rho_U)}}\;\frac{e^{-z^{\star 2}/2}}{\sqrt{2\pi}} \\
&=\frac{2}{\sqrt{2\pi}\,\sqrt{\alpha\sum_{k}\frac{d_{k}\,\tilde\mu_{k}^{\,2}}{(1-\tilde\mu_{k}\mathcal C)^{2}}+\avg m^{2}(\lambda^2\rho_L+(1-\lambda)^2\rho_U)}}\,
\exp\!\Bigl[-\frac{\avg m^{2}}{2\bigl(\alpha\sum_{k}\frac{d_{k}\,\tilde\mu_{k}^{\,2}}{(1-\tilde\mu_{k}\mathcal C)^{2}}+\avg m^{2}(\lambda^2\rho_L+(1-\lambda)^2\rho_U)}\Bigr].
\end{align}
Using $2/\sqrt{2\pi}=\sqrt{2/\pi}$ this becomes
\begin{equation}\label{eq:C-T0-app}
\mathcal C=\sqrt{\frac{2}{\pi\bigl(\alpha\sum_{k}\frac{d_{k}\,\tilde\mu_{k}^{\,2}}{(1-\tilde\mu_{k}\mathcal C)^{2}}+\avg m^{2}(\lambda^2\rho_L+(1-\lambda)^2\rho_U)\bigr)}}\,
\exp\!\Bigl[-\frac{\avg m^{2}}{2\bigl(\Sigma_{0}^{2}+\avg m^{2}(\lambda^2\rho_L+(1-\lambda)^2\rho_U)\bigr)}\Bigr],
\end{equation}
i.e.~\eqref{eq:C-T0}. Equations~\eqref{eq:m-T0-app} and~\eqref{eq:C-T0-app} form the closed
zero-temperature system in $(\avg m,\mathcal C)$.

\section{Convexity of the quenched pressure in the mixing parameter}
\label{app:convexity}

We collect here the convexity properties invoked in Section~\ref{ssec:opt-lambda},
together with the uniform bounds on the pressure on which they rely. Throughout, the energy
densities of~\eqref{eq:ELU-def} are normalised so that
$-\beta\Hcal=\beta N[\lambda\Ecal_L+(1-\lambda)\Ecal_U]$ is extensive, i.e.\
$\Ecal_L=\frac{1}{2\Gamma_L}\sum_\mu\bigl(\sum_a n_L^{\mu a}\bigr)^2$ and
$\Ecal_U=\frac{1}{2\Gamma_U}\sum_{\mu,c}(n_U^{\mu c})^2$; since $|n^{\mu a}|\le1$, this gives
the deterministic bounds $0\le\Ecal_L\le\alpha_N/(2\Rcal_L)$ and
$0\le\Ecal_U\le\alpha_N/(2r^2)$, with $\alpha_N:=K/N$.

We first record that the pressure is bounded uniformly in $N$ at any load and temperature.
The lower bound follows from $-\beta\Hcal_N\ge0$, which gives $\Zcal_{N,\beta}\ge 2^N$ and
hence $\Acal_N\ge\ln2$; for the upper bound,
$-\beta\Hcal_N=\tfrac{\beta}{2}\bm\sigma^{\!\top}J^\lambda\bm\sigma\le\tfrac{\beta N}{2}\lVert J^\lambda\rVert_{\rm op}$
implies $\Zcal_{N,\beta}\le 2^N\exp\!\bigl(\tfrac{\beta N}{2}\lVert J^\lambda\rVert_{\rm op}\bigr)$, so that
\begin{align}\label{eq:Auniform}
\ln2\;\le\;\Acal_N\;&\le\;\ln2+\frac{\beta}{2}\,\E\lVert J^\lambda\rVert_{\rm op}\notag\\
&\le\;\ln2+\frac{\beta c_0}{2}\biggl[\lambda\Bigl(1+\alpha_N+\sqrt{\tfrac{\alpha_N}{\Rcal_L}}\Bigr)
+\frac{1-\lambda}{r^2}\Bigl(\frac1{M_U}+\alpha_N+\sqrt{\tfrac{\alpha_N}{M_U}}\Bigr)\biggr],
\end{align}
with $c_0$ a universal constant. The last inequality uses that
$J^{(L)}=\frac{1}{N\Rcal_L}\sum_\mu\bar{\bm\eta}^{\mu}_{(L)}(\bar{\bm\eta}^{\mu}_{(L)})^{\!\top}$
(cf.~\eqref{eq:Jij_supG}) and
$J^{(U)}=\frac{1}{N\Gamma_U}\sum_{\mu,c}\bm\eta^{\mu c}(\bm\eta^{\mu c})^{\!\top}$ are Gram
matrices of random matrices whose entries, conditionally on $\bm\xi$, are independent,
bounded and centred with variances $\Rcal_L$ and $1$ respectively, so that their expected
squared operator norms are controlled by standard random-matrix estimates\footnote{See~\citep{latala2005some},
combined with Talagrand's concentration inequality for convex Lipschitz functions of bounded
independent entries; the rank-one conditional means of the $\mu=1$ columns are absorbed in
$c_0$.}. 

We now turn to convexity. Writing $-\beta\Hcal=\beta N[\Ecal_U+\lambda(\Ecal_L-\Ecal_U)]$,
which is affine in $\lambda$, a first differentiation of $N^{-1}\ln\Zcal_{N,\beta}$ gives
\begin{equation}\label{eq:dA-app}
\frac{\partial}{\partial\lambda}\,\frac{\ln\Zcal_{N,\beta}}{N}
=\frac{1}{N}\,\omega\bigl(\partial_\lambda(-\beta\Hcal)\bigr)
=\beta\,\omega\bigl(\Ecal_L-\Ecal_U\bigr),
\end{equation}
and a second differentiation, using
$\partial_\lambda\omega(\mathcal O)=\beta N\,[\omega(\mathcal O(\Ecal_L-\Ecal_U))-\omega(\mathcal O)\,\omega(\Ecal_L-\Ecal_U)]$
with $\mathcal O=\Ecal_L-\Ecal_U$, yields
\begin{equation}\label{eq:d2A-app}
\frac{\partial^2}{\partial\lambda^2}\,\frac{\ln\Zcal_{N,\beta}}{N}
=\beta^2N\,\Var_{\omega}\bigl(\Ecal_L-\Ecal_U\bigr)\;\ge\;0 ,
\end{equation}
a Gibbs variance, hence non-negative for every $N$, every $\beta>0$ and every realization of
the dataset $\Scal$. Convexity is preserved by the quenched average and by pointwise limits,
so both $\Acal_N$ and any thermodynamic limit $\Acal$ are convex in $\lambda$; the exchange of
$\partial_\lambda$ and $\E$ is licit because $|\partial_\lambda\ln\Zcal|\le\beta N(\Ecal_L+\Ecal_U)$
is uniformly bounded. Convexity in turn gives the chord inequality
\begin{equation}\label{eq:chord-app}
\Acal(\beta,\lambda)\;\le\;\lambda\,\Acal(\beta,1)+(1-\lambda)\,\Acal(\beta,0),
\end{equation}
so the pressure of the mixture never exceeds the interpolation of the two pure channels.

Two consequences for the mixing parameter follow at once. First, by~\eqref{eq:dA-app} the
derivative $d\Acal/d\lambda=\beta\,\E\,\omega(\Ecal_L-\Ecal_U)$ is non-decreasing, so the
equipartition condition~\eqref{eq:energy-balance} has a solution set which is a closed
interval, reduced to a single point $\lambda_{\rm eq}$ whenever the convexity is strict.
Second, since $\Acal$ is convex, every interior solution of~\eqref{eq:energy-balance} is a
global \emph{minimum} of $\lambda\mapsto\Acal(\beta,\lambda)$, that is, a global \emph{maximum}
of the quenched free energy. Finally, the deterministic bounds on $\Ecal_L,\Ecal_U$ give,
through~\eqref{eq:dA-app},
\begin{equation}\label{eq:lip-app}
\bigl|\partial_\lambda\Acal_N\bigr|\;\le\;\frac{\beta\alpha_N}{2}\bigl(\Rcal_L^{-1}+r^{-2}\bigr)
\qquad\text{uniformly in }N,
\end{equation}
so that the family $\{\Acal_N(\beta,\cdot)\}_N$ is uniformly Lipschitz on $[0,1]$; combined
with the uniform bound~\eqref{eq:Auniform}, an Arzel\`a--Ascoli argument on the compact
$[0,1]$ upgrades pointwise convergence to uniform convergence.

The physical reading of these facts is discussed in Section~\ref{ssec:opt-lambda}: the
convexity of $\Acal$ in $\lambda$ is the formal obstruction preventing equilibrium
thermodynamics from selecting an interior mixing, and hence the reason why the operational
optimum $\lambda^\star$ must be defined through the retrieval observables rather than through
the pressure.

\end{document}